\documentclass[useAMS,usenatbib]{mn2e}
\usepackage{times}
\usepackage{graphics,epsfig}
\usepackage{graphicx}
\usepackage{amssymb}
\usepackage{lscape}

\title[GMRT observations of ``normal" FR\,II radio galaxies]
{``Normal" FR\,II Radio Galaxies as a Probe of the Nature
  of X-Shaped Radio Sources} 
\author[D. V. Lal, M. J. Hardcastle and R. P. Kraft]{Dharam V. Lal$^{1,2}$\thanks{
E-mail: dharam@mpifr-bonn.mpg.de}, Martin J. Hardcastle$^3$ and Ralph P. Kraft$^4$ \\
\\
$^1$National Centre for Radio Astrophysics (NCRA--TIFR),
Pune University campus, Ganeshkhind, Pune - 411 007, India \\
$^2$Max-Planck-Institut f\"ur Radioastronomie,
Auf dem H\"ugel 69, 53121 Bonn, Germany \\
$^3$School of Physics, Astronomy \& Mathematics, Univ of Hertfordshire,
College Lane, Hatfield, AL10 9AB, UK \\
$^4$Harvard--Smithsonian Center for Astrophysics, 60 Garden Street, Cambridge,
MA 02138, USA}
\begin{document}

\date{Draft of \today}

\pagerange{\pageref{firstpage}--\pageref{lastpage}} \pubyear{2008}

\maketitle

\label{firstpage}

\begin{abstract}
We present a multiwavelength radio study of a sample of nearby
Fanaroff-Riley class II (FR\,II) radio galaxies,
matched with the sample of known X-shaped radio sources
in size, morphological properties and redshift, using new Giant
Metrewave Radio Telescope (GMRT) data and archival data from the Very
Large Array (VLA). Our principal aim in this paper is to provide a
control sample for earlier studies of samples of `X-shaped' radio
sources, which have similar luminosities and small-scale radio
structures to our targets but exhibit large-scale extensions to their
lobes that more typical FR\,II sources lack; earlier spectral work with
the GMRT has suggested that these `wings' sometimes have flat spectral
indices at low frequencies, in contrast to expectations from models
in which the wings are formed hydrodynamically or by jet
reorientation. In our new observations we find that almost all of our
target FR\,II radio galaxies show standard spectral steepening as a
function of distance from the hotspot at the low frequencies (610~MHz and
240 MHz) provided by the GMRT data, even when transverse extensions to
the lobes are present. However, one source, 3C\,321, has a
low-surface-brightness extension to one lobe that shows a flatter
spectral index than the high-surface-brightness hotspots$/$lobes, as
found in X-shaped sources.
\end{abstract}


\section{Introduction}
\label{introduction}

A small but significant fraction of low-luminosity radio galaxies
have a pair of large low-surface-brightness lobes oriented at an angle
to the high-surface-brightness `active' radio lobes, giving the total
source an `X' shape.
This peculiar and small subclass of extragalactic radio sources
is commonly referred to as X-shaped, or `winged' sources.
Typically, both sets of lobes are symmetrically aligned through
the centre of the associated elliptical host galaxy and are
approximately equal in linear extent.
\citet{MerrittEkers} noted that the majority of these sources
are of Fanaroff-Riley type II (FR\,II) \citep{FanaroffRiley}
and the rest are either FR\,Is or have an intermediate classification.

Several authors have attempted to explain the unusual structure in
X-shaped sources. One proposal is that these objects are produced by a
central engine that has been reoriented, perhaps due to a minor merger
(\citealt{MerrittEkers}; \citealt{Dennett}; \citealt{Krishna}).
Alternatively, they may also result from two pairs of jets, which are
associated with a pair of unresolved AGNs (\citealt{LalRao2005},
2007), although this model does not as yet explain (i) why is
there always only one pair of hotspots and (ii) why, when a jet is
observed, is it always in the same pair of lobes? Additional models
for the morphologies of the X-shaped sources include a hydrodynamic
origin (\citealt{LeahyWilliams}; \citealt{Worralletal};
\citealt{Capettietal}; \citealt{Kraftetal}) while some authors have
suggested a conical precession of the jet axis (\citealt{Rees};
\citealt{Parmaetal}; \citealt{Macketal}). See \citet{LalRao2007} and
\citet{Cheung2007} for more detailed discussion of the competing
models.

\begin{table*}
\caption{The observing log for all the observed sample radio sources.}
\centering
\begin{tabular}{l|cccrcllcc}
\hline 
       & RA & Dec & Observing & \multicolumn{2}{c}{Bandwidth} & Centre & \multicolumn{2}{c}{Calibrator} & t$_{\rm integration}$ \\
       & & & date & \multicolumn{2}{l}{Nominal ~~~Effective} & frequency & flux & phase & (on-source) \\
       &\multicolumn{2}{c}{(J2000)}& & \multicolumn{3}{c}{---------- \hfill 610~MHz~/~240~MHz \hfill ----------} & density &  & (hour) \\
\hline 
3C\,033   &01:08:50.5&$+$13:18:31&16 Jun 2005 &16 / 8& 14.25 / 6.750& 606.25 / 237.69 & 3C\,048 & 3C\,48      & 3.51 \\
3C\,098   &03:58:54.4&$+$10:26:03&16 Jun 2005 &16 / 8& 14.25 / 6.750& 606.25 / 240.56 & 3C\,147 & 0521$+$166 & 3.58 \\
3C\,285   &13:21:17.8&$+$42:35:15&16 Jun 2005 &16 / 8& 14.25 / 6.750& 606.25 / 240.06 & 3C\,147 & 3C\,286     & 2.72 \\
3C\,321   &15:31:43.4&$+$24:04:19&30 May 2005 &16 / 8& 14.25 / 6.750& 606.25 / 240.19 & 3C\,048 & 1419$+$064 & 3.58 \\
3C\,382   &18:35:02.1&$+$32:41:50&16 Jun 2005 &16 / 8& 14.25 / 6.750& 606.25 / 236.44 & 3C\,147 & 1829$+$487 & 2.85 \\
3C\,390.3 &18:42:09.0&$+$79:46:17&29 May 2005 &16 / 6& 14.25 / 4.375& 606.25 / 240.12 & 3C\,048 & 1459$+$716 & 4.12 \\
3C\,388   &18:44:02.4&$+$45:33:30&29 May 2005 &16 / 6& 14.25 / 4.375& 606.25 / 237.19 & 3C\,147 & 1459$+$716 & 3.73 \\
3C\,452   &22:45:48.8&$+$39:41:16&30 May 2005 &16 / 8& 14.25 / 6.750& 606.25 / 240.06 & 3C\,048 & 2350$+$646 & 4.47 \\
\hline
\end{tabular}
\label{observ}
\end{table*}

Spectral studies of the lobes of X-shaped sources provide a crucial
test of the different models for these objects. In either the
hydrodynamic, reorientation or precession models, the `wings' -- which
never show evidence for compact jet-related or hotspot structures in
the radio -- would be expected to contain older plasma than the lobes
that do show hotspots and jets, assumed to be the currently active
lobes. In standard synchrotron ageing models (e.g.
\citealt{JaffePerola}) this would give rise to steeper radio spectra.
In early work in this area, \citet{Dennett} showed that there was no
evidence for spectral ageing at high frequencies in the wings of two
well-studied X-shaped sources, 3C\,223.1 and 3C\,403. More recently
\citet{LalRao2007} showed that even at low frequencies a significant
fraction of a sample of X-shaped sources had wings with flatter, or at
least comparable, spectral indices to those in the brighter `active
lobes'. This motivated their suggestion that these sources could be
powered by a pair of associated, unresolved AGN. Another possibility
is simply that our current understanding of spectral ageing in radio
lobes, particularly at low frequencies, is incorrect.

In this paper we present Giant Metrewave Radio Telescope (GMRT) and
Very Large Array (VLA) observations of a control sample of
low-redshift {\it normal} FR\,II radio galaxies. These objects lack the
twin extended wings that are characteristic of the X-shaped sources,
though many of them have low-luminosity transverse extensions of the
lobes on smaller scales. Using observations carried out in the same
manner as those of \citet{LalRao2007}, we are able to investigate
whether standard spectral ageing models provide an adequate
description of these normal FR\,II sources, and to discuss the
implications of our results for models of X-shaped sources.

Throughout the paper we define the spectral index $\alpha$ in the
sense that $S_\nu \propto \nu^\alpha$.

\section{Sample}
\label{sample}

Our sample (Table~\ref{observ}) is selected on the basis of
similarity with known X-shaped sources and consists of all nearby ($z
< 0.1$) normal FR\,II sources from the 3CRR catalogue. These sources
have radio luminosities similar to that of the X-shaped sources, which
lie close to the FR\,I/FR\,II divide. We impose an angular size cutoff
(based on high-frequency radio maps) on the target sample to exclude
the giant radio galaxies and ensure that our sample sources are of
similar angular size to typical X-shaped sources.
In addition, many of the sources show diffuse, low-surface-brightness
transverse extensions of the lobes close to the core,
implying that the sample sources could be similar to known X-shaped sources
but with the wings being shorter either as a result of projection or
(in hydrodynamical models) because they have not yet grown to lengths
comparable to those of the active lobes.

The luminosity and size cutoffs imposed ensure that our sample is a
good qualitative match to the sample of known X-shaped sources,
mentioned in \citet{MerrittEkers} and compiled by \citet{LeahyParma},
which were observed by \citet{LalRao2007} using the GMRT.

\section{GMRT Observations}
\label{observations}

\begin{table*}
\centering
\caption{The total intensity for all the sources.
The total flux densities quoted are in Jy along with corresponding error-bars
(1$\sigma$). The 240 MHz and 610 MHz are our GMRT measurements. The labels denote:
$^a$\citet{Kellermann} and \citet{LaingPeacock};
$^b$\citet{Ekers1969} and \citet{Kuhretal1981};
$^c$\citet{Kuhretal1981};
$^d$Green Bank, Northern Sky Survey \citep{WhiteBecker,BeckerWhite};
$^e$\citet{Collaetal} and \citet{LaingPeacock};
$^f$\citet{Ficarraetal}.
}
\begin{tabular}{l|rrrrrrrr}
\hline
\\
         & 178~MHz            & 240~MHz         & 408~MHz            & 610~MHz        & 750~MHz           & 1400~MHz           & 2695~MHz           & 4850~MHz \\
\\
\hline
\\
3C\,33    & 59.30$\pm$2.97$^a$ & 44.97 $\pm$0.12 &29.60 $\pm$5.44$^b$ &23.84 $\pm$0.09 &20.00$\pm$1.00$^c$ &12.82 $\pm$1.92$^d$ & 7.92 $\pm$0.40$^c$ & 5.04 $\pm$0.76$^c$ \\
3C\,98    & 51.45$\pm$2.57$^a$ & 32.27 $\pm$0.09 &24.20 $\pm$4.63$^b$ &19.82 $\pm$0.03 &16.00$\pm$0.80$^c$ &~9.90 $\pm$0.50$^c$ & 7.01 $\pm$0.35$^c$ & 4.97 $\pm$0.25$^c$ \\
3C\,285   & 11.40$\pm$1.71$^c$ & ~8.64 $\pm$0.06 &                    &~4.44 $\pm$0.01 & 3.14$\pm$0.17$^a$ &~2.16 $\pm$0.32$^d$ & 1.23 $\pm$0.06$^c$ & 0.58 $\pm$0.09$^d$ \\
3C\,321   & 14.72$\pm$1.47$^a$ & 11.71 $\pm$0.07 &~8.64 $\pm$0.71$^c$ &~6.81 $\pm$0.03 &~5.90$\pm$0.29$^c$ &~3.57 $\pm$0.08$^a$ & 2.05 $\pm$0.04$^a$ & 1.21 $\pm$0.12$^c$ \\
3C\,382   &                    & 21.93 $\pm$0.05 &14.70 $\pm$0.74$^e$ &11.10 $\pm$0.02 &~9.29$\pm$0.50$^a$ &~5.60 $\pm$0.84$^c$ & 3.49 $\pm$0.17$^c$ & 2.20 $\pm$0.07$^a$ \\
3C\,390.3 & 52.70$\pm$4.20$^c$ & 41.05 $\pm$0.14 &                    &19.04 $\pm$0.07 &17.60$\pm$0.90$^c$ &11.26 $\pm$0.58$^a$ & 6.76 $\pm$0.34$^c$ & 4.48 $\pm$0.22$^c$ \\
3C\,388   & 26.81$\pm$1.34$^a$ & 20.60 $\pm$0.16 &14.42 $\pm$0.29$^f$ &11.14 $\pm$0.10 &~9.80$\pm$0.50$^c$ &~5.74 $\pm$0.24$^a$ & 3.14 $\pm$0.05$^c$ & 1.76 $\pm$0.09$^c$ \\
3C\,452   & 59.30$\pm$2.97$^a$ & 48.02 $\pm$0.04 &32.67 $\pm$2.87$^c$ &21.73 $\pm$0.02 &19.32$\pm$0.44$^a$ &10.54 $\pm$0.27$^a$ & 5.93 $\pm$0.08$^a$ & 3.24 $\pm$0.32$^c$ \\
\\
\hline
\end{tabular}
\label{flux_density}
\end{table*}

We adopted an observing strategy nearly identical to our earlier
observations for X-shaped radio sources \citep{LalRao2007}. The key
difference was that we observed two sources in each full synthesis
observing run, typically of $\sim$10~h, as against one source during
our earlier observations. The 240 MHz and 610~MHz feeds of GMRT are
coaxial feeds and therefore it was possible to perform simultaneous
dual frequency observations at these two frequencies; accordingly we
observed all the FR\,II sources in our sample at these two frequencies,
using the GMRT in the standard spectral line mode with a spectral
resolution of 125 kHz (program ID 08MHa01). The primary beams are
$\sim$108 arcmin and $\sim$ 43 arcmin at 240 MHz and 610~MHz,
respectively. Table~\ref{observ} gives the details of the
observations.

The GMRT has a hybrid configuration \citep{Swarupetal} with 14 of
its 30 antennas located in a central compact array with size $\sim$1.1~km
and the remaining antennas distributed in a roughly `Y' shaped
configuration, giving a maximum baseline length of $\sim$25~km.
The baselines obtained from antennas in the central
square are similar in length to the VLA~$D$-array
configuration, while the baselines between the arm antennas are
comparable in length to the VLA~$B$-array configuration.
The maximum and minimum baseline lengths for the VLA~$D$-array
configuration, and for the VLA~$B$-array configuration are 1.03~km and
0.035~km, and 11.4~km and 0.21~km, respectively without foreshortening.  The
typical maximum and minimum baseline lengths for the GMRT during observations
were $\sim$17.6~km and $\sim$0.12~km, and the actual maximum and minimum
baseline lengths without foreshortening are $\sim$26.0~km and
$\sim$0.1~km.  Hence, a single observation with the GMRT
samples the ($u,v$) plane adequately on both short and long baselines,
and provides good angular resolution of
$\sim$5~arcsec and $\sim$12~arcsec at 610~MHz and 240~MHz, respectively.
The array can map detailed
source structure with reasonably good sensitivity.

\section{GMRT Data Reduction}
\label{data_reduction}

The visibility data were converted to FITS and analyzed using {\sc
aips} in the standard way. The flux density calibrators 3C~48, 3C~147 and
3C~286 were observed, depending on their availability, at the
beginning or at the end, both as amplitude calibrators and to
estimate and correct for the bandpass shape. For the
flux-density scale we used an extension of the \citet{Baarsetal} scale to low
frequencies, using the coefficients in the {\sc aips} task `SETJY'.
Secondary phase calibrators were observed at intervals of
$\sim$30~min. The error in the estimated flux density, both due to
calibration and systematic errors, is $\lesssim$ 5 per cent. As is standard
at these low frequencies, the data suffered from scintillations and
intermittent radio frequency interference (RFI). In addition to normal
editing of the data, we identified and flagged
time ranges affected by scintillation and channels affected by
RFI, after which the central channels were averaged
using the {\sc aips} task `SPLAT' to reduce the data volume. To avoid
bandwidth smearing, the effective band at 240 MHz and 610~MHz was reduced
to 6 channels and 3 channels, respectively.

We used the {\sc aips} task `IMAGR' to map the full field at both
frequencies, using 49 facets spread across a $\sim$1.8$^\circ
\times1.8^\circ$ field at 240~MHz and 9 facets covering slightly less
than a 0.7$^\circ\times0.7^\circ$ field at 610~MHz. To obtain
high-resolution images that were also sensitive to extended structure,
we employed the SDI CLEANing algorithm \citep{Steeretal}. We used
uniform weighting and the 3D option for $w$-term correction throughout
our analysis. The presence of a large number of point sources in the
field allowed us to do phase self-calibration to improve the image.
After two or three cycles of phase self-calibration, a final
self-calibration of both amplitude and phase was made and a final
image was produced. At each iteration of self-calibration, the Fourier
transform of the image and the visibilities were compared to check for
improvement in the source model. The final images were stitched
together using the {\sc aips} task `FLATN' and corrected for the
primary beam of the GMRT antennas.

\section{VLA data}

To complement our GMRT observations we obtained VLA images at 1.4 GHz
to 1.5 GHz. The majority of these were taken from the online 3CRR
Atlas\footnote{Available at http://www.jb.man.ac.uk/atlas/ .} which
provides well-calibrated, well-sampled images for most of our targets.
However, superior maps were available to us for two objects, 3C~285
(the images of \citep{Hetal2007}) and 3C~321 (image from
\citealt{Evans2008}). For 3C~98 the 3CRR Atlas provides only a 5~GHz
image; we extracted data from the VLA archive in B, C and D
configurations (observing IDs and dates are respectively AB403 on 1986
Jul 17, AB403 on 1986 Oct 29 and AR440 on 2000 Sep 09) and reduced
them in {\sc aips} in the standard way to make a map suitable for our
analysis. The VLA maps had a resolution of $\sim$4 arcsec for all our
target sources, which is comparable to the GMRT resolution at 610 MHz.

\section{Results}
\label{results}

The radio images, shown in Figs.~\ref{full_syn_33} to~\ref{full_syn_452},
have good ($u,v$) coverage, angular resolutions
of $\sim$4 arcsec, $\sim$5 arcsec and
$\sim$13 arcsec, and 
rms~noise in the maps in the range $\sim$0.02 to 0.7 mJy~beam$^{-1}$,
$\sim$0.2 to 2.1 mJy~beam$^{-1}$ and $\sim$1.5--13.9 mJy~beam$^{-1}$ at
1.4$/$1.5~GHz, 610~MHz and 240~MHz, respectively.
The dynamic ranges in the maps are in the range 400--7000, 300--4400
and 300 to 3300 at 1.4 to 1.5~GHz, 610~MHz and 240~MHz, respectively.
Consequently, in the vicinity of strong sources,
the local noise was sometimes higher
than the noise in empty regions. The selection of contours shown in
the figures is based on the rms noise in the immediate vicinity of the source,
with the first contour level being 3 to 5 times this rms noise.
In addition, the peak surface brightness in the full field of view
is sometimes higher than the peak surface brightness mentioned in each panel,
which corresponds to the region of the figure displayed.

To make further comparisons of the morphology and flux densities,
the final calibrated ($u,v$) data at 610~MHz
were mapped using a ($u,v$)~range of 0 to 22~k$\lambda$,
which is similar to that of the 240~MHz data, and then restored
using a restoring beam corresponding to that of the 240~MHz map.
{\bf The resolution of archive VLA 1.4 to 1.5~GHz maps was slightly better
than the resolution of GMRT 610 MHz maps.  Therefore each of the VLA 1.4
to 1.5~GHz maps was matched with the resolution of the corresponding 610~MHz
map using {\sc aips} task `CONVL'.
The VLA contour maps matched to the resolution of 610~MHz,
the 610~MHz contour maps, the 610~MHz contour maps matched to the
resolution of 240~MHz, and the 240 MHz contour maps for all the observed
sources are shown in Figs.~\ref{full_syn_33} to \ref{full_syn_452}.}
The sequence of maps is ordered in right ascension. The ellipse in the box
in the lower left-hand corner of each map shows the shape of the
synthesized beam (FWHM). All positions are given in J2000 coordinates.

\subsection{Radio morphology and low frequency radio spectra}

The observations described allow us to investigate in detail the
morphologies and spectral index distributions of all sources. The
restored and matched maps at the three frequencies were used further
for the spectral analysis for each of these sources. We determine the
spectral index distribution using the standard direct method of
determining the spectral index between maps $S_{\nu_1}(x,y)$ and
$S_{\nu_2}(x,y)$ at two frequencies ${\nu_1}$ and ${\nu_2}$, where the
spectral index $\alpha$ is given by the ratio of ${\rm log}~(S_{\nu_1}
(x,y)/S_{\nu_2} (x,y))$ and ${\rm log}~(\nu_1/\nu_2)$.

\begin{figure}
\begin{center}
\begin{tabular}{l}
\includegraphics[width=7.1cm]{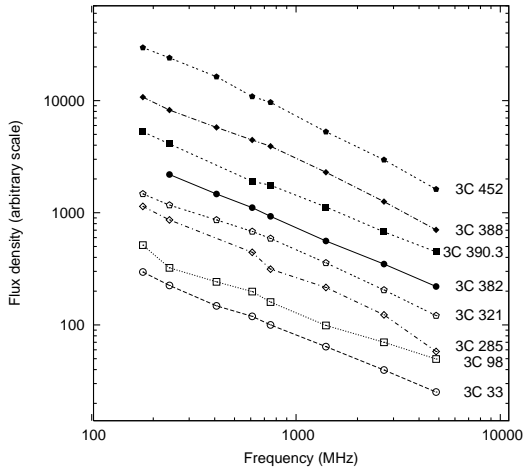}
\end{tabular}
\end{center}
\caption{Integrated flux density spectra for the sample of
radio sources.  Various measurements along with the error-bars
(not plotted) are explained in Table~\ref{flux_density}.
The spectra are shifted with respect to one another for clarity.
}
\label{fluxint}
\end{figure}

\begin{table*}
\caption{Flux densities of all the distinct representative regions at
1.4 to 1.5~GHz and 610~MHz (high resolution, upper panel maps) and
at 610~MHz and 240 MHz (low resolution, lower panel maps).
Representative regions are described in the text: note that they do
not include all of the flux density from a particular named region.
To make (nearly) similar comparison with X-shaped sources,
the faint diffuse, low-surface-brightness extensions in our FR\,II sample sources
are treated as ``wing" (like) features.
Each of the regions is given a label which is marked on the relevant figure
 (Figs.~\ref{full_syn_33} to~\ref{full_syn_452}).
The two spectral indices quoted are `real' and `conservative' estimates,
based on
the error bars for these sources quoted in the respective figure captions,
and the conservative error estimates determined from the fluctuations in the
region being averaged.
The corresponding spectral indices along with the error bars for these two
sets of measurements for each of these sources are presented in the last two
sets of columns.
Labels denotes: `$^\dag$' -- the upper limits on
the flux densities quoted are five times the `conservative' rms noise levels;
`$^\ddag$' -- limits on the spectral indices based on such measurements.}
\centering
\begin{tabular}{l|lcrr|rrcrrrr}
\hline \\
   & & & \multicolumn{4}{c}{Flux density}      &   &  \multicolumn{4}{c}{Spectral index} \\ \cline{4-7}  \cline{9-12}
   & & & \multicolumn{2}{c}{(high resolution)} & \multicolumn{2}{c}{(low resolution)} & &  \multicolumn{2}{c}{(high resolution)} &  \multicolumn{2}{c}{(low resolution)} \\
   & & & \multicolumn{2}{r}{1.5 GHz~~610 MHz}   & \multicolumn{2}{r}{610 MHz~~240 MHz} & &  \multicolumn{2}{c}{$\alpha^{1.5}_{\rm 0.61~GHz}$} &  \multicolumn{2}{c}{$\alpha^{610}_{\rm 240~MHz}$} \\
   & & & \multicolumn{2}{c}{(mJy)}             & \multicolumn{2}{c}{(mJy)} & & (real) & (conservative) & (real) & (conservative) \\
\hline \\
3C\,33   &North lobe       &(A)&2187.2 & 3444.9    & 4246.6 & 6161.2        & &$-$0.50~$\pm$0.01 &$-$0.61~$\pm$0.51 &$-$0.37~$\pm$0.01             &$-$0.48~$\pm$0.25 \\
         &South lobe       &(E)&6127.5 &12047.0    &11874.0 &16775.0        & &$-$0.75~$\pm$0.01 &$-$0.79~$\pm$0.30 &$-$0.40~$\pm$0.01             &$-$0.46~$\pm$0.41 \\
%        &Core             &   &  39.6 &   39.8    &        &               & &$ $0.00~$\pm$0.06 &                  &                              &                  \\
         &North Left wing  &(C)&  19.9 &   33.0    &  229.6 &$<$565.3$^\dag$& &$-$0.56~$\pm$0.08 &$-$0.80~$\pm$0.24 &$>$ $-$0.97~$\pm$0.19$^\ddag$ &                  \\
         &South Left wing  &(D)&  30.2 &   69.8    &  372.3 &$<$165.3$^\dag$& &$-$0.93~$\pm$0.04 &$-$0.89~$\pm$0.38 &$>$ $-$0.87~$\pm$0.20$^\ddag$ &                  \\
3C\,98   &North lobe       &(A)& 399.3 &  804.5    & 3522.5 & 4914.8        & &$-$0.77~$\pm$0.01 &$-$0.76~$\pm$0.08 &$-$0.36~$\pm$0.01             &$-$0.40~$\pm$0.14 \\
         &South lobe       &(E)& 275.7 &  517.1    & 2440.9 & 3038.1        & &$-$0.82~$\pm$0.01 &$-$0.77~$\pm$0.10 &$-$0.27~$\pm$0.01             &$-$0.37~$\pm$0.14 \\
%        &Core             &   &  23.0 &   48.3    &        &               & &$-$0.69~$\pm$0.01 &                  &                              &                  \\
         &North Left wing  &(C)& 153.3 &  302.2    &        &               & &$-$0.75~$\pm$0.01 &$-$0.75~$\pm$0.12 &                              &                  \\
         &Bottom Left wing &(D)&  24.5 &   58.7    &        &               & &$-$0.97~$\pm$0.01 &$-$1.05~$\pm$0.20 &                              &                  \\
         &Top Right wing   &(B)&  28.7 &   77.1    &        &               & &$-$1.09~$\pm$0.01 &$-$1.14~$\pm$0.11 &                              &                  \\
3C\,285  &East lobe        &(A)& 114.6 &  217.8    &  724.3 & 1360.3        & &$-$0.70~$\pm$0.01 &$-$0.78~$\pm$0.08 &$-$0.68~$\pm$0.01             &$-$0.69~$\pm$0.07 \\
         &West lobe        &(E)&  83.0 &  188.7    &  451.6 &  807.9        & &$-$0.90~$\pm$0.01 &$-$0.88~$\pm$0.12 &$-$0.63~$\pm$0.01             &$-$0.64~$\pm$0.06 \\
%        &Core             &   &  24.9 &   62.9    &        &               & &$-$1.02~$\pm$0.01 &                  &                              &                  \\
         &North Left wing  &(C)&  41.1 &  100.3    &  402.3 &  816.3        & &$-$0.98~$\pm$0.01 &$-$1.02~$\pm$0.23 &$-$0.76~$\pm$0.01             &$-$0.80~$\pm$0.17 \\
         &South Left wing  &(F)&  48.6 &  120.7    &        &               & &$-$1.00~$\pm$0.01 &$-$1.03~$\pm$0.24 &                              &                  \\
         &North Right wing &(D)&  46.4 &  105.0    &        &               & &$-$0.89~$\pm$0.01 &$-$0.90~$\pm$0.13 &                              &                  \\
         &South Right wing &(B)&  31.6 &   84.4    &        &               & &$-$1.08~$\pm$0.01 &$-$1.06~$\pm$0.11 &$-$0.67~$\pm$0.01             &$-$0.65~$\pm$0.16 \\
3C\,321  &North-West lobe  &(A)& 310.4 &  704.4    &  588.3 &  881.6        & &$-$0.89~$\pm$0.01 &$-$0.87~$\pm$0.90 &$-$0.44~$\pm$0.03             &$-$0.69~$\pm$0.29 \\
         &South-East lobe  &(E)&1855.9 & 4091.2    & 2955.3 & 4034.8        & &$-$0.86~$\pm$0.01 &$-$0.85~$\pm$0.51 &$-$0.34~$\pm$0.03             &$-$0.65~$\pm$0.31 \\
%        &Core             &   &  27.3 &   11.1    &  162.5 &  243.4        & &$-$0.99~$\pm$0.08 &                  &$-$0.44~$\pm$0.02             &                  \\
         &North wing       &(C)&  24.5 &   32.1    &   17.2 &$<$53.9$^\dag$ & &$-$0.30~$\pm$0.03 &$-$0.31~$\pm$0.30 &$>$ $-$1.22~$\pm$0.21$^\ddag$ &                  \\
3C\,382  &South-West lobe  &(E)&  90.5 &  276.1    & 1811.3 & 3537.0        & &$-$1.24~$\pm$0.03 &$-$1.32~$\pm$0.12 &$-$0.72~$\pm$0.01             &$-$0.77~$\pm$0.15 \\
         &North-East lobe  &(A)& 381.0 & 1142.0    & 1829.6 & 3789.0        & &$-$1.22~$\pm$0.04 &$-$1.19~$\pm$0.17 &$-$0.79~$\pm$0.01             &$-$0.82~$\pm$0.23 \\
%        &Core             &   & Undef &  219.5    &        &               & &                  &                  &                              &                  \\
%    &Bottom right         &(B)& Undef &   21.0    &   25.9 &   21.0        & &                  &                  &                              &                  \\
         &South-East wing  &(C)&  12.8 &   47.2    &  186.3 &  434.7        & &$-$1.48~$\pm$0.01 &$-$1.45~$\pm$0.18 &$-$0.91~$\pm$0.01             &$-$0.89~$\pm$0.21 \\
3C\,390.3&North-West lobe  &(A)& 534.8 & 1040.5    &  791.8 & 1382.5        & &$-$0.73~$\pm$0.01 &$-$0.80~$\pm$0.07 &$-$0.60~$\pm$0.01             &$-$0.84~$\pm$0.50 \\
         &South-East lobe  &(E)&1803.0 & 3744.5    & 2468.7 & 4942.1        & &$-$0.81~$\pm$0.01 &$-$0.81~$\pm$0.14 &$-$0.75~$\pm$0.01             &$-$0.90~$\pm$0.30 \\
%        &Core             &   & 185.4 &  199.2    &        &               & &$-$0.79~$\pm$0.01 &                  &                              &                  \\
         &South Left wing  &(D)&  56.2 &  100.9    &  102.1 &  242.7        & &$-$0.64~$\pm$0.01 &$-$0.70~$\pm$0.13 &$-$0.93~$\pm$0.02             &$-$0.90~$\pm$0.18 \\
3C\,388  &North-East lobe  &(A)& 545.0 & 1118.2    & 1538.8 & 3173.5        & &$-$0.85~$\pm$0.01 &$-$0.90~$\pm$0.12 &$-$0.78~$\pm$0.01             &$-$0.81~$\pm$0.18 \\
         &South-West lobe  &(E)& 428.7 &  815.0    & 2213.6 & 3552.5        & &$-$0.76~$\pm$0.01 &$-$0.72~$\pm$0.08 &$-$0.51~$\pm$0.01             &$-$0.56~$\pm$0.20 \\
3C\,452  &East lobe        &(A)& 263.5 &  512.7    & 1988.5 & 4106.4        & &$-$0.79~$\pm$0.01 &$-$0.77~$\pm$0.07 &$-$0.78~$\pm$0.01             &$-$0.79~$\pm$0.13 \\
         &West lobe        &(E)& 234.5 &  460.2    & 2470.1 & 4555.2        & &$-$0.80~$\pm$0.01 &$-$0.76~$\pm$0.06 &$-$0.66~$\pm$0.01             &$-$0.72~$\pm$0.13 \\
%        &Core             &   & 140.4 &  161.2    &        &               & &$-$0.16~$\pm$0.01 &                  &                              &                  \\
         &North wing       &(B)&  18.8 &   53.1    &  586.4 & 1353.6        & &$-$1.23~$\pm$0.02 &$-$1.25~$\pm$0.18 &$-$0.90~$\pm$0.02             &$-$0.96~$\pm$0.13 \\
         &South wing       &(C)&   8.6 &   25.9    &  201.6 &  598.2        & &$-$1.30~$\pm$0.04 &$-$1.22~$\pm$0.17 &$-$1.17~$\pm$0.02             &$-$1.17~$\pm$0.20 \\
\\
\hline
\end{tabular}
\label{fdregions}
\end{table*}

\begin{figure*}
\begin{center}
\begin{tabular}{lll}
\includegraphics[width=6.3cm]{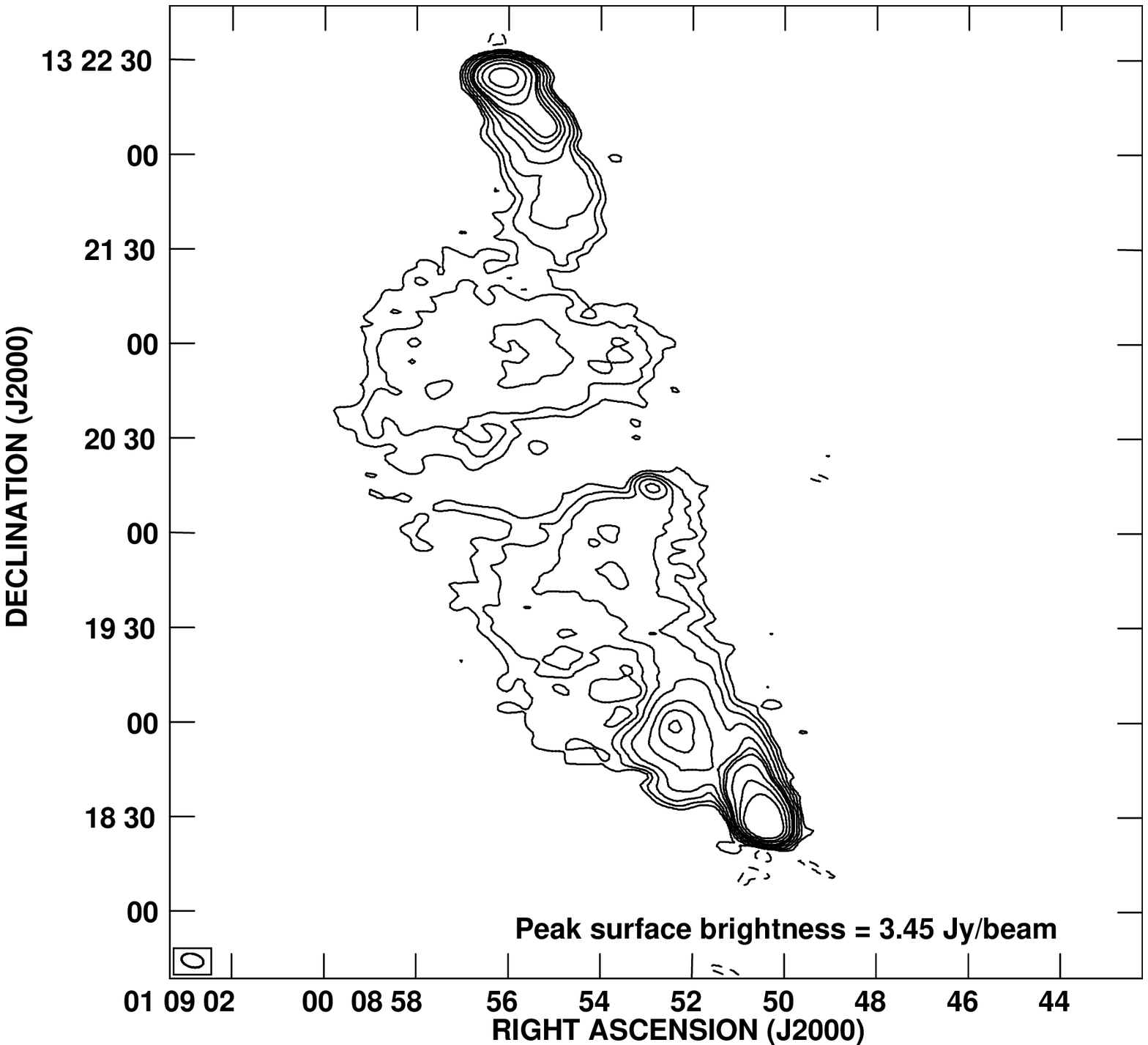} &
\hspace*{-0.2cm}\includegraphics[width=5.24cm]{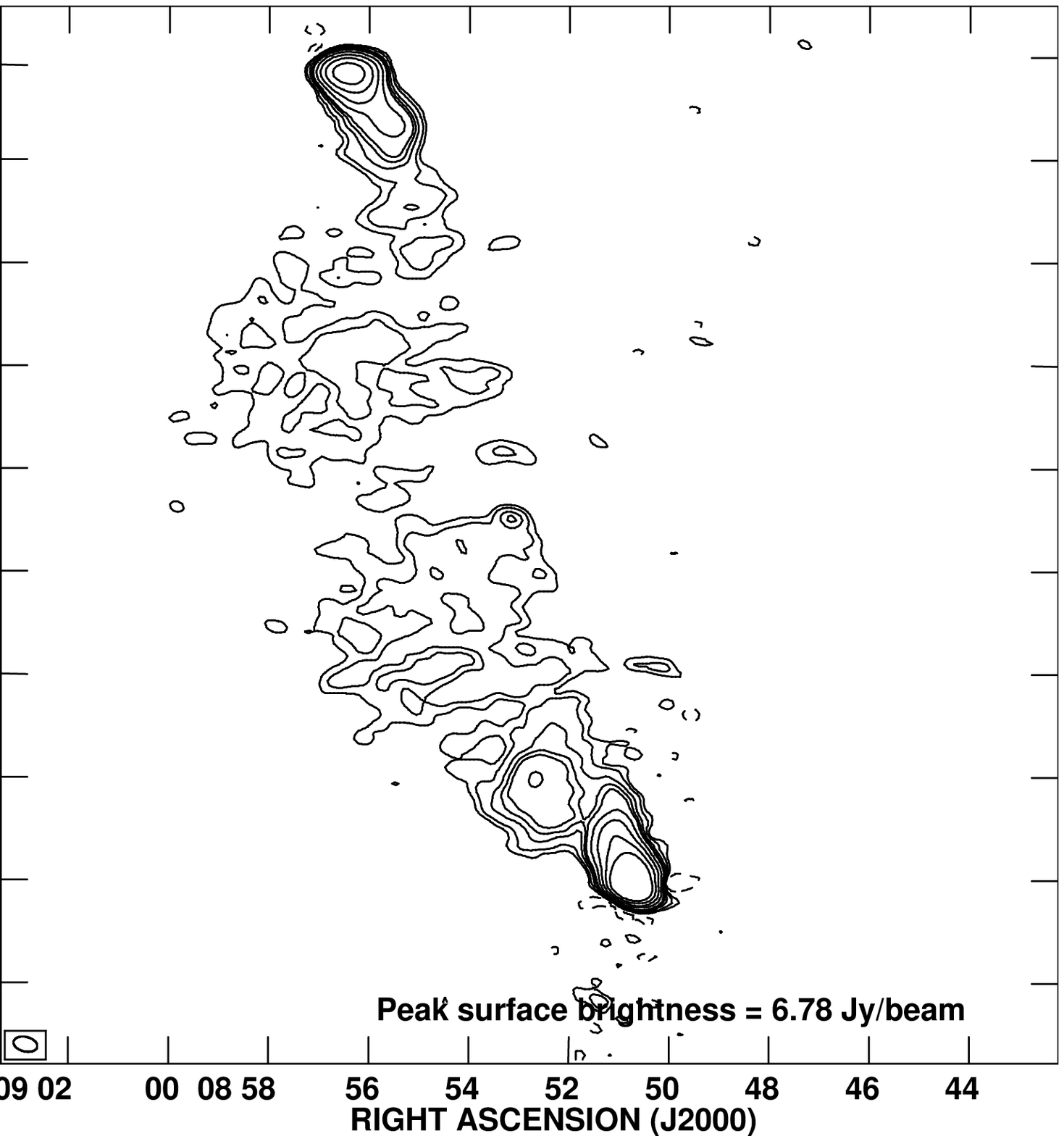} &
\hspace*{-0.2cm}\includegraphics[width=5.24cm]{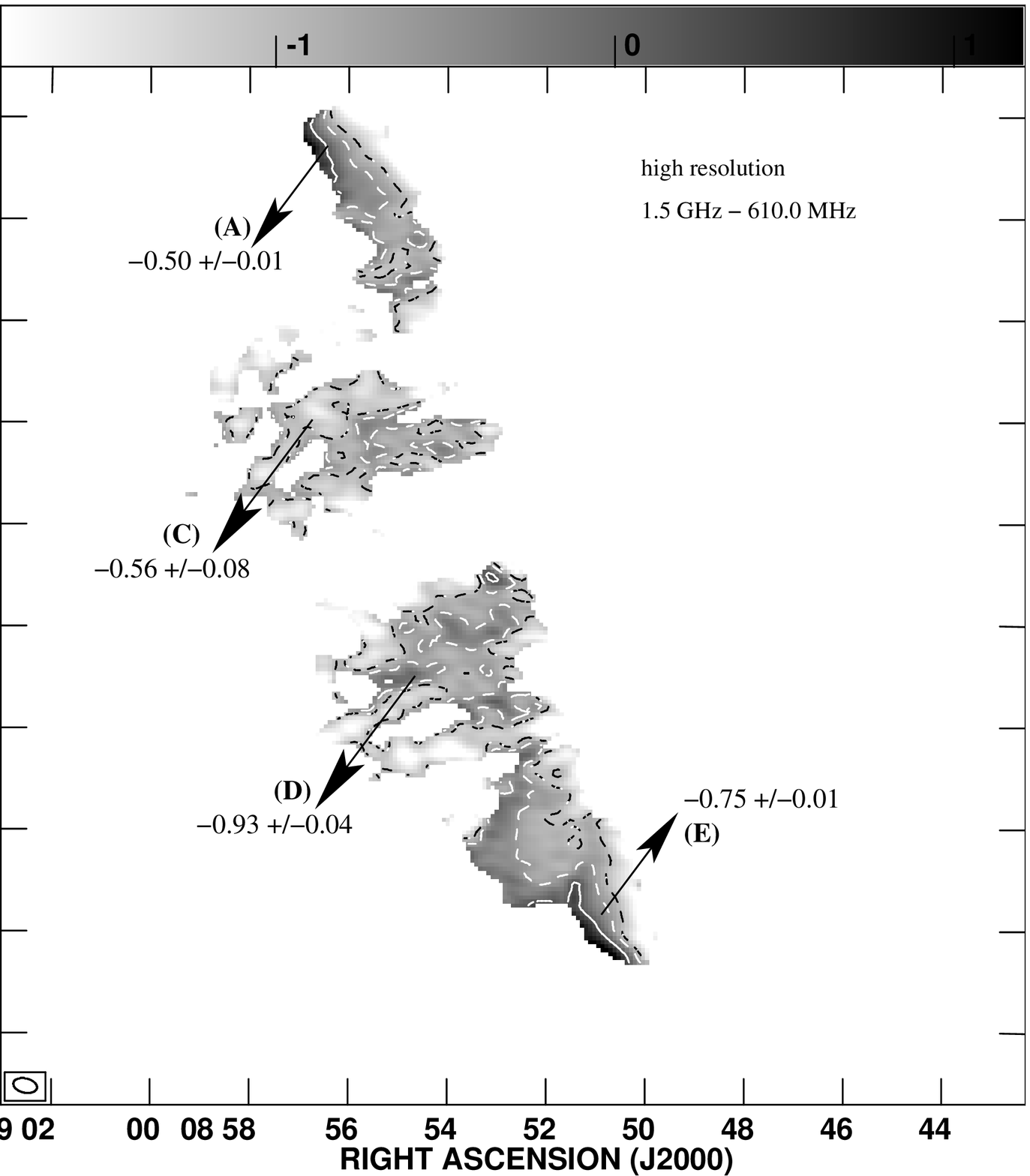} \\
\includegraphics[width=6.3cm]{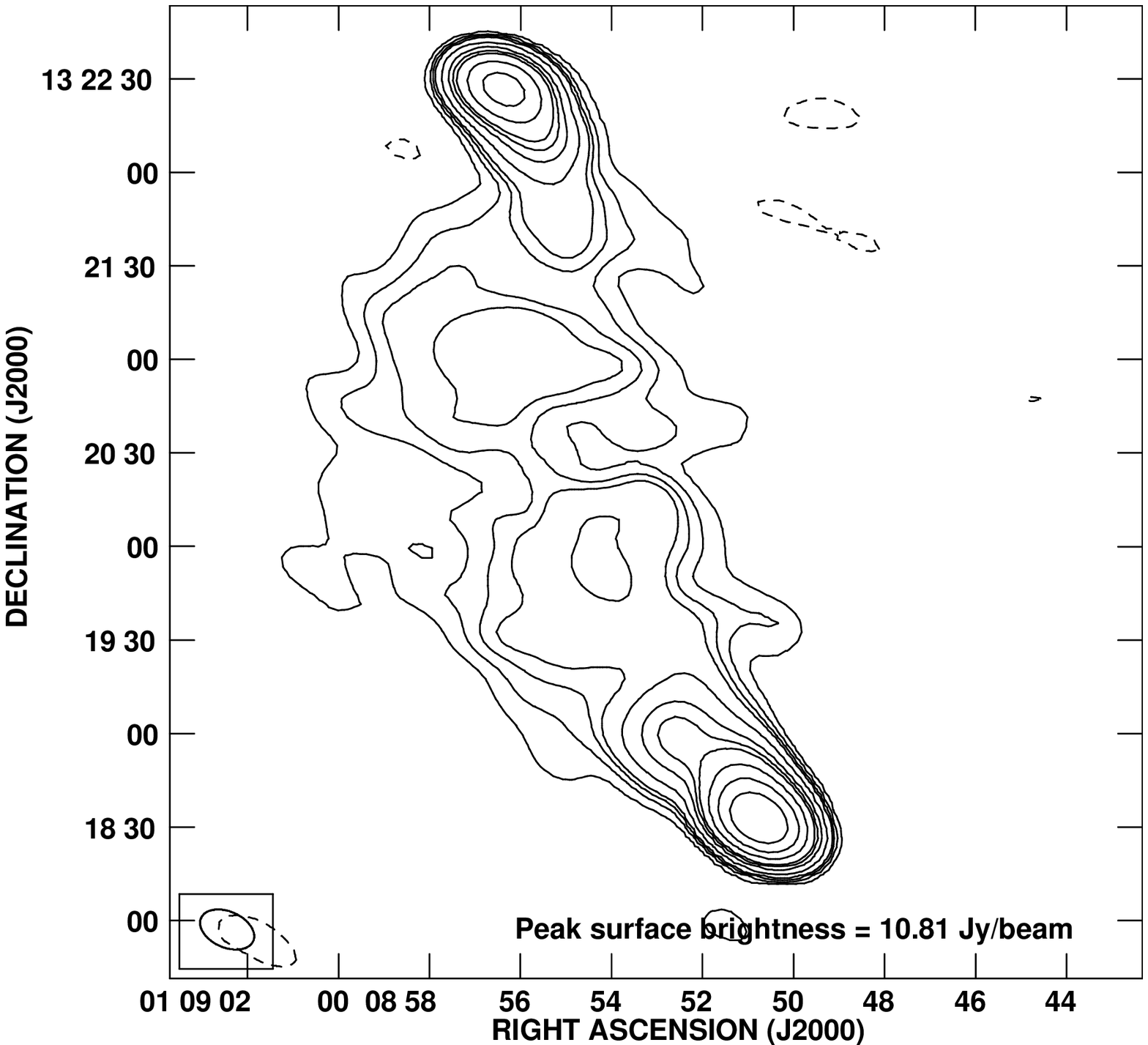} &
\hspace*{-0.2cm}\includegraphics[width=5.24cm]{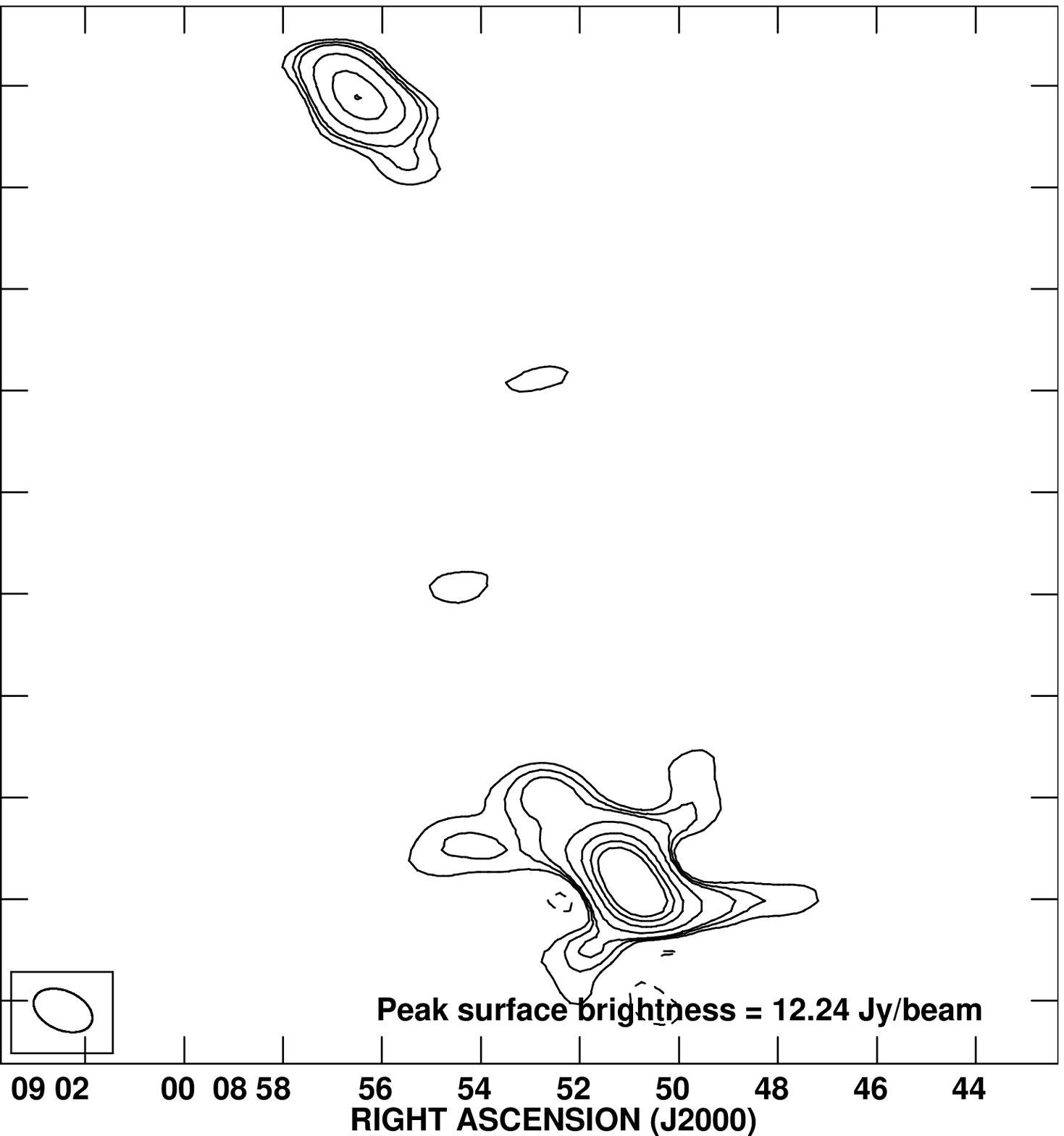} &
\hspace*{-0.2cm}\includegraphics[width=5.24cm]{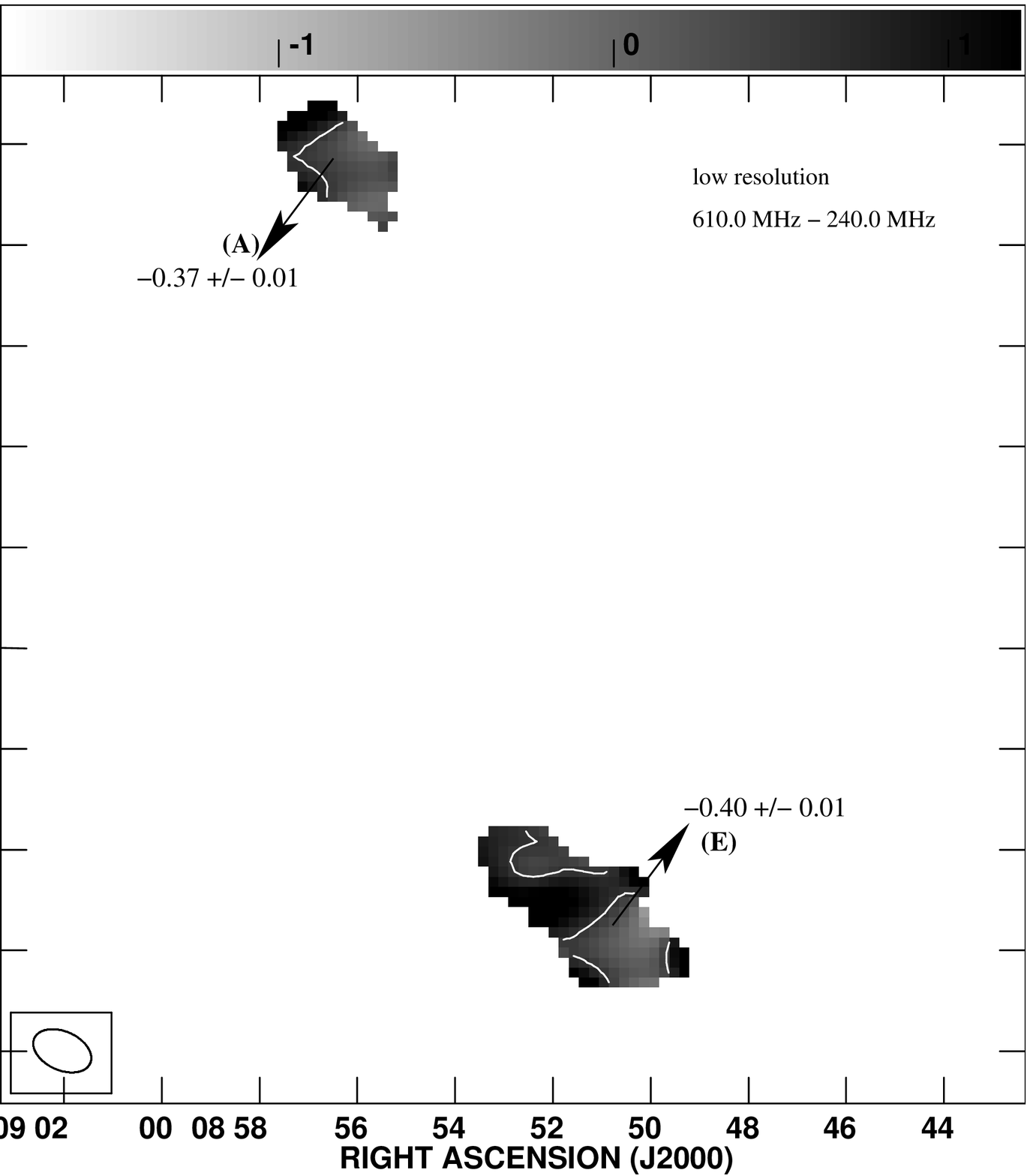} \\
\end{tabular}
\end{center}
\caption{Radio maps of 3C 33.
Upper left: The VLA map of 3C\,33 at 1.5 GHz
matched with the resolution of 610~MHz; the contour levels in the map are
($-$2, 2, 4, 8, 16, 32, 48, 64, 128, 256) mJy~beam$^{-1}$.
Upper middle: The GMRT map of 3C\,33 at 610 MHz; the contour levels in the map are
($-$9, 9, 16, 32, 48, 64, 128, 256, 512) mJy~beam$^{-1}$.
Lower left: The GMRT map of 3C\,33 at 610 MHz 
matched with the resolution of 240~MHz; the contour levels in the map are
($-$20, 20, 40, 60, 80, 160, 320, 480, 640, 1280) mJy~beam$^{-1}$.
Lower middle: The GMRT map of 3C\,33 at 240 MHz; the contour levels in the map are
($-$400, 400, 600, 800, 1600, 3200, 4800) mJy~beam$^{-1}$.
Upper right and Lower right panels: The distribution of the spectral index,
between 1.5~GHz and 610 MHz (upper right),
and 240~MHz and 610 MHz (lower right), for the source.
The spectral index range displayed in the two maps are
$-$1.8 and 1.2 (upper right), and $-$1.8 and 1.2 (lower right), respectively.
The spectral index contours are at $-$1.0,~$-$0.6, 0.2 and
$-$0.8,~0.6, respectively in the two maps.
The spectral indices listed for various regions are
tabulated in Table~\ref{fdregions}.
The r.m.s. noise values in the radio images found at a source free location
are $\sim$0.5, $\sim$2.1 and $\sim$5.9~mJy~beam$^{-1}$ at
1.5~GHz, 610~MHz and 240~MHz, respectively.
The uniformly weighted CLEAN beams for upper and lower panel maps are
7.1~arcsec $\times$~4.2 arcsec
at a P.A. of $+$76.7$^{\circ}$
and
18.3~arcsec $\times$~11.5 arcsec
at a P.A. of $+$66.1$^{\circ}$, respectively.
}
\label{full_syn_33}
\end{figure*}

\begin{figure*}
\begin{center}
\begin{tabular}{lll}
\includegraphics[width=6.0cm]{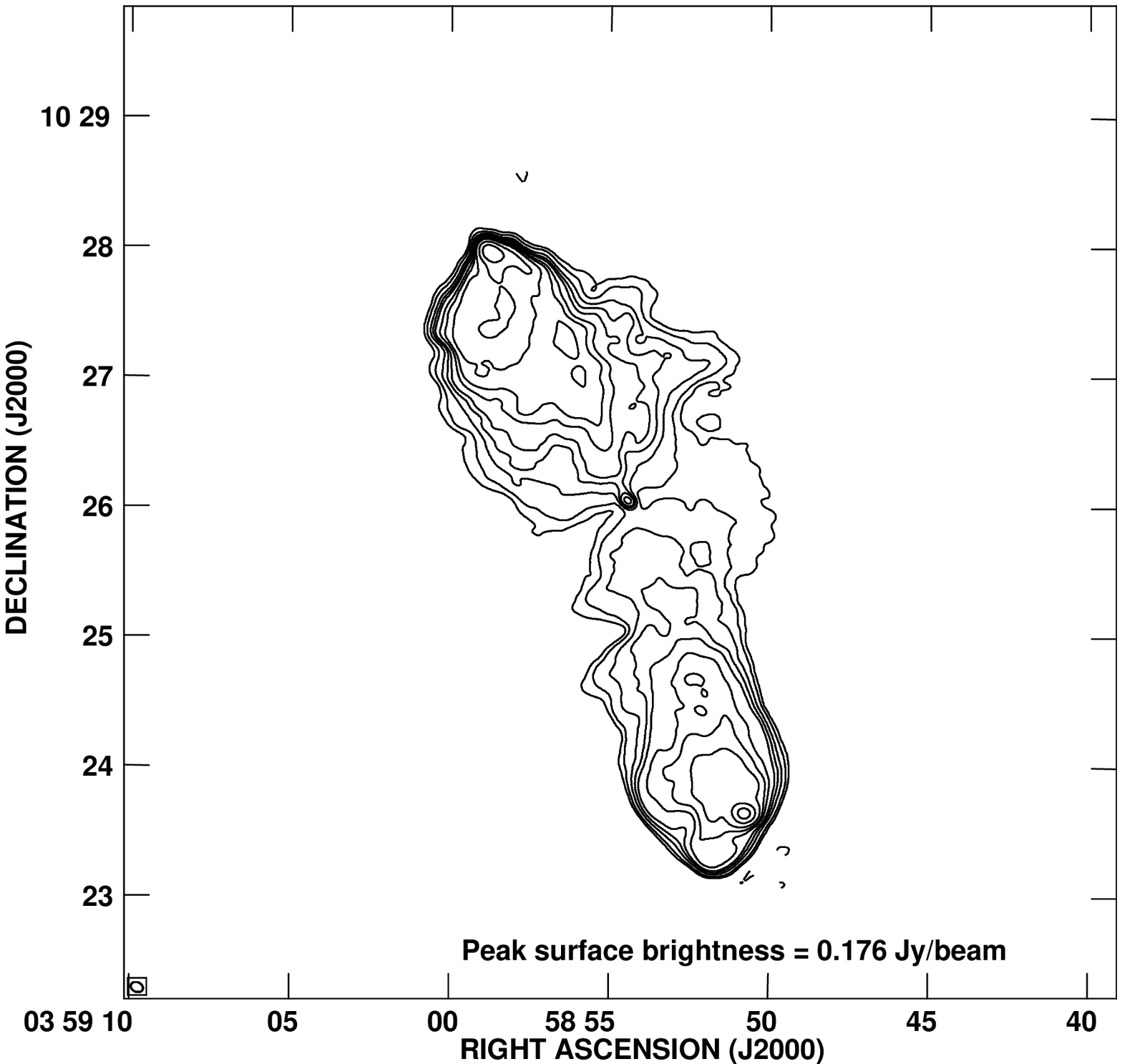} &
\hspace*{-0.2cm}\includegraphics[width=5.24cm]{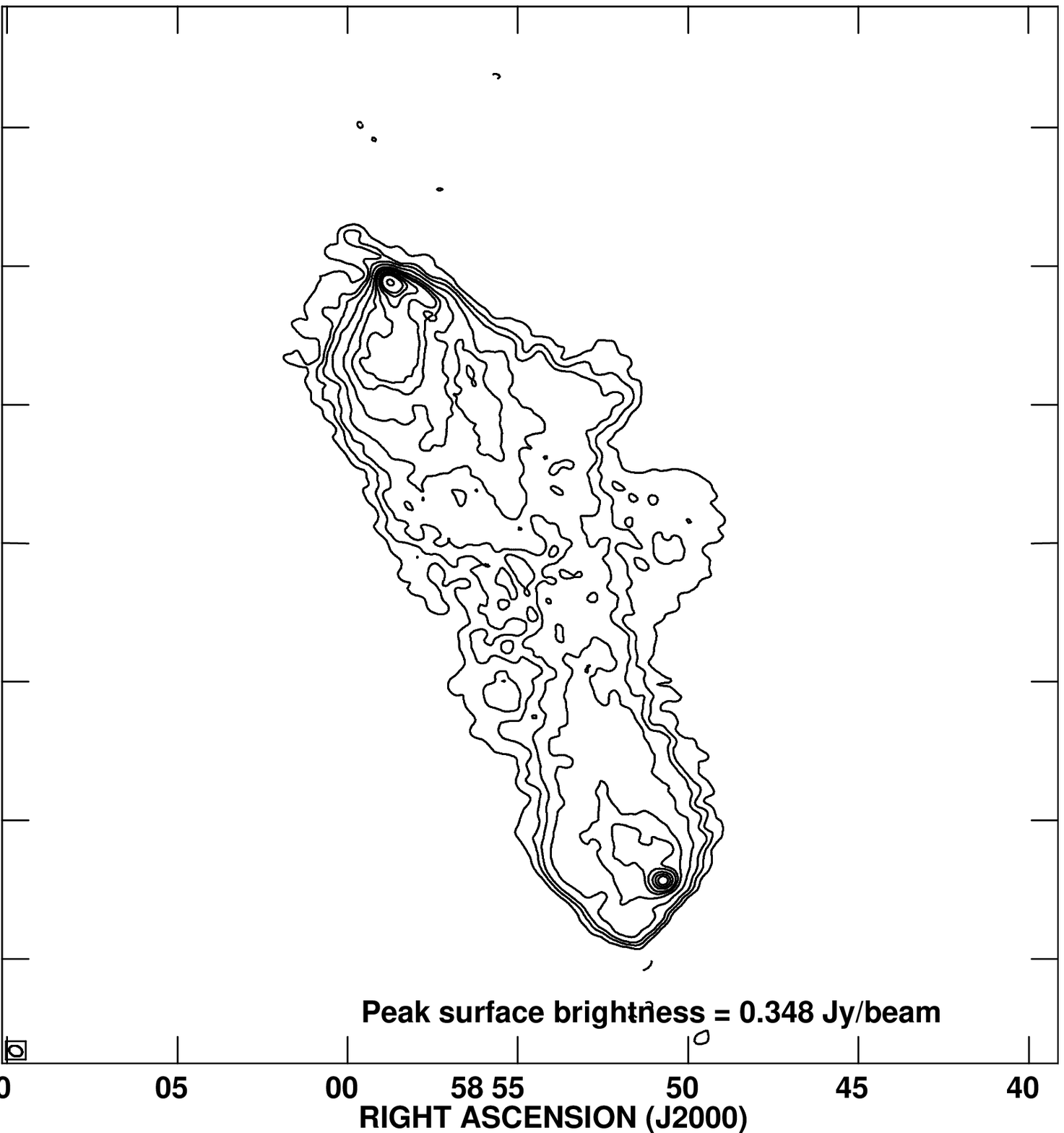} &
\hspace*{-0.2cm}\includegraphics[width=5.24cm]{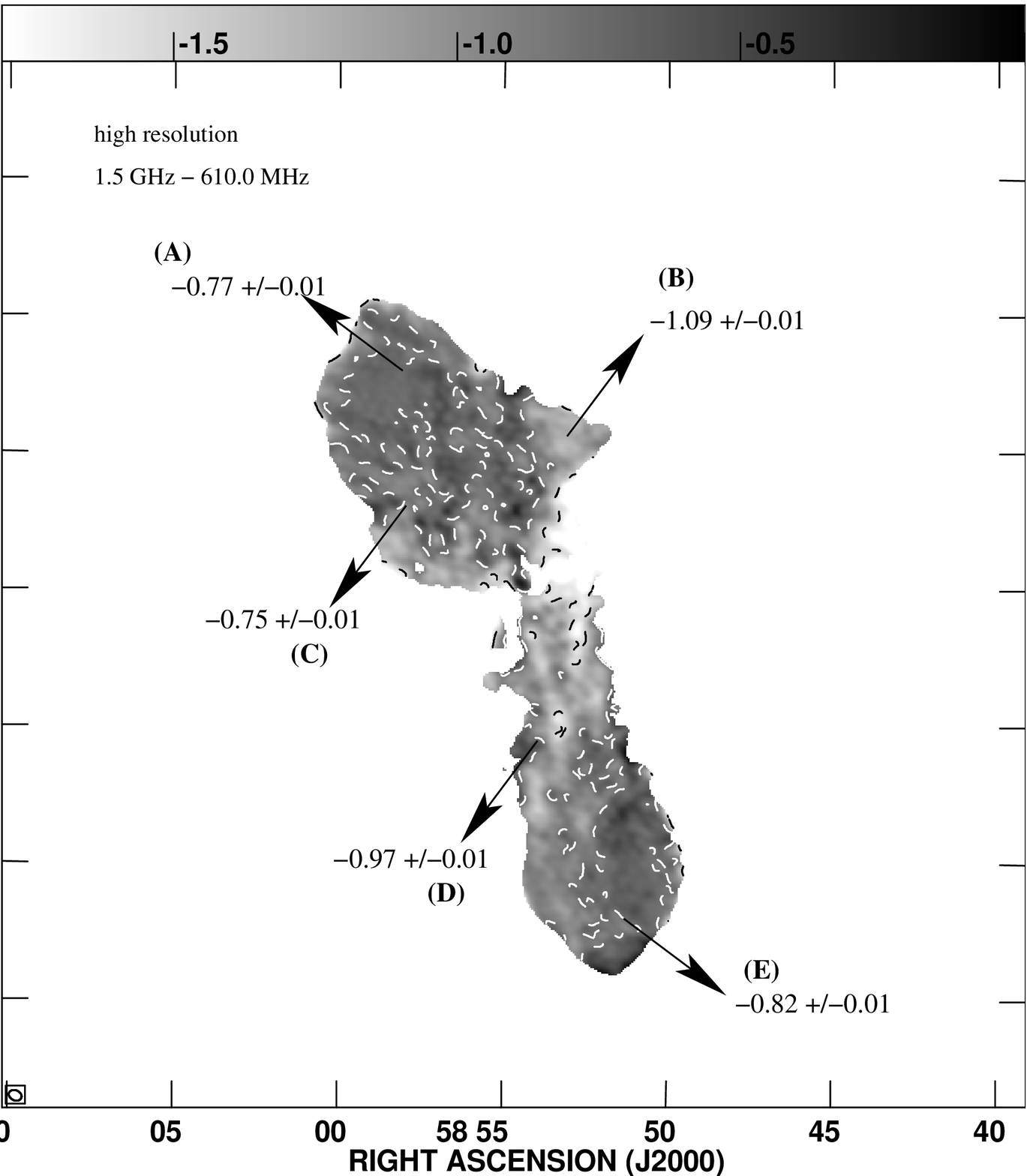} \\
\includegraphics[width=6.0cm]{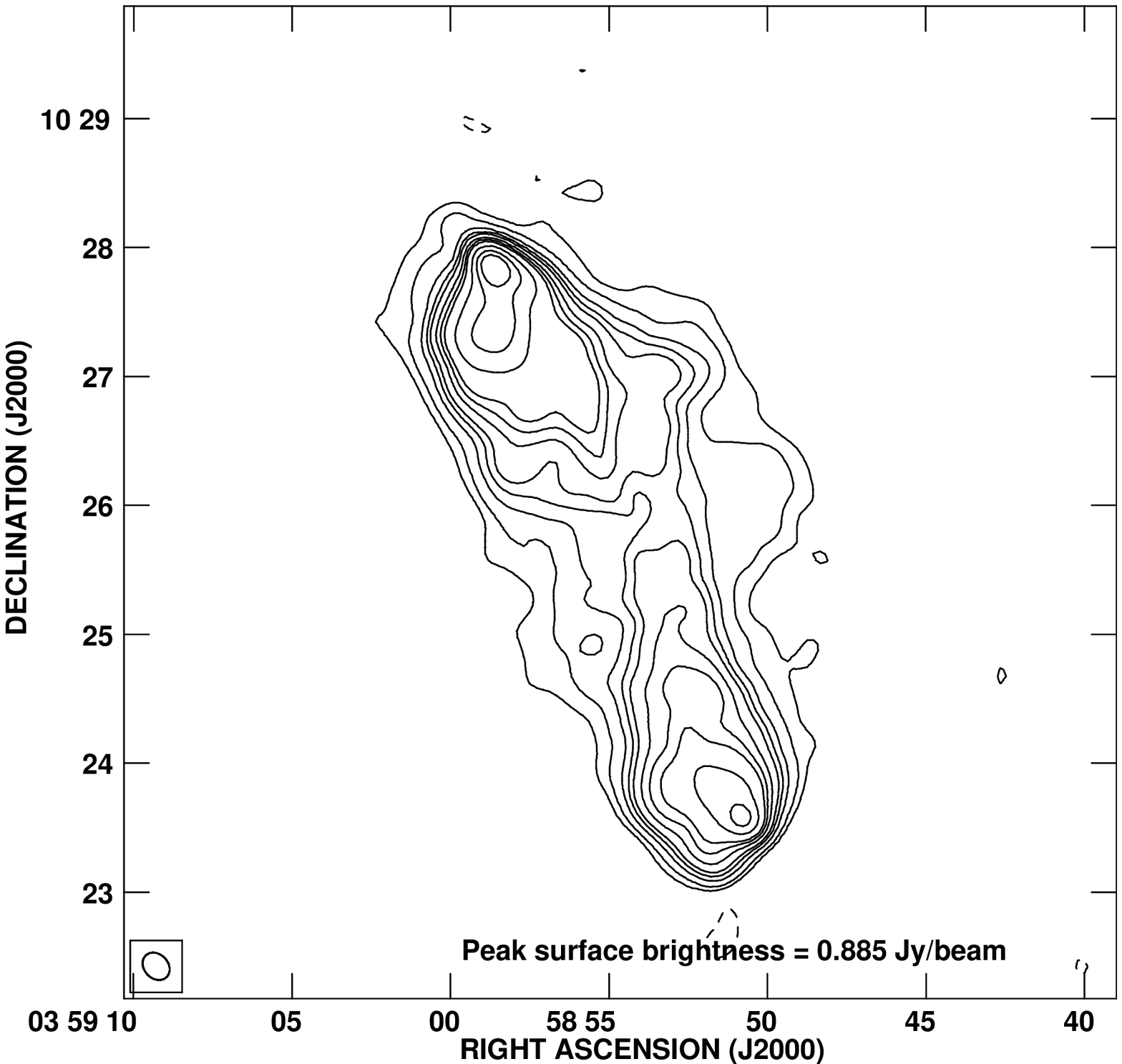} &
\hspace*{-0.2cm}\includegraphics[width=5.24cm]{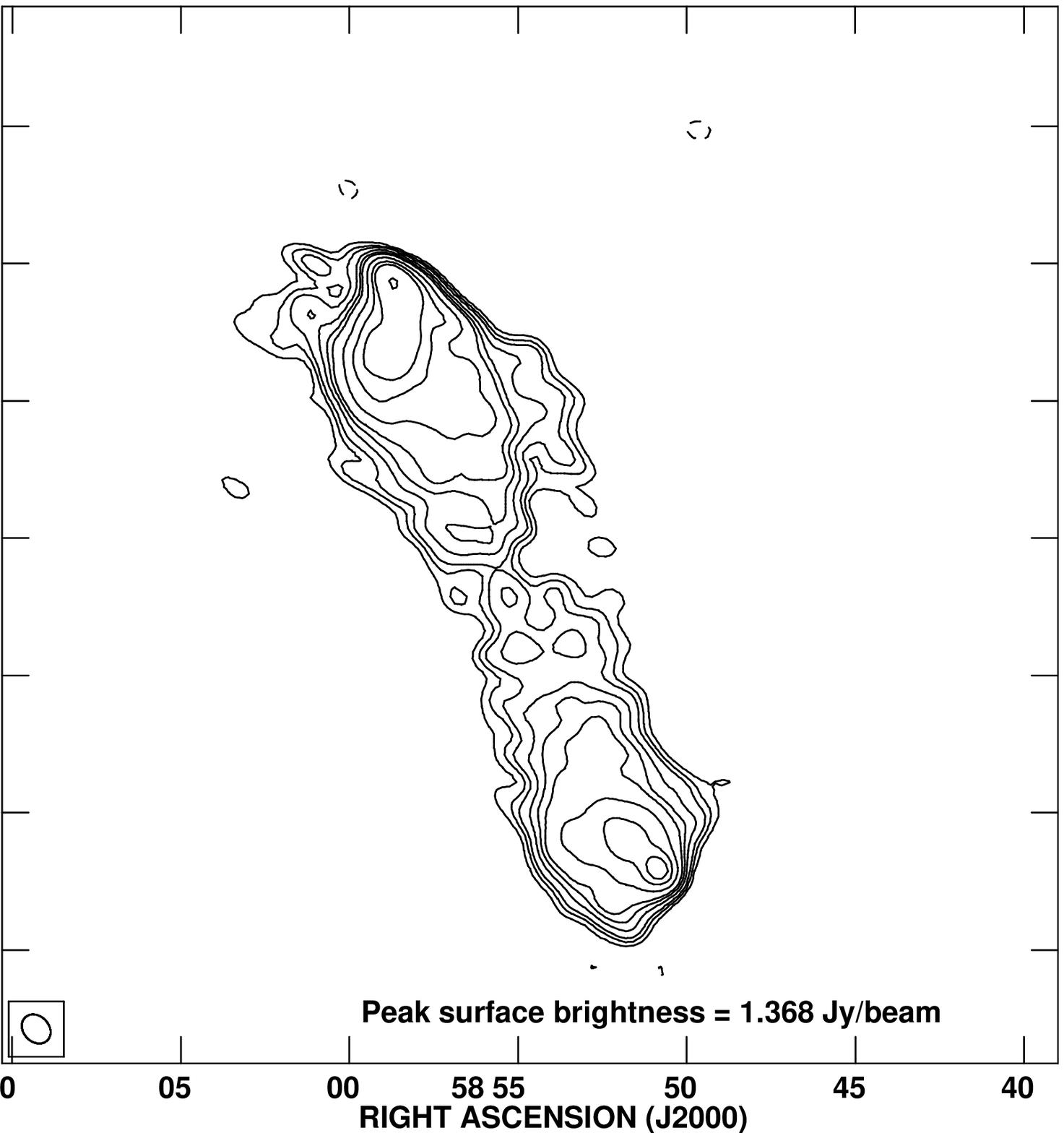} &
\hspace*{-0.2cm}\includegraphics[width=5.24cm]{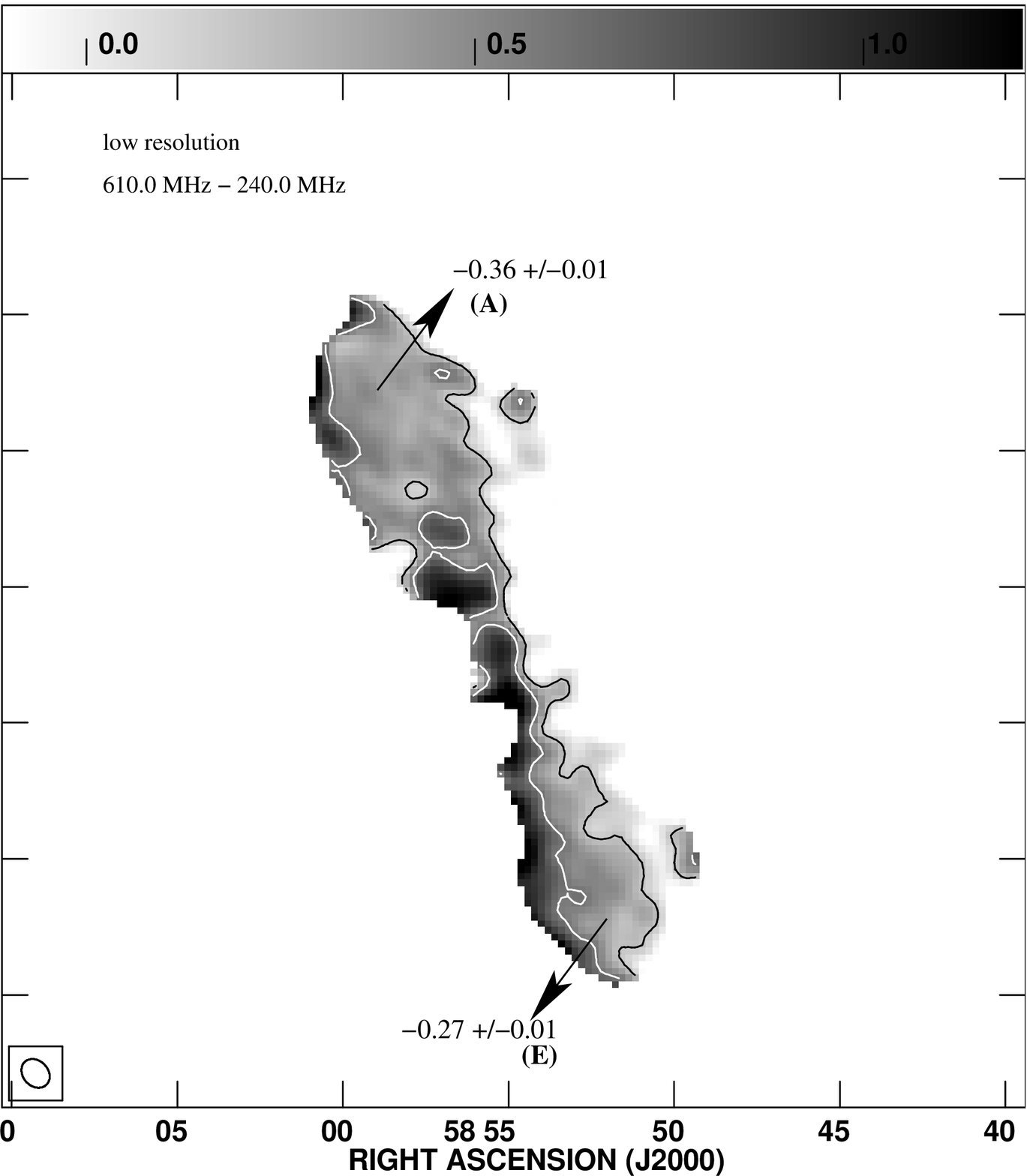} \\
\end{tabular}
\end{center}
\caption{Radio maps of 3C\,98.
Upper left: The VLA map of 3C\,98 at 1.5 GHz
matched with the resolution of 610~MHz; the contour levels in the map are
($-$1, 1, 2, 4, 6, 8, 12, 16, 24, 48, 96) mJy~beam$^{-1}$.
Upper middle: The GMRT map of 3C\,98 at 610 MHz; the contour levels in the map are
($-$3, 3, 6, 10, 20, 40, 60, 80, 120, 160) mJy~beam$^{-1}$.
Lower left: The GMRT map of 3C\,98 at 610 MHz 
matched with the resolution of 240~MHz; the contour levels in the map are
($-$10, 10, 20, 40, 60, 80, 120, 160, 200, 320, 480) mJy~beam$^{-1}$.
Lower middle: The GMRT map of 3C\,98 at 240 MHz; the contour levels in the map are
($-$40, 40, 60, 80, 120, 160, 200, 320, 480, 640) mJy~beam$^{-1}$.
Upper right and Lower right panels: The distribution of the spectral index,
between 1.5~GHz and 610 MHz (upper right),
and 240~MHz and 610 MHz (lower right), for the source.
The spectral index range displayed in the two maps are
$-$1.8 and 0.0 (upper right), and $-$0.8 and 1.2 (lower right), respectively.
The spectral index contours are at $-$1.5,~$-$0.8,~0 and
0.2,~0.6, 1.0 respectively in the two maps.
The spectral indices listed for various regions are
tabulated in Table~\ref{fdregions}.
The r.m.s. noise values in the radio images found at a source free location
are $\sim$0.08, $\sim$0.4 and $\sim$1.5~mJy~beam$^{-1}$ at
1.5~GHz, 610~MHz and 240~MHz, respectively.
The uniformly weighted CLEAN beams for upper and lower panel maps are
8.9~arcsec $\times$~4.4 arcsec
at a P.A. of $+$66.0$^{\circ}$
and
14.0~arcsec $\times$~11.2 arcsec
at a P.A. of $+$43.5$^{\circ}$, respectively.
}
\label{full_syn_98}
\end{figure*}

\begin{figure*}
\begin{center}
\begin{tabular}{lll}
\includegraphics[width=6.0cm]{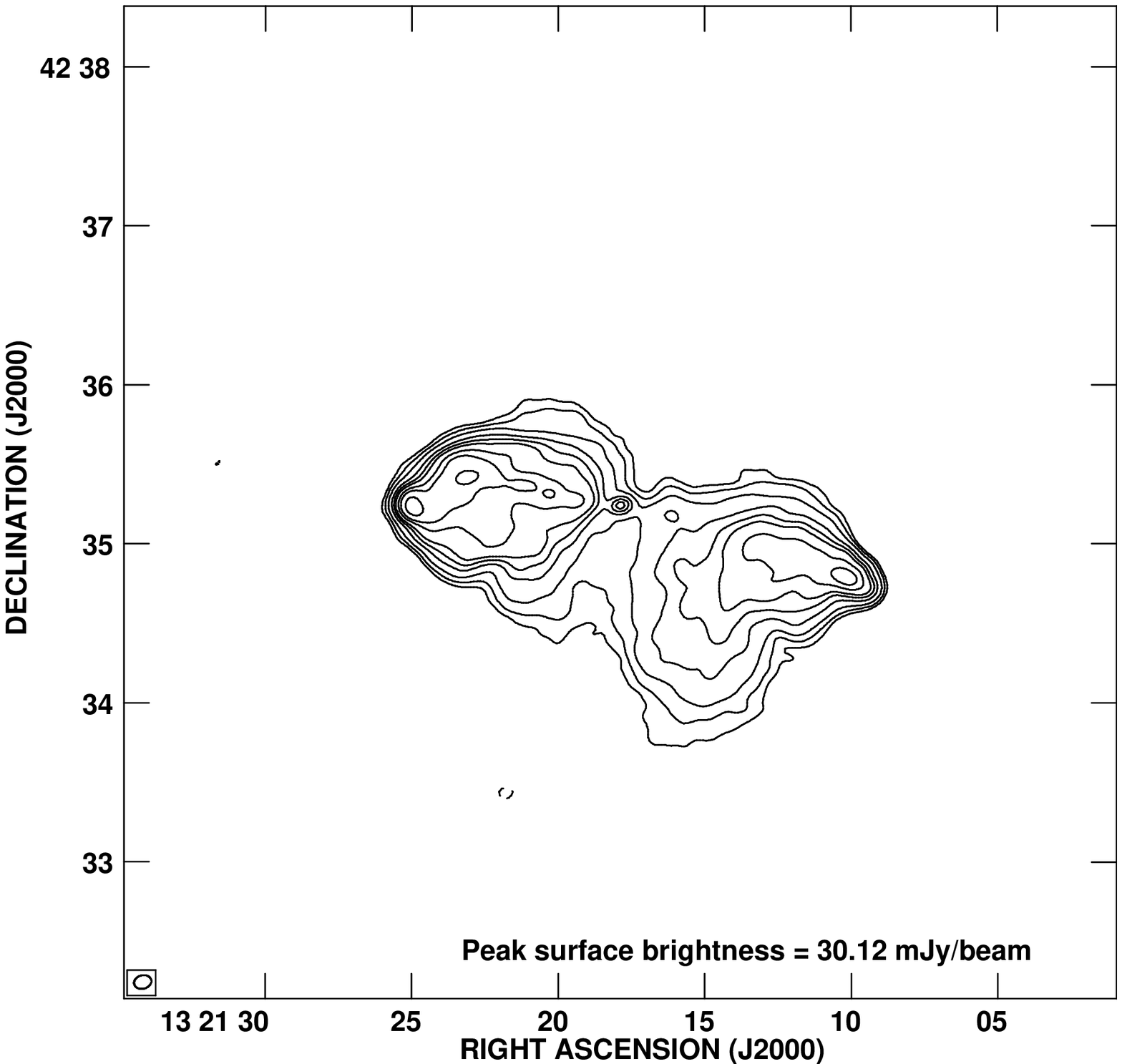} &
\hspace*{-0.2cm}\includegraphics[width=5.24cm]{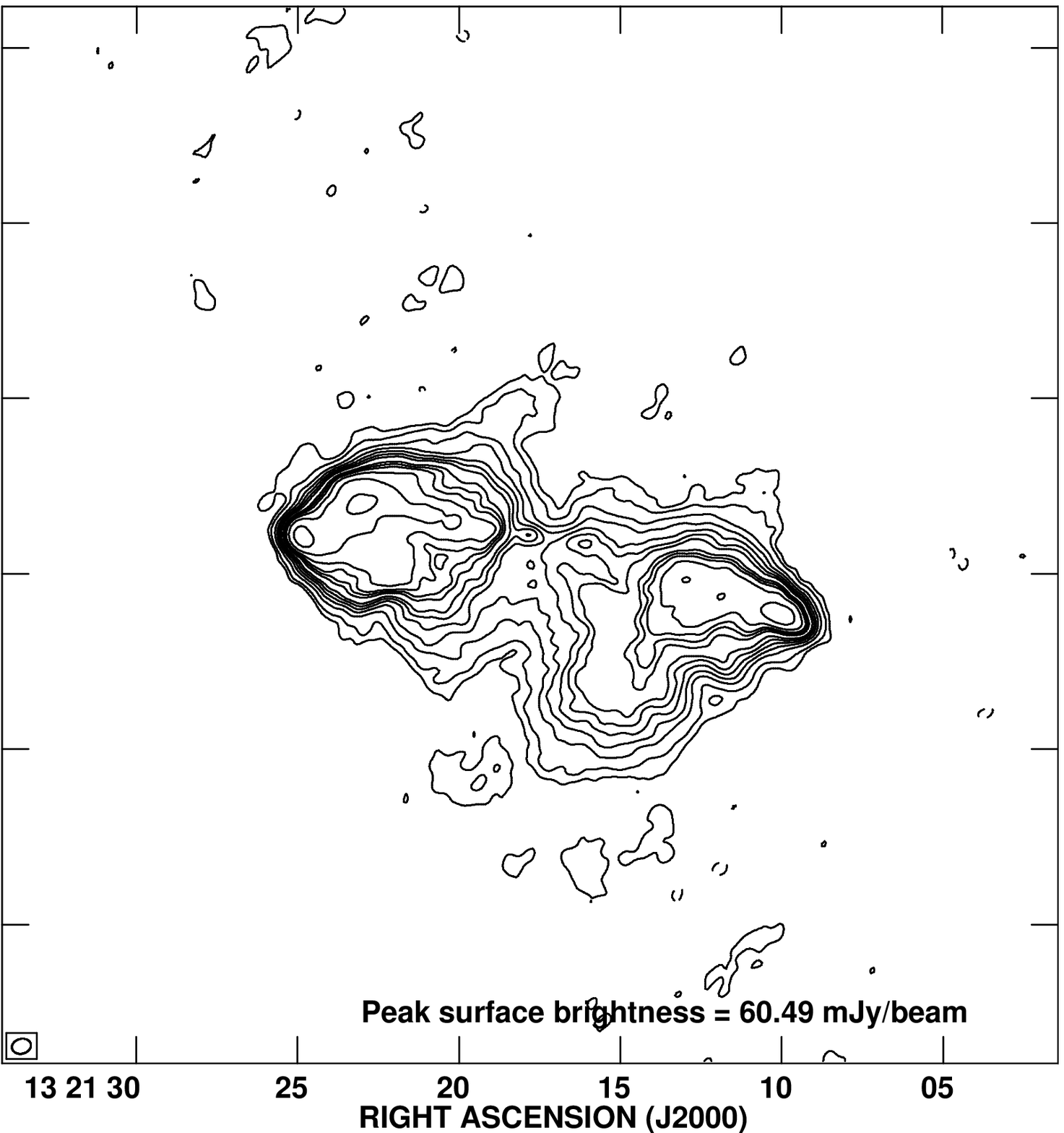} &
\hspace*{-0.2cm}\includegraphics[width=5.24cm]{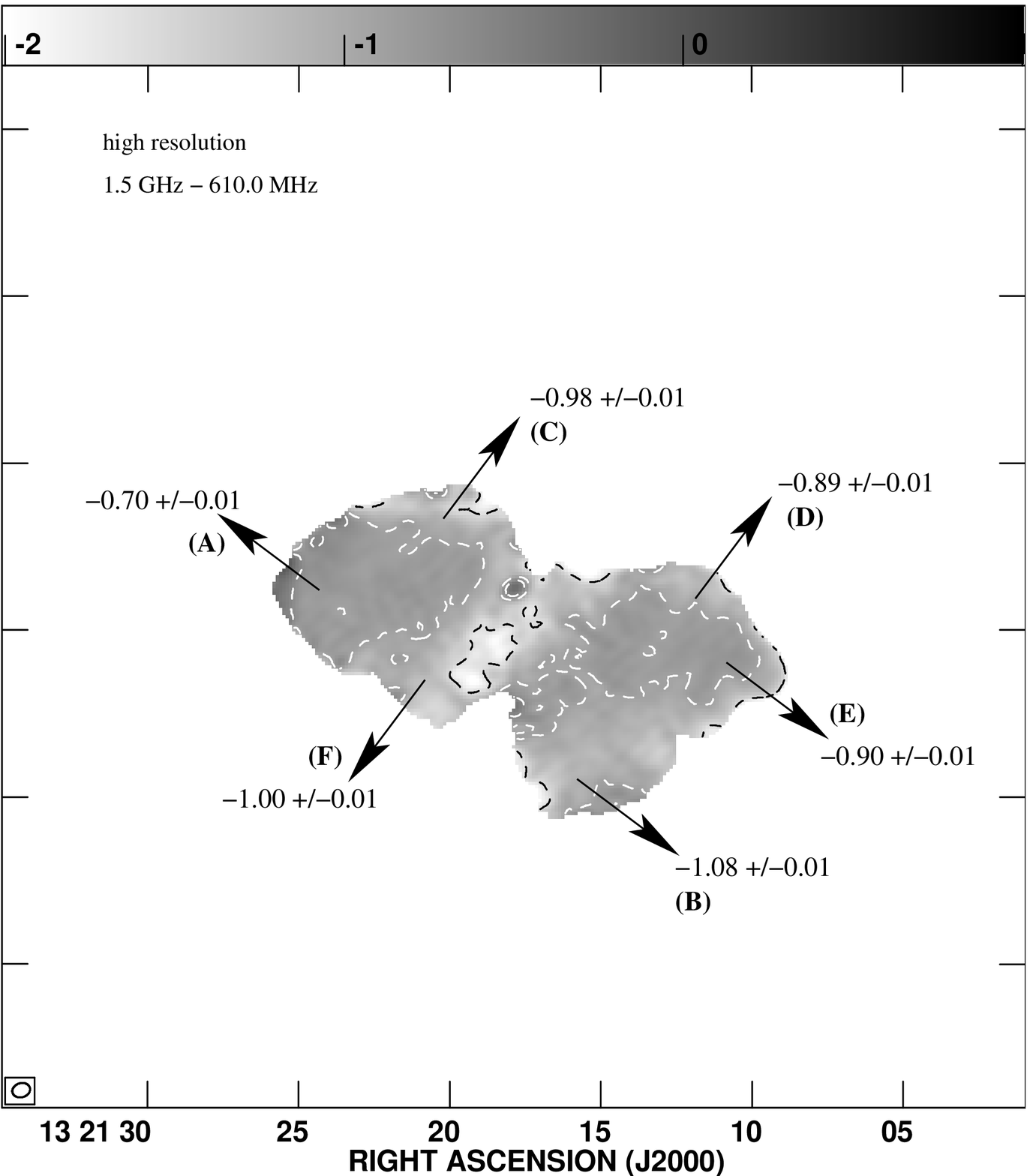} \\
\includegraphics[width=6.0cm]{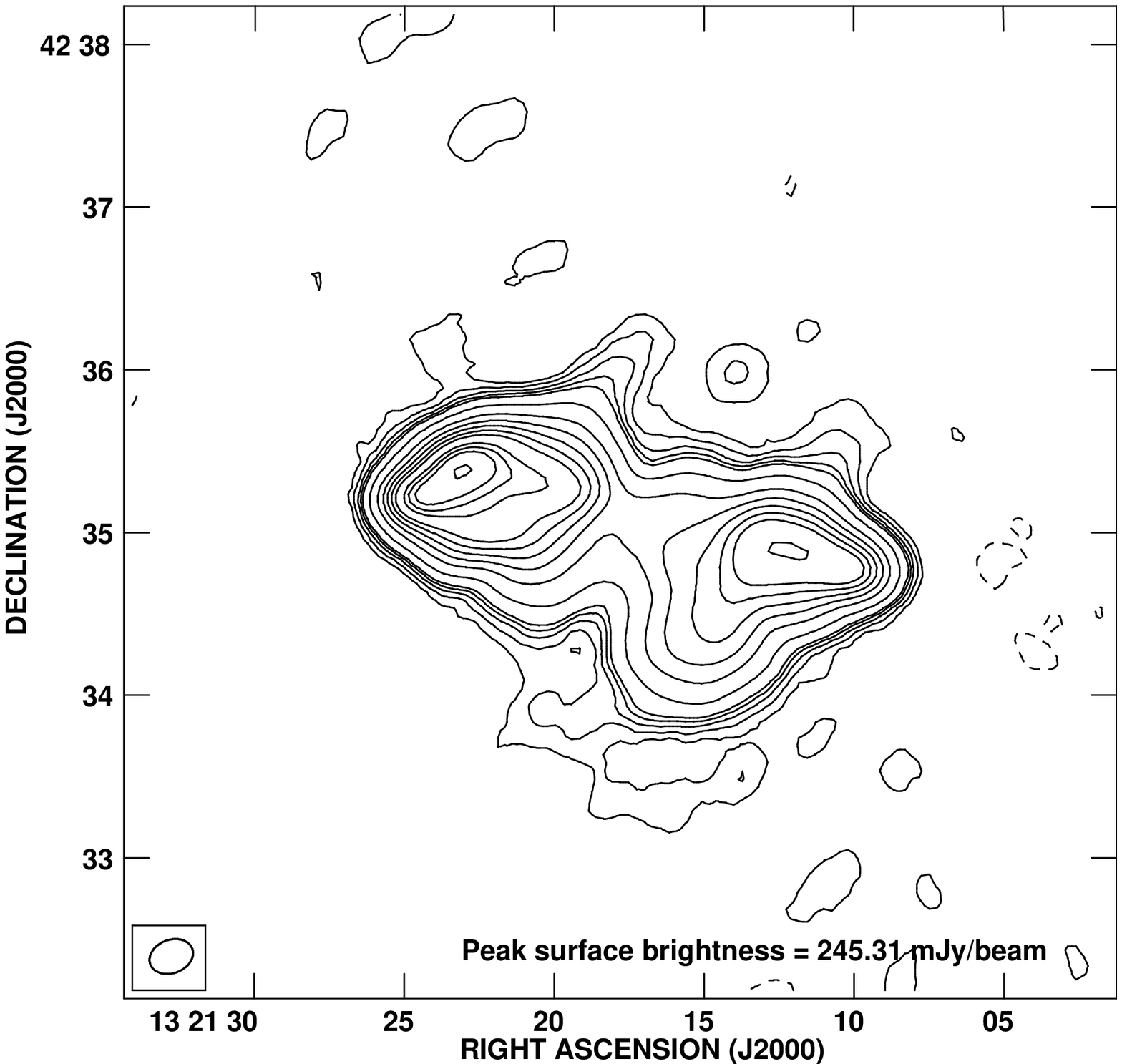} &
\hspace*{-0.2cm}\includegraphics[width=5.24cm]{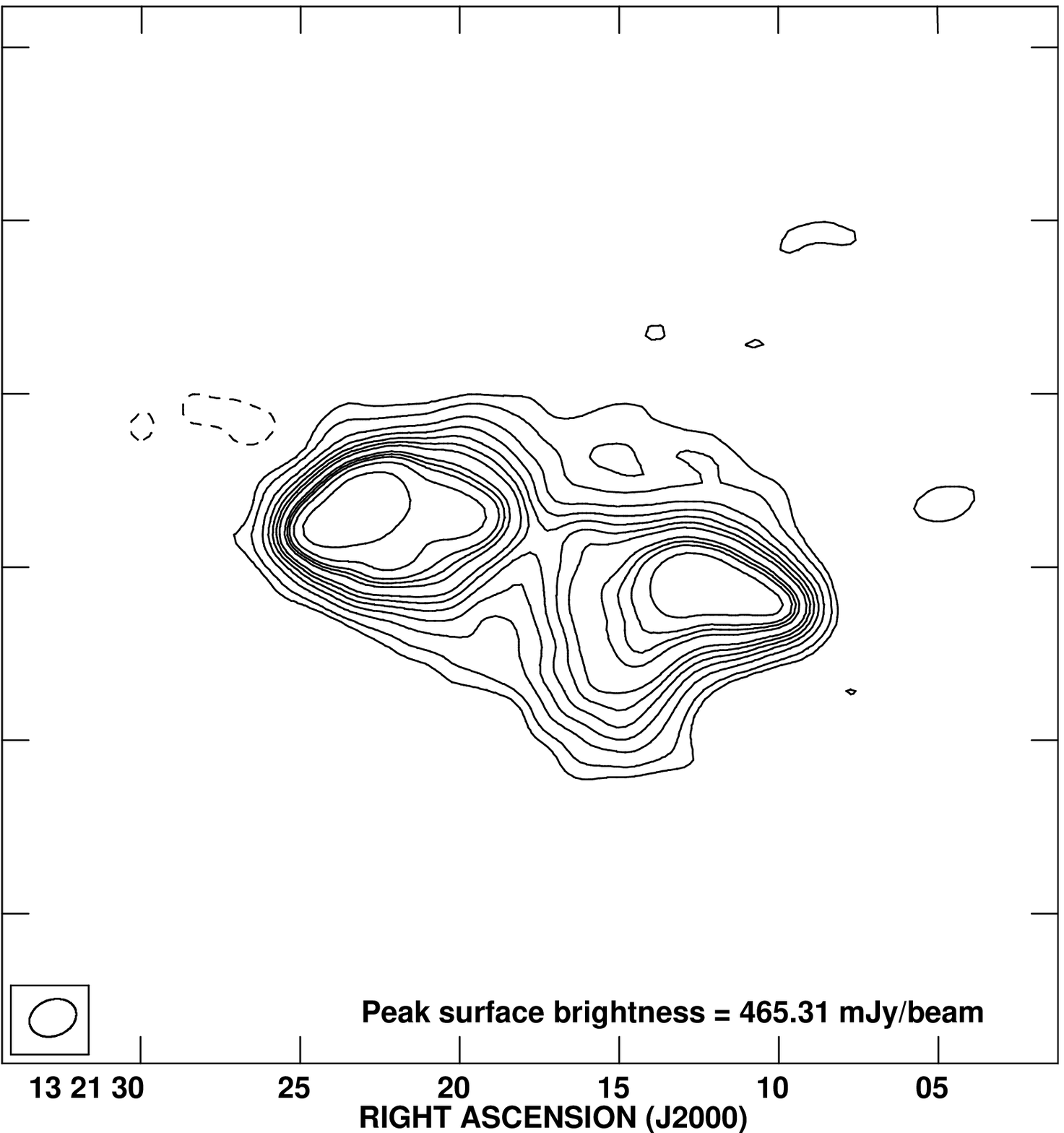} &
\hspace*{-0.2cm}\includegraphics[width=5.24cm]{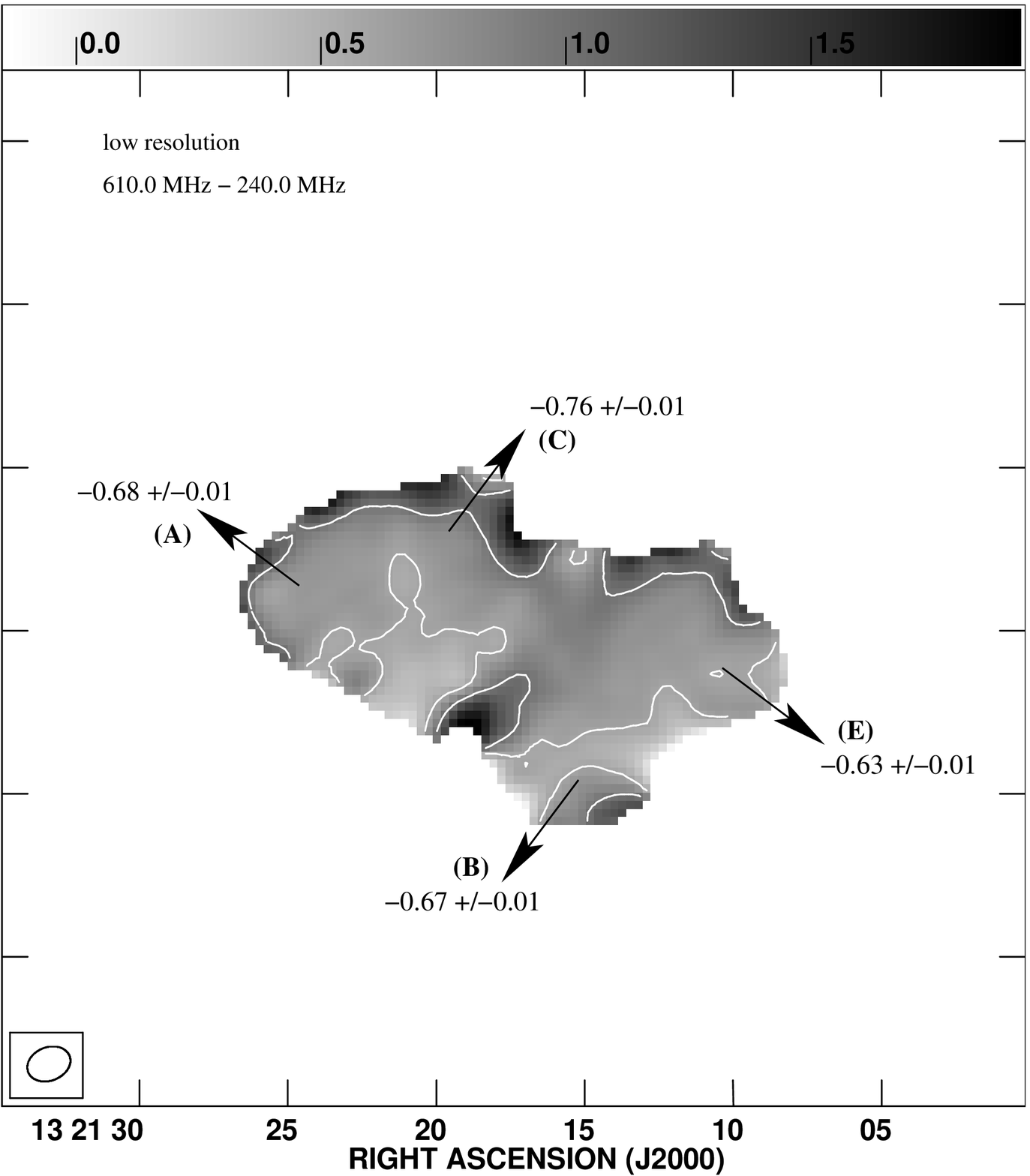} \\
\end{tabular}
\end{center}
\caption{Radio maps of 3C\,285.
Upper left: The VLA map of 3C\,285 at 1.5 GHz
matched with the resolution of 610~MHz; the contour levels in the map are
($-$0.5, 0.5, 1, 2, 4, 6, 8, 12, 16, 24, 48) mJy~beam$^{-1}$.
Upper middle: The GMRT map of 3C\,285 at 610 MHz; the contour levels in the map are
($-$1.2, 1.2, 2, 4, 6, 8, 10, 12, 16, 18, 20) mJy~beam$^{-1}$.
Lower left: The GMRT map of 3C\,285 at 610 MHz 
matched with the resolution of 240~MHz; the contour levels in the map are
($-$2.4, 2.4, 4, 6, 8, 10, 20, 40, 60, 80, 100, 120, 160) mJy~beam$^{-1}$.
Lower middle: The GMRT map of 3C\,285 at 240 MHz; the contour levels in the map are
($-$10, 10, 20, 40, 60, 80, 100, 120, 160, 180, 200, 240) mJy~beam$^{-1}$.
Upper right and Lower right panels: The distribution of the spectral index,
between 1.5~GHz and 610 MHz (upper right),
and 240~MHz and 610 MHz (lower right), for the source.
The spectral index range displayed in the two maps are
$-$2.0 and 1.0 (upper right), and $-$0.1 and 1.9 (lower right), respectively.
The spectral index contours are at $-$1.6, $-$0.9, $-$0.6,~0.0 and
0.6,~1.0, respectively in the two maps.
The spectral indices listed for various regions are
tabulated in Table~\ref{fdregions}.
The r.m.s. noise values in the radio images found at a source free location
are $\sim$0.08, $\sim$0.2 and $\sim$1.5~mJy~beam$^{-1}$ at
1.5~GHz, 610~MHz and 240~MHz, respectively.
The uniformly weighted CLEAN beams for upper and lower panel maps are
6.6~arcsec $\times$~5.1 arcsec
at a P.A. of $-$75.4$^{\circ}$
and
16.5~arcsec $\times$~12.3 arcsec
at a P.A. of $-$69.4$^{\circ}$, respectively.
}
\label{full_syn_285}
\end{figure*}

\begin{figure*}
\begin{center}
\begin{tabular}{lll}
\includegraphics[width=6.0cm]{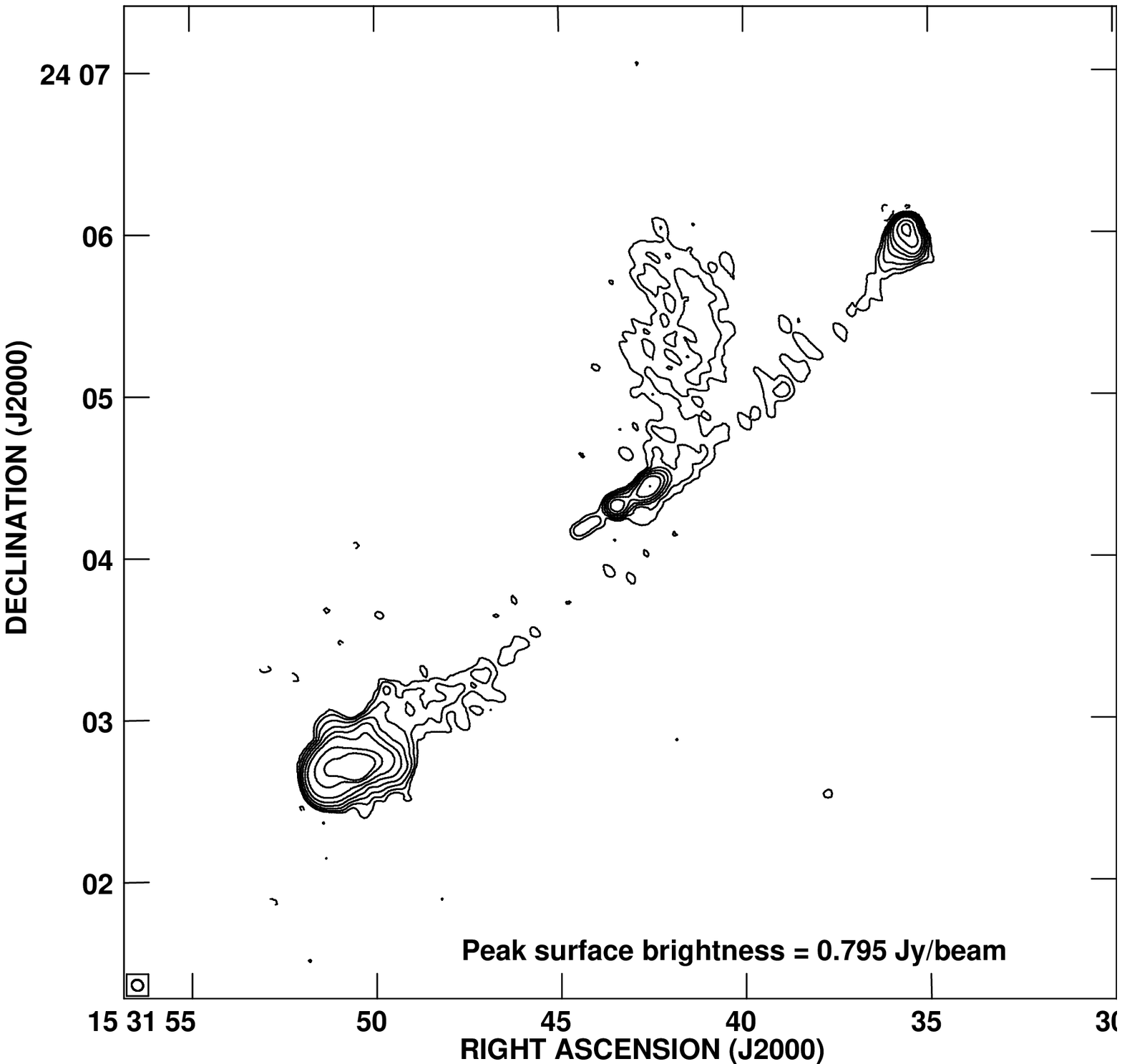} &
\hspace*{-0.2cm}\includegraphics[width=5.24cm]{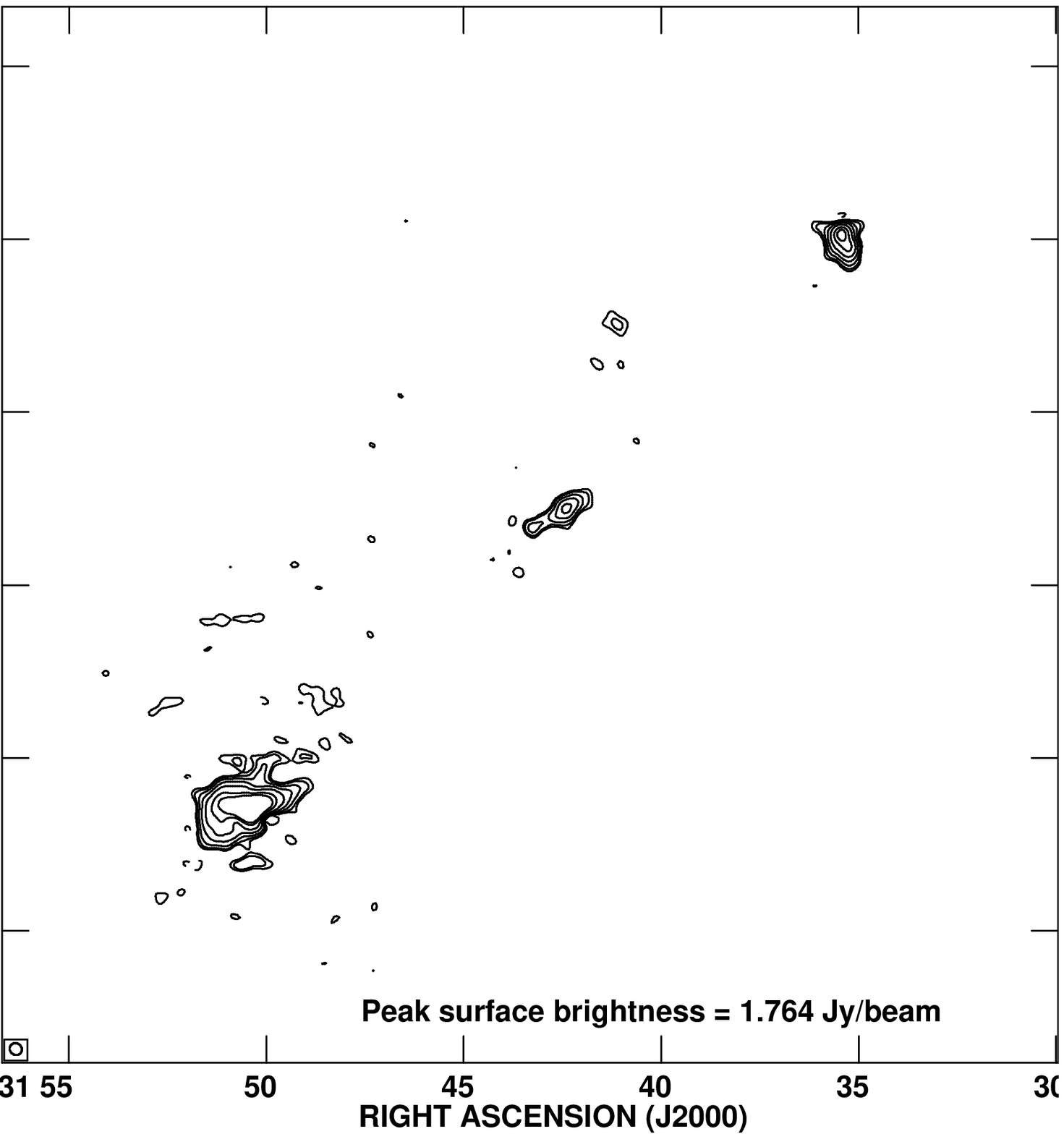} &
\hspace*{-0.2cm}\includegraphics[width=5.24cm]{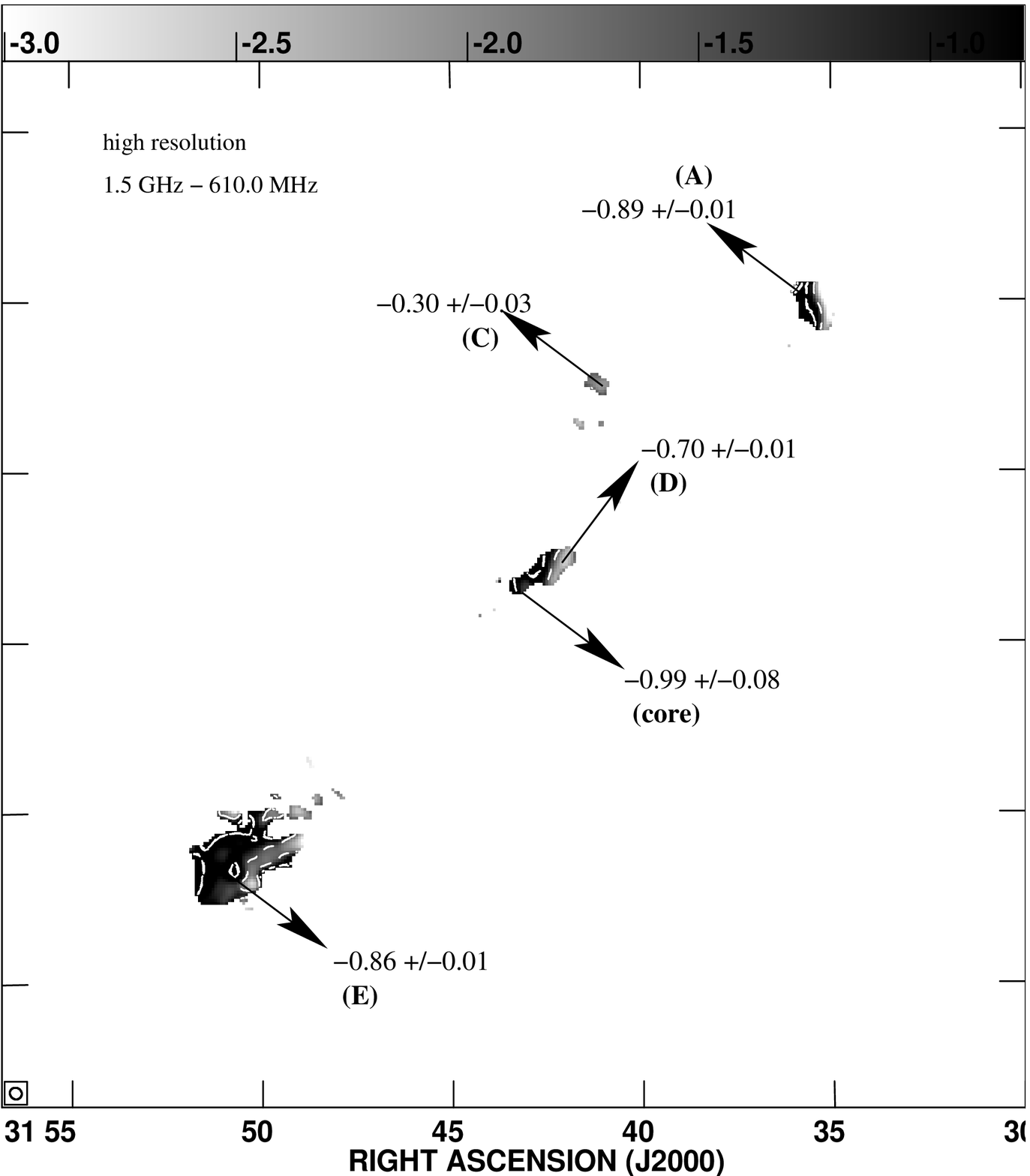} \\
\includegraphics[width=6.0cm]{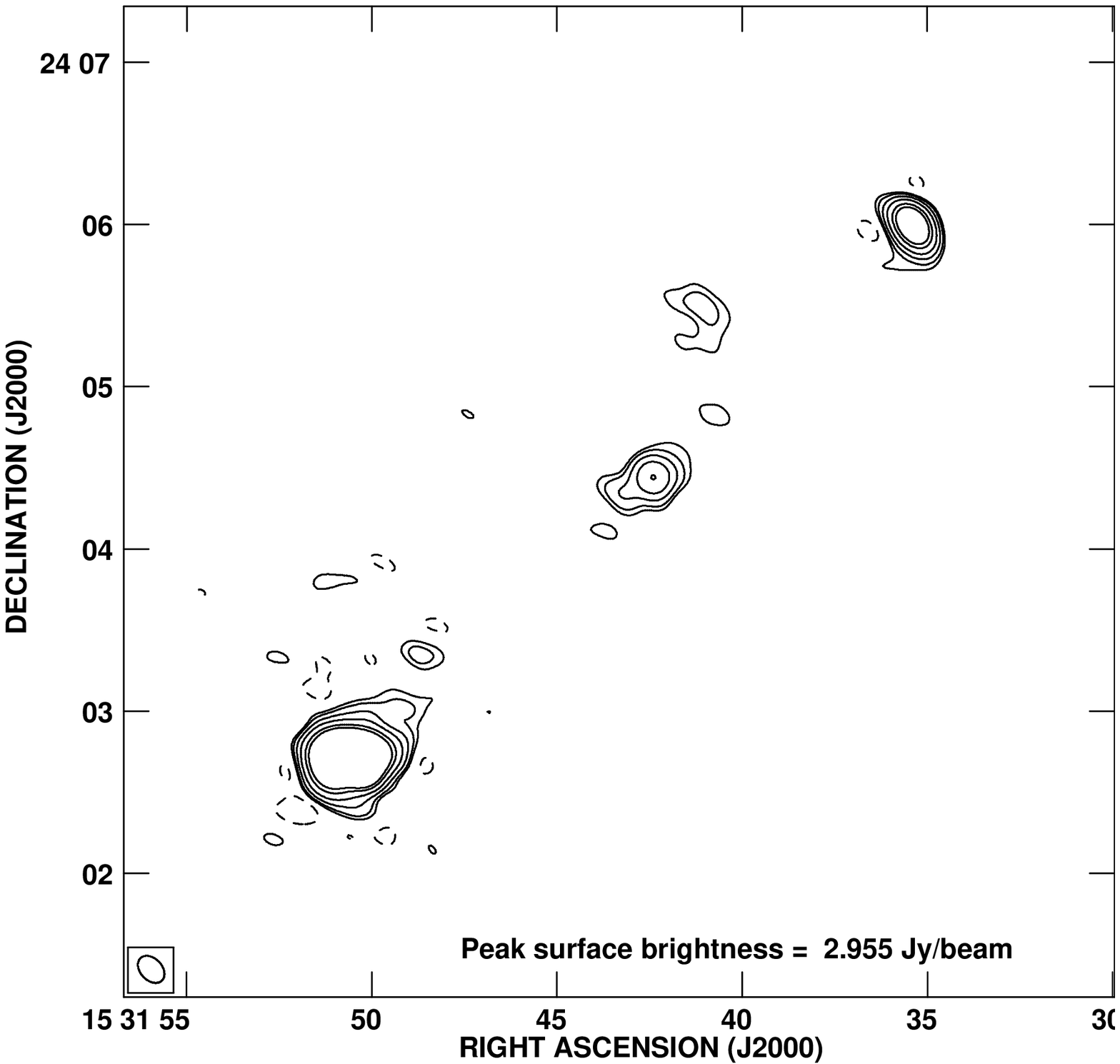} &
\hspace*{-0.2cm}\includegraphics[width=5.24cm]{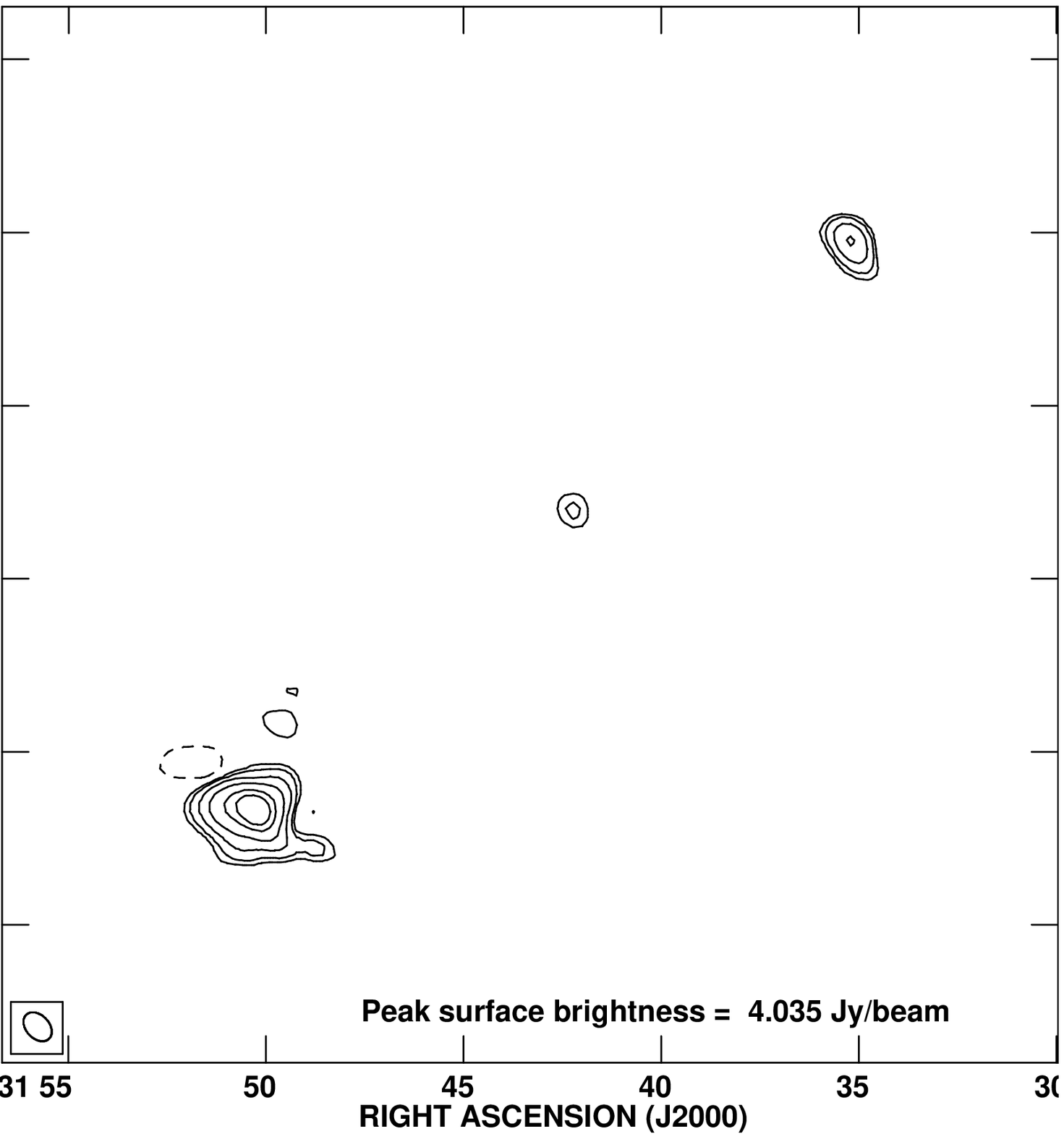} &
\hspace*{-0.2cm}\includegraphics[width=5.24cm]{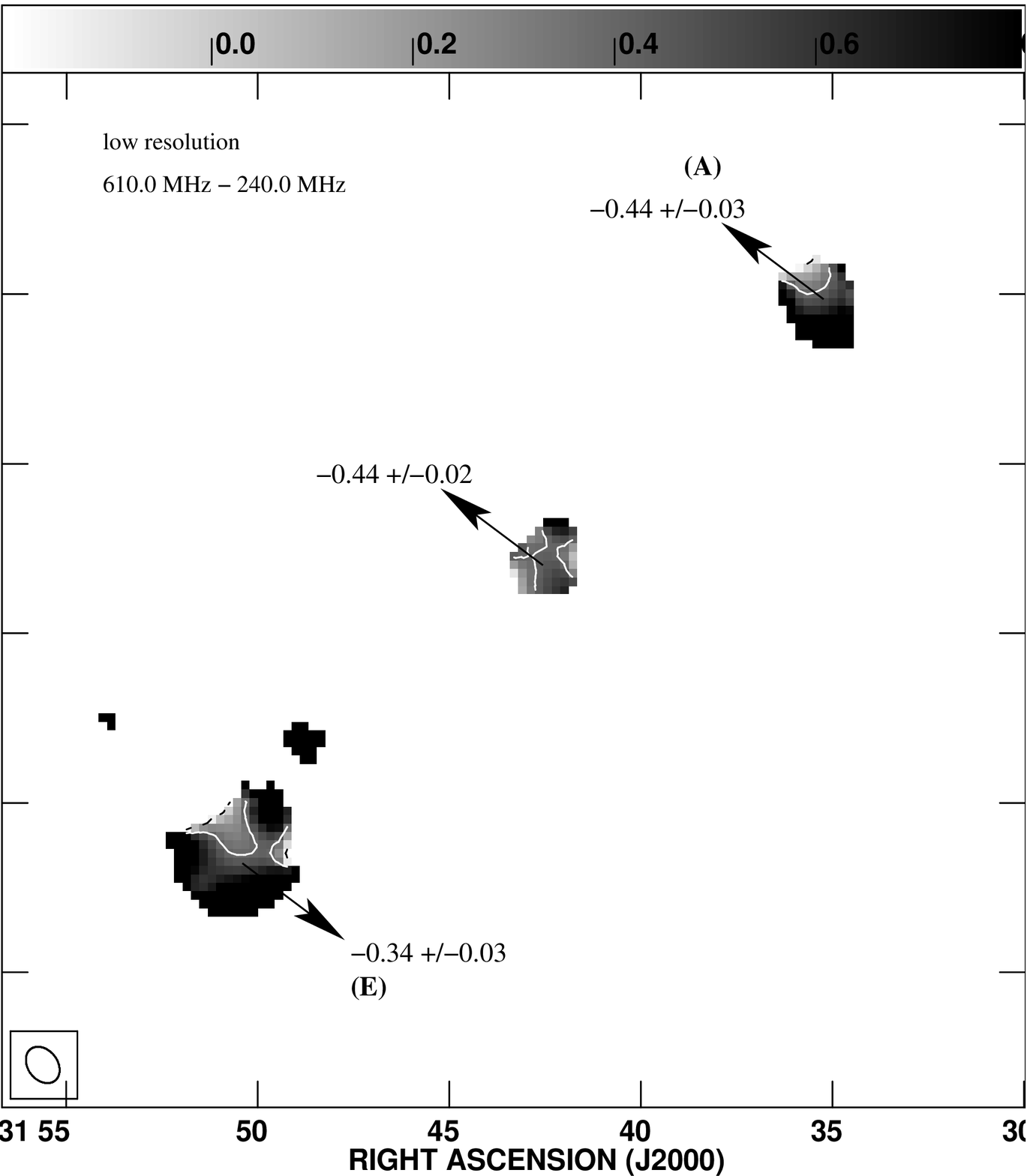} \\
\end{tabular}
\end{center}
\caption{Radio maps of 3C\,321.
Upper left: The VLA map of 3C\,321 at 1.5 GHz
matched with the resolution of 610~MHz; the contour levels in the map are
($-$0.5, 0.5, 1, 2, 4, 8, 16, 48, 96) mJy~beam$^{-1}$.
Upper middle: The GMRT map of 3C\,321 at 610 MHz; the contour levels in the map are
($-$6, 6, 10, 20, 40, 80, 160, 240) mJy~beam$^{-1}$.
Lower left: The GMRT map of 3C\,321 at 610 MHz 
matched with the resolution of 240~MHz; the contour levels in the map are
($-$12, 12, 20, 40, 80, 160, 240) mJy~beam$^{-1}$.
Lower middle: The GMRT map of 3C\,321 at 240 MHz; the contour levels in the map are
($-$120, 120, 200, 400, 800, 1600, 2400) mJy~beam$^{-1}$.
Upper right and Lower right panels: The distribution of the spectral index,
between 1.5~GHz and 610 MHz (upper right),
and 240~MHz and 610 MHz (lower right), for the source.
The spectral index range displayed in the two maps are
$-$3.0 and $-$0.8 (upper right), and $-$2.0 and 0.8 (lower right), respectively.
The spectral index contours are at $-$1.5, $-$0.1,~0.0 and
$-$0.2,~0.4, respectively in the two maps.
The spectral indices listed for various regions are
tabulated in Table~\ref{fdregions}.
The r.m.s. noise values in the radio images found at a source free location
are $\sim$0.08, $\sim$0.8 and $\sim$3.3~mJy~beam$^{-1}$ at
1.5~GHz, 610~MHz and 240~MHz, respectively.
The uniformly weighted CLEAN beams for upper and lower panel maps are
4.3~arcsec $\times$~4.0 arcsec
at a P.A. of $+$54.2$^{\circ}$
and
11.5~arcsec $\times$~8.0 arcsec
at a P.A. of $+$43.3$^{\circ}$, respectively.
}
\label{full_syn_321}
\end{figure*}

\begin{figure*}
\begin{center}
\begin{tabular}{lll}
\includegraphics[width=6.3cm]{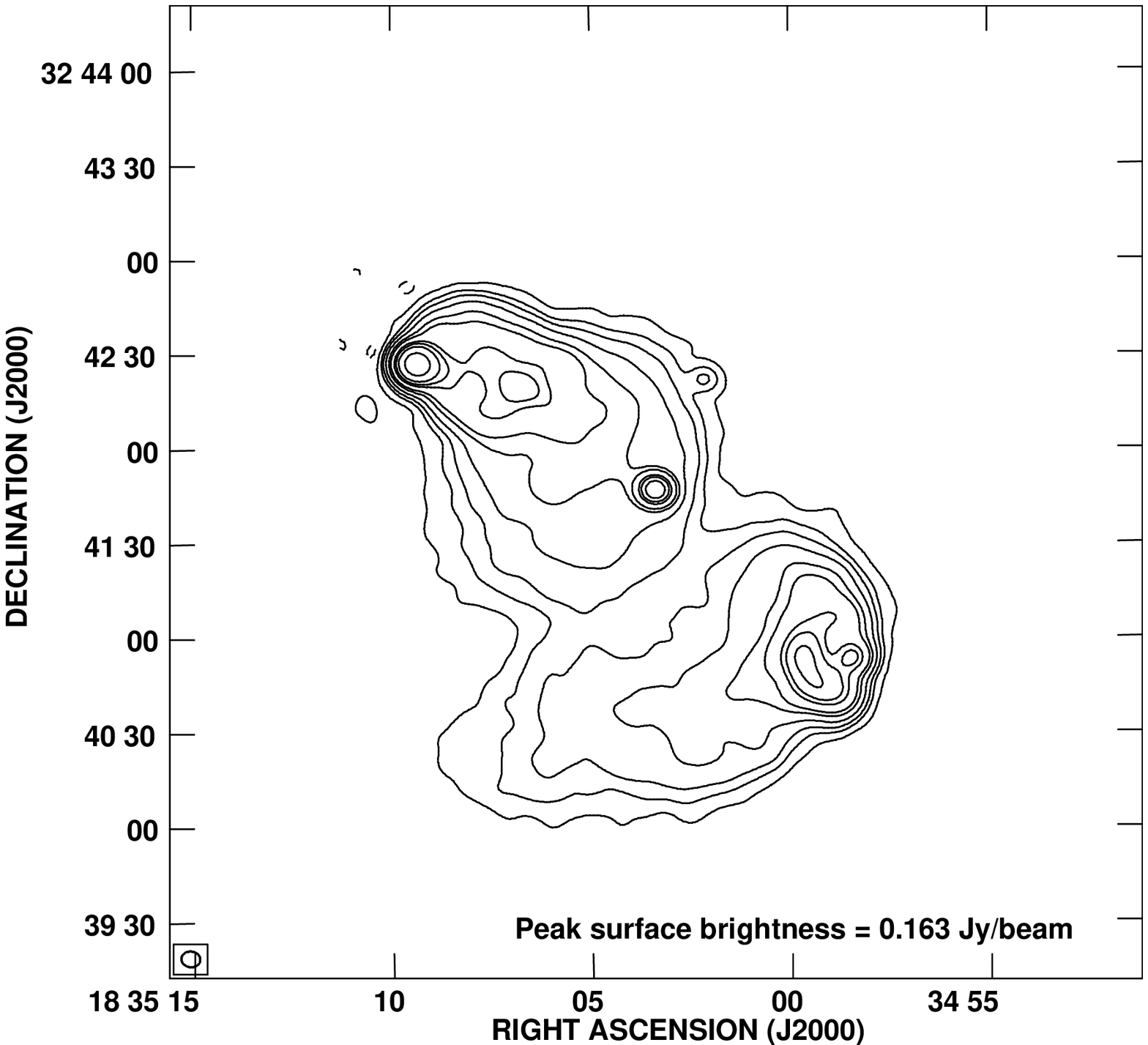} &
\hspace*{-0.2cm}\includegraphics[width=5.24cm]{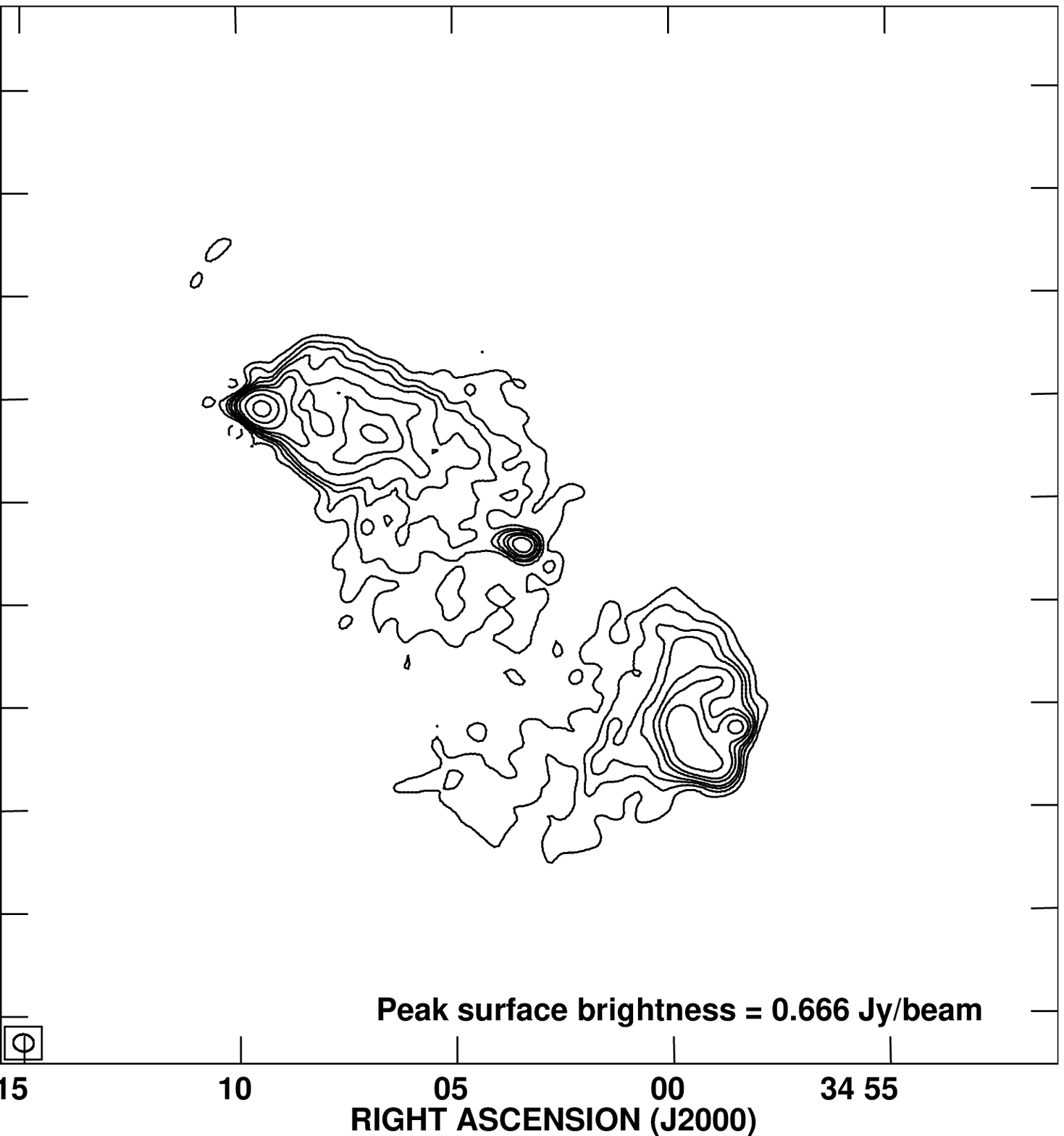} &
\hspace*{-0.2cm}\includegraphics[width=5.24cm]{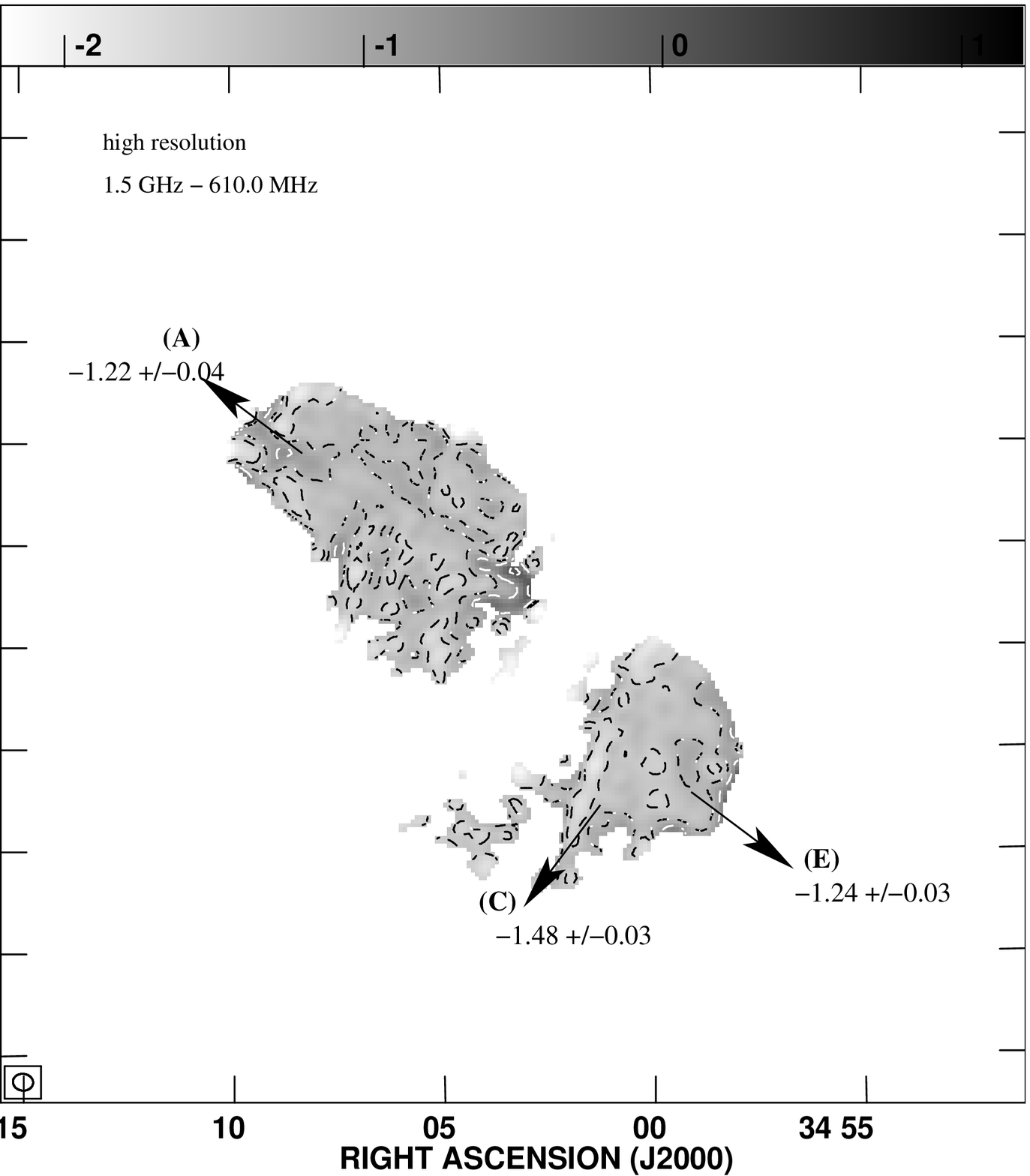} \\
\includegraphics[width=6.3cm]{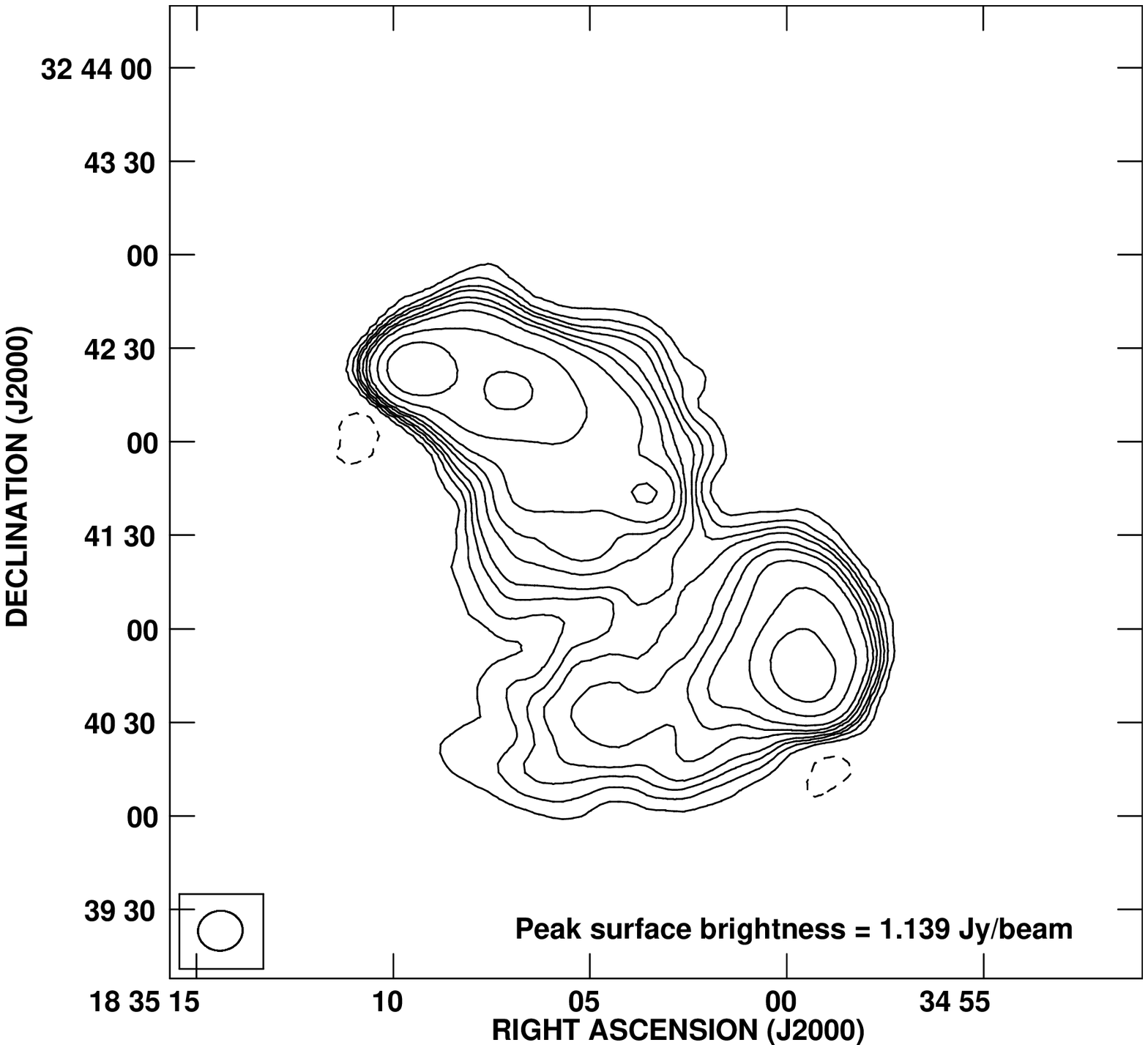} &
\hspace*{-0.2cm}\includegraphics[width=5.24cm]{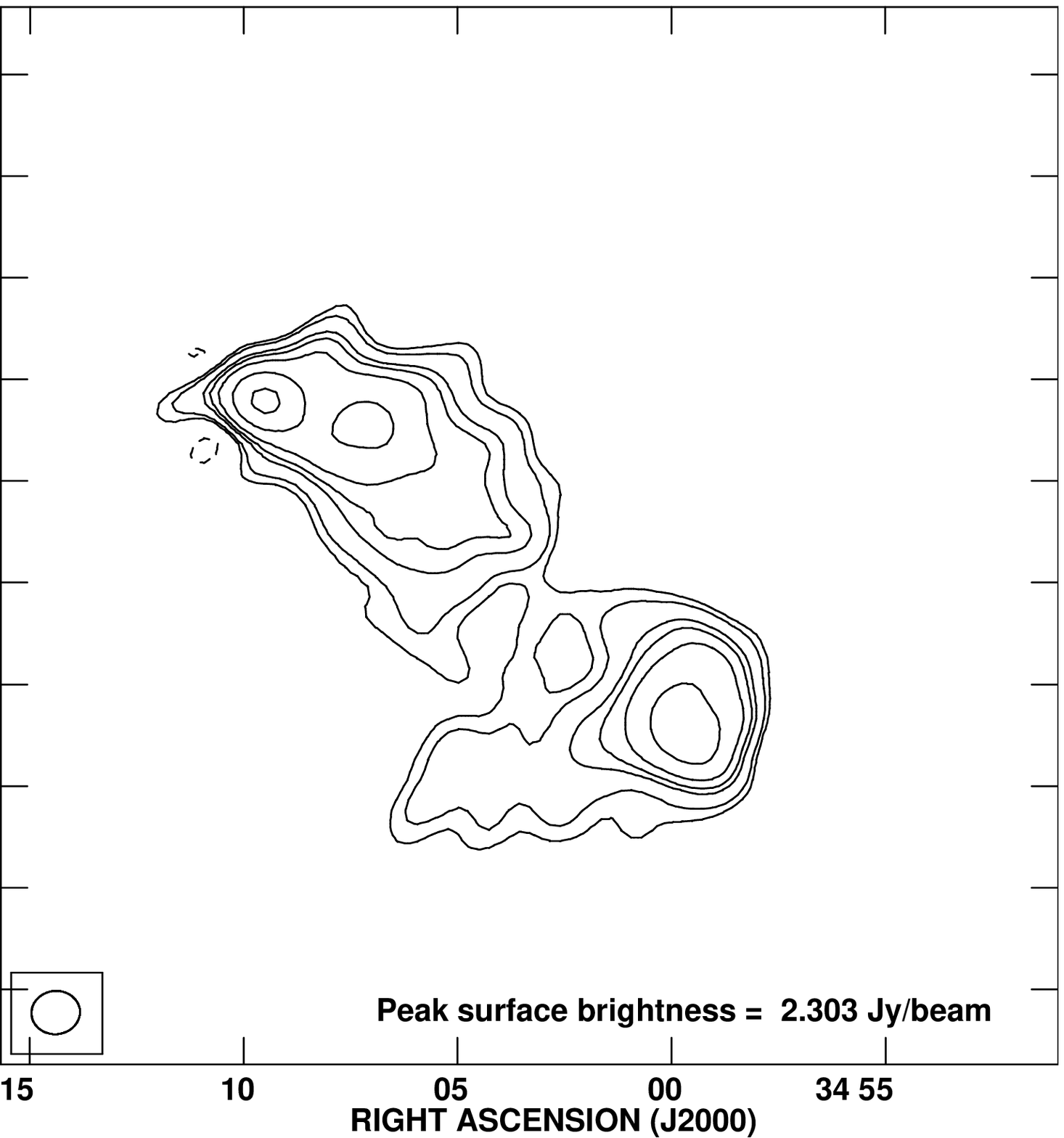} &
\hspace*{-0.2cm}\includegraphics[width=5.24cm]{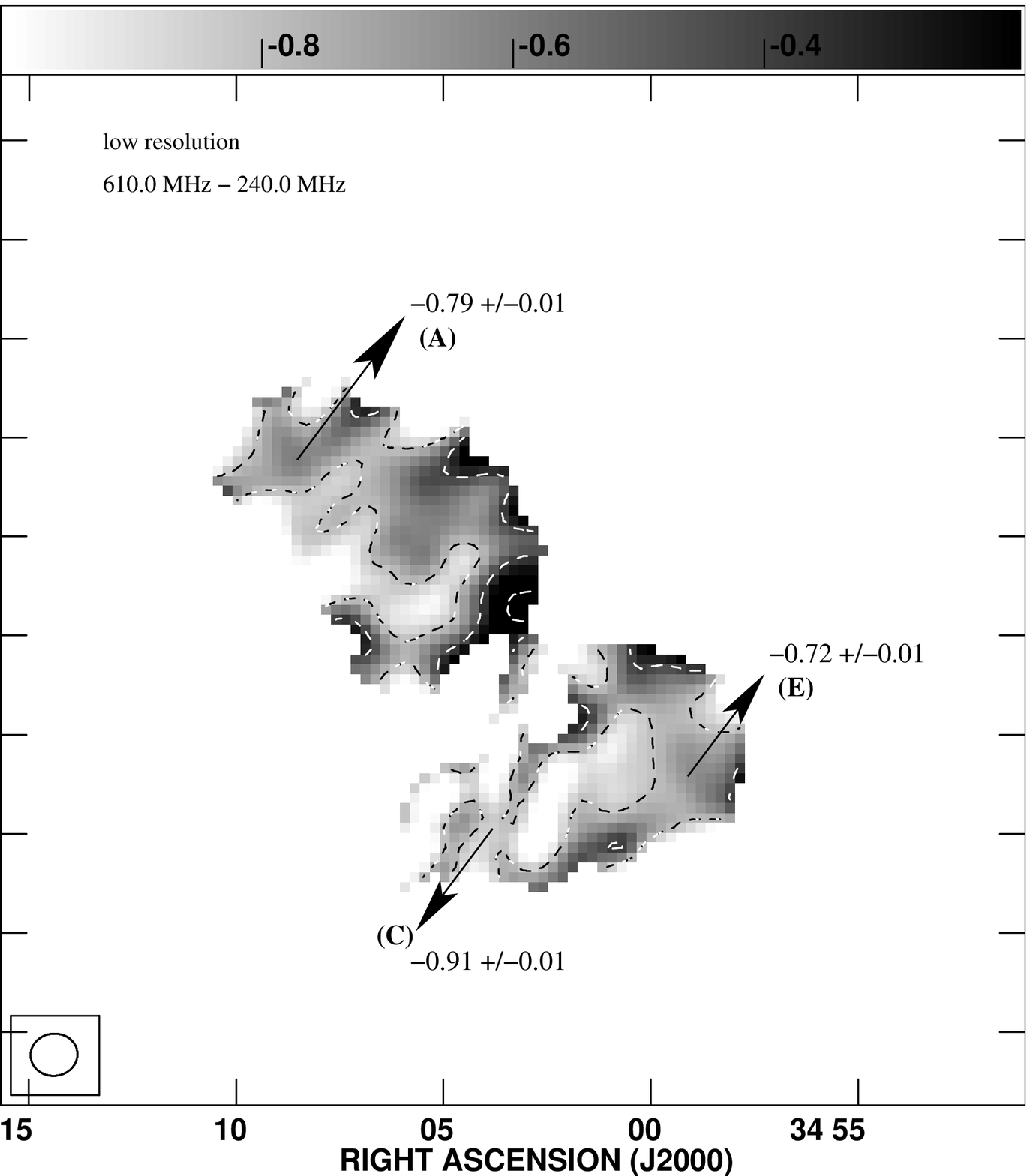} \\
\end{tabular}
\end{center}
\caption{Radio maps of 3C\,382.
Upper left: The VLA map of 3C\,382 at 1.5 GHz
matched with the resolution of 610~MHz; the contour levels in the map are
($-$0.2, 0.2, 1, 2, 4, 8, 12, 24, 32, 48, 96) mJy~beam$^{-1}$.
Upper middle: The GMRT map of 3C\,382 at 610 MHz; the contour levels in the map are
($-$10, 10, 20, 30, 40, 60, 80, 120, 240, 480) mJy~beam$^{-1}$.
Lower left: The GMRT map of 3C\,382 at 610 MHz 
matched with the resolution of 240~MHz; the contour levels in the map are
($-$10, 10, 20, 30, 40, 60, 80, 120, 240, 480) mJy~beam$^{-1}$.
Lower middle: The GMRT map of 3C\,382 at 240 MHz; the contour levels in the map are
($-$80, 80, 120, 240, 320, 480, 960, 1920) mJy~beam$^{-1}$.
Upper right and Lower right panels: The distribution of the spectral index,
between 1.5~GHz and 610 MHz (upper right),
and 240~MHz and 610 MHz (lower right), for the source.
The spectral index range displayed in the two maps are
$-$2.2 and 1.2 (upper right), and $-$0.2 and 1.2 (lower right), respectively.
The spectral index contours are at $-$1.5, $-$1.2, $-$0.8,~0.0 and
0.2,~0.8, respectively in the two maps.
The spectral indices listed for various regions are
tabulated in Table~\ref{fdregions}.
The r.m.s. noise values in the radio images found at a source free location
are $\sim$0.02, $\sim$0.6 and $\sim$2.9~mJy~beam$^{-1}$ at
1.5~GHz, 610~MHz and 240~MHz, respectively.
The uniformly weighted CLEAN beams for upper and lower panel maps are
6.0~arcsec $\times$~5.0 arcsec
at a P.A. of $+$83.1$^{\circ}$
and
14.3~arcsec $\times$~12.3 arcsec
at a P.A. of $-$84.5$^{\circ}$, respectively.
}
\label{full_syn_382}
\end{figure*}

\begin{figure*}
\begin{center}
\begin{tabular}{lll}
\includegraphics[width=6.3cm]{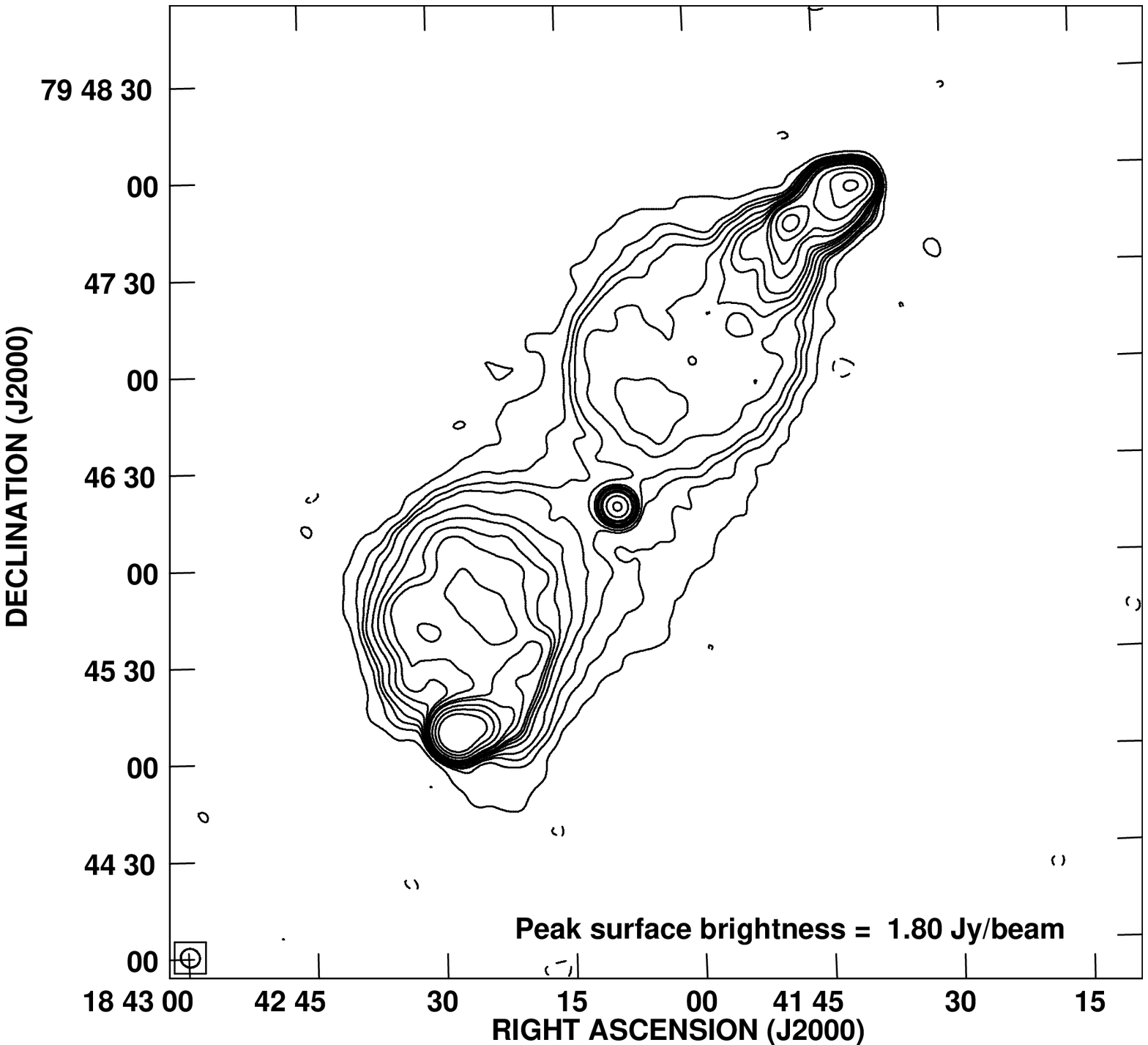} &
\hspace*{-0.2cm}\includegraphics[width=5.24cm]{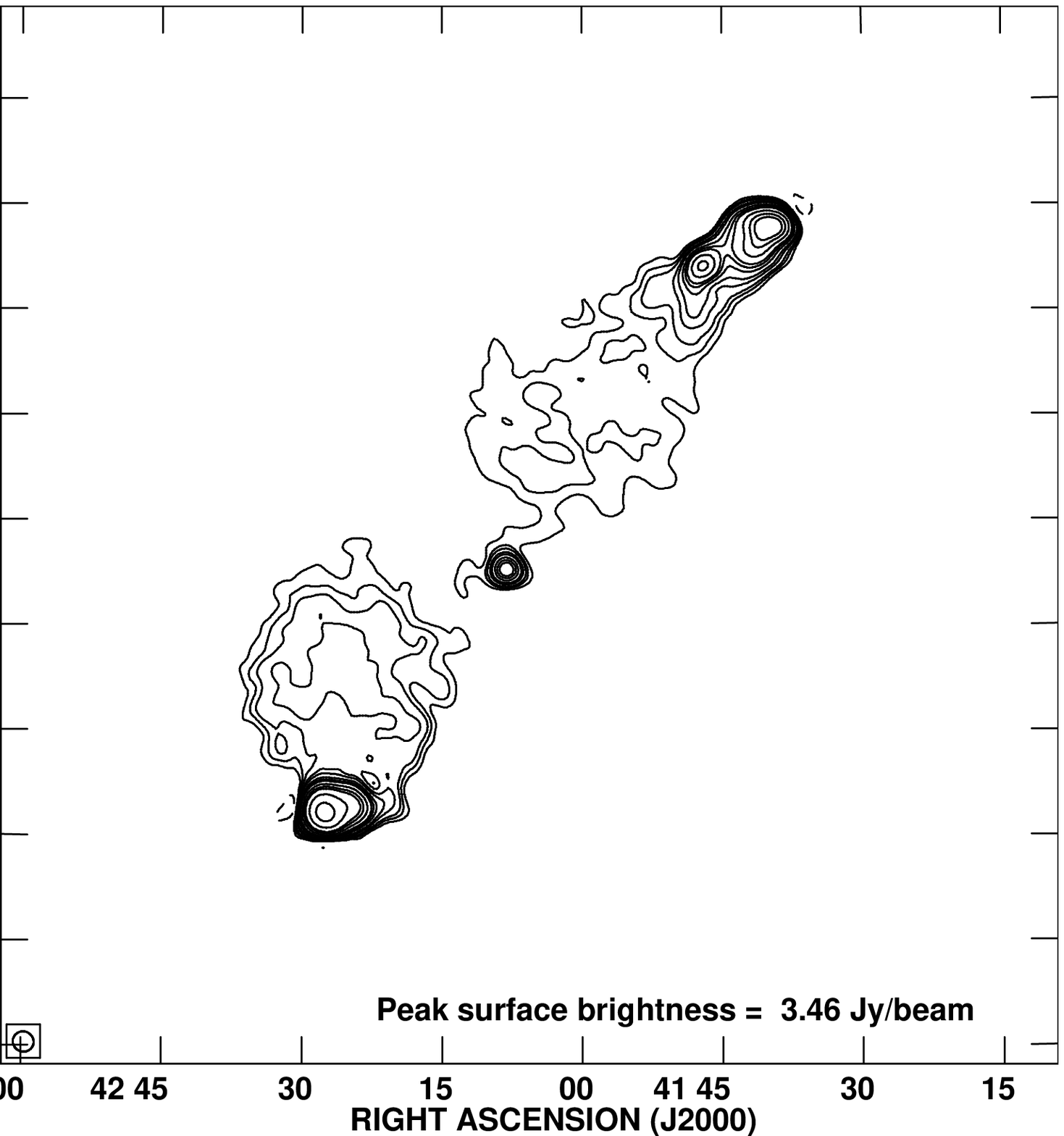} &
\hspace*{-0.2cm}\includegraphics[width=5.24cm]{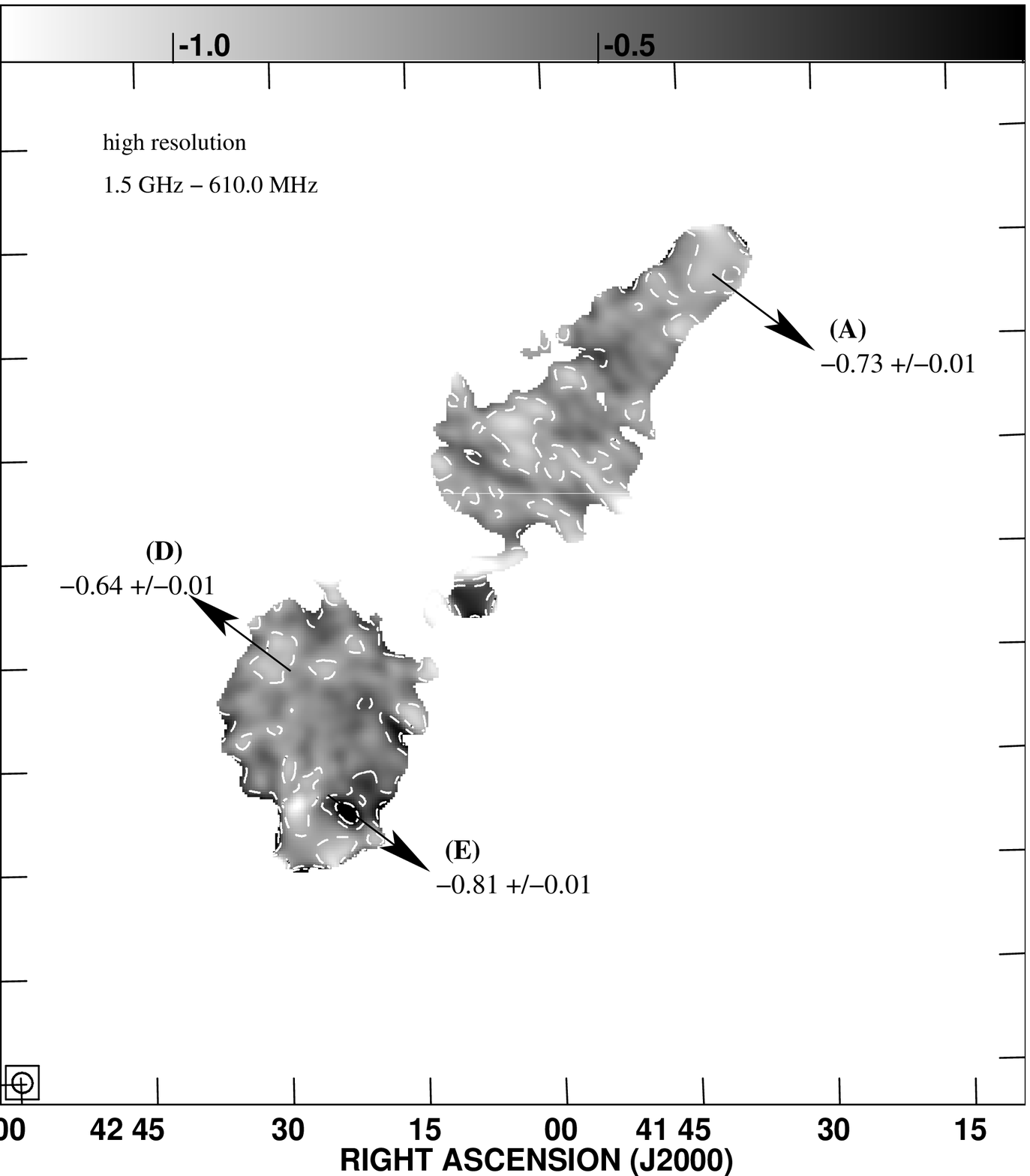} \\
\includegraphics[width=6.3cm]{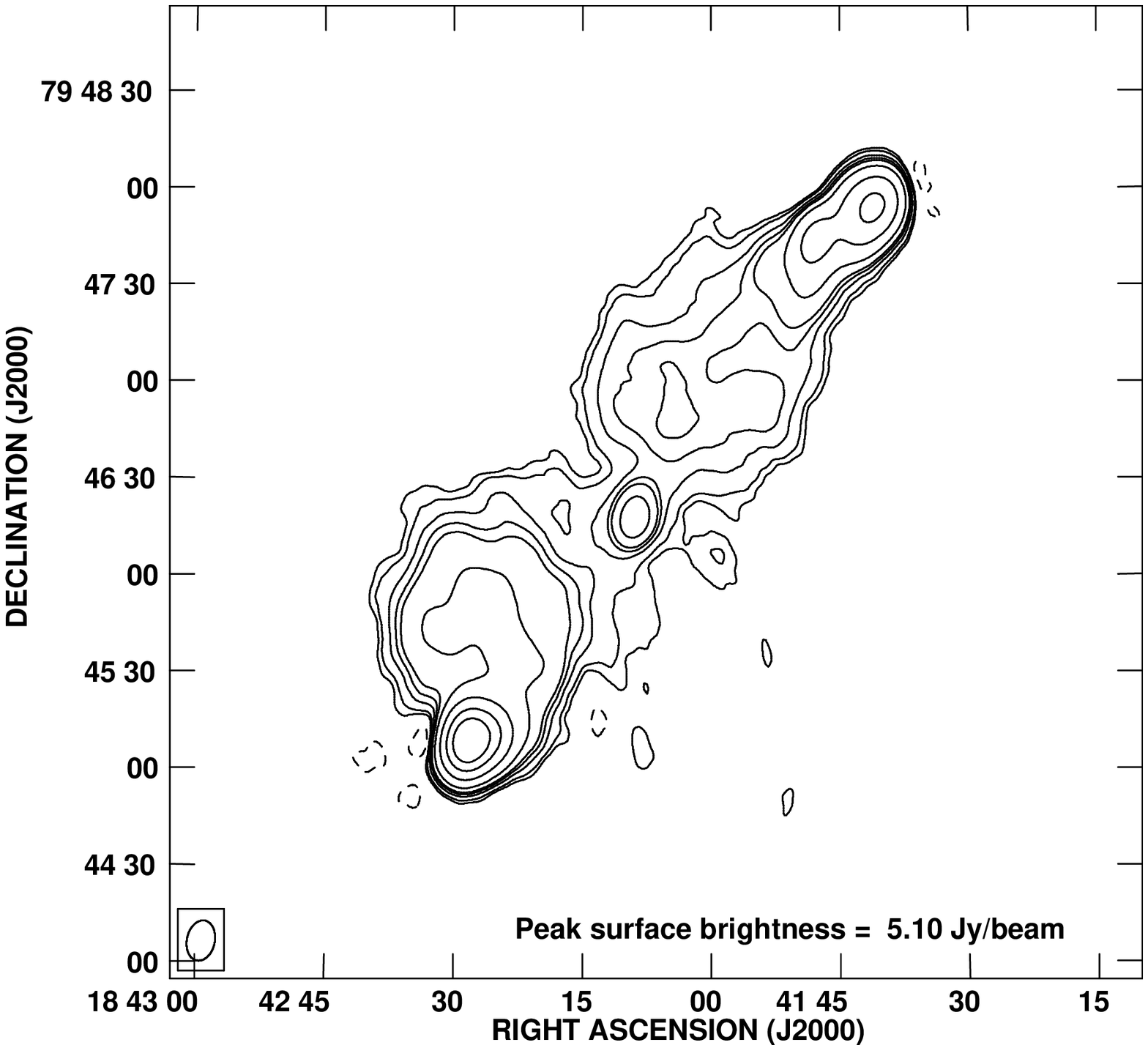} &
\hspace*{-0.2cm}\includegraphics[width=5.24cm]{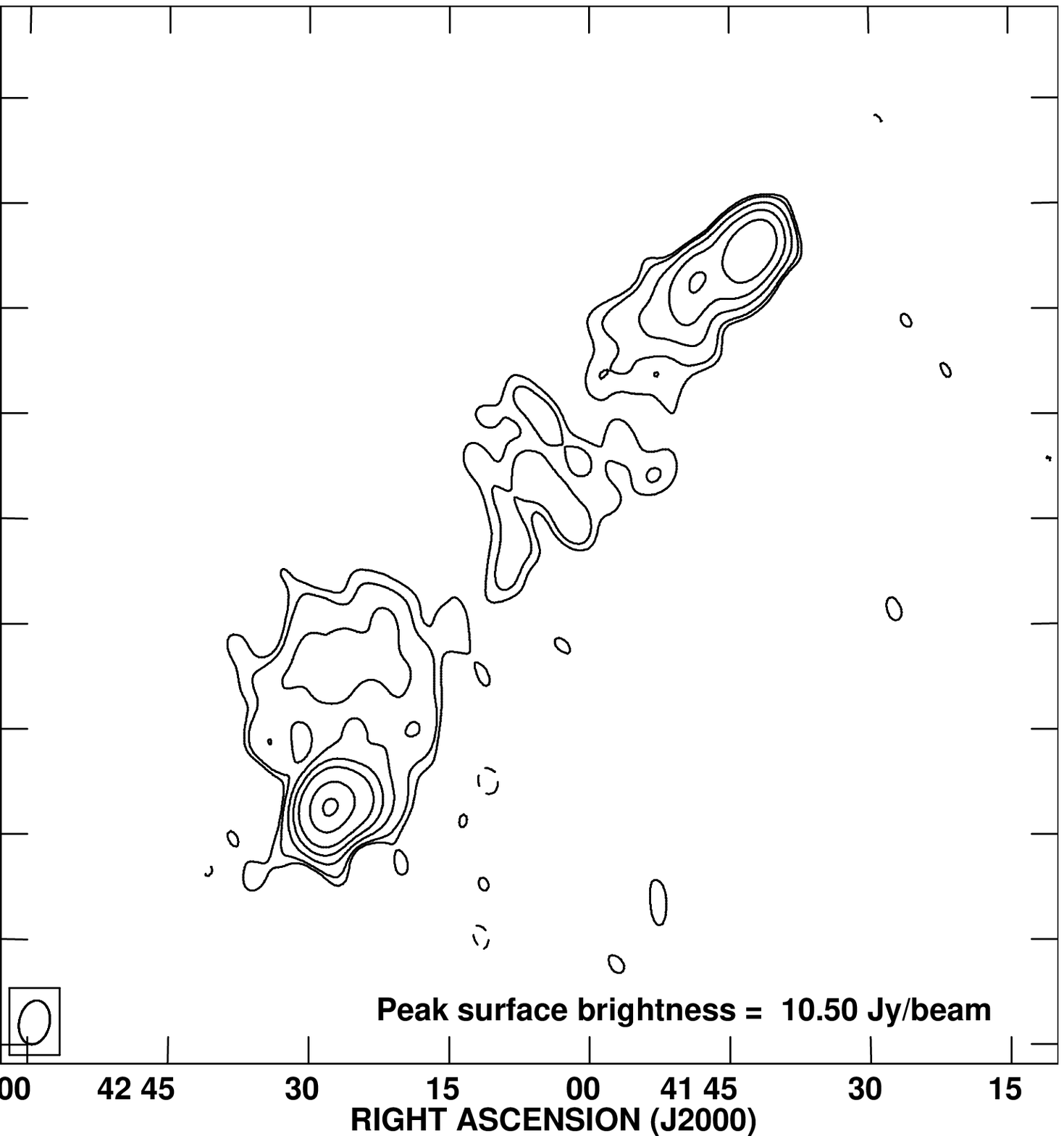} &
\hspace*{-0.2cm}\includegraphics[width=5.24cm]{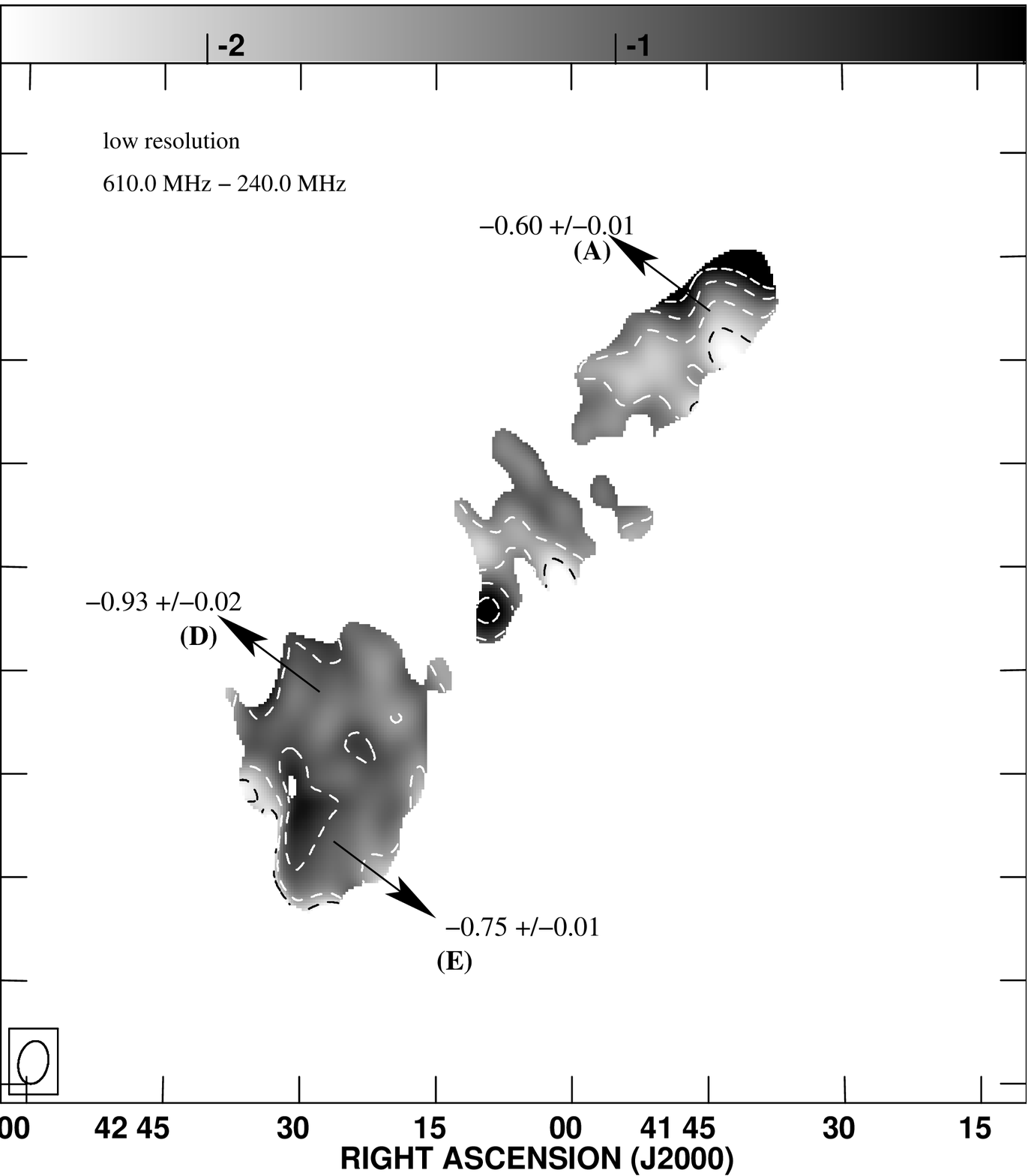} \\
\end{tabular}
\end{center}
\caption{Radio maps of 3C\,390.3.
Upper left: The VLA map of 3C\,390.3 at 1.5 GHz
matched with the resolution of 610~MHz; the contour levels in the map are
($-$1.2, 1.2, 4, 6, 8, 12, 16, 20, 24, 36, 48, 96, 192) mJy~beam$^{-1}$.
Upper middle: The GMRT map of 3C\,390.3 at 610 MHz; the contour levels in the map are
($-$20, 20, 30, 40, 60, 80, 120, 160, 200, 320, 480, 640) mJy~beam$^{-1}$.
Lower left: The GMRT map of 3C\,390.3 at 610 MHz 
matched with the resolution of 240~MHz; the contour levels in the map are
($-$12, 12, 20, 40, 60, 80, 160, 480, 1280) mJy~beam$^{-1}$.
Lower middle: The GMRT map of 3C\,390.3 at 240 MHz; the contour levels in the map are
($-$140, 140, 200, 400, 800, 1600, 4800, 9600) mJy~beam$^{-1}$.
Upper right and Lower right panels: The distribution of the spectral index,
between 1.5~GHz and 610 MHz (upper right),
and 240~MHz and 610 MHz (lower right), for the source.
The spectral index range displayed in the two maps are
$-$1.2 and 0.0 (upper right), and $-$2.5 and 0.0 (lower right), respectively.
The spectral index contours are at $-$0.8, $-$0.4,~0.0
and $-$2.2, $-$1.5, $-$0.7,~0.0, respectively in the two maps.
The spectral indices listed for various regions are
tabulated in Table~\ref{fdregions}.
The r.m.s. noise values in the radio images found at a source free location
are $\sim$0.3, $\sim$1.2 and $\sim$2.4~mJy~beam$^{-1}$ at
1.5~GHz, 610~MHz and 240~MHz, respectively.
The uniformly weighted CLEAN beams for upper and lower panel maps are
6.0~arcsec $\times$~5.7 arcsec
at a P.A. of $+$63.0$^{\circ}$
and
12.6~arcsec $\times$~8.7 arcsec
at a P.A. of $-$12.1$^{\circ}$, respectively.
}
\label{full_syn_390_3}
\end{figure*}

\begin{figure*}
\begin{center}
\begin{tabular}{lll}
\includegraphics[width=6.3cm]{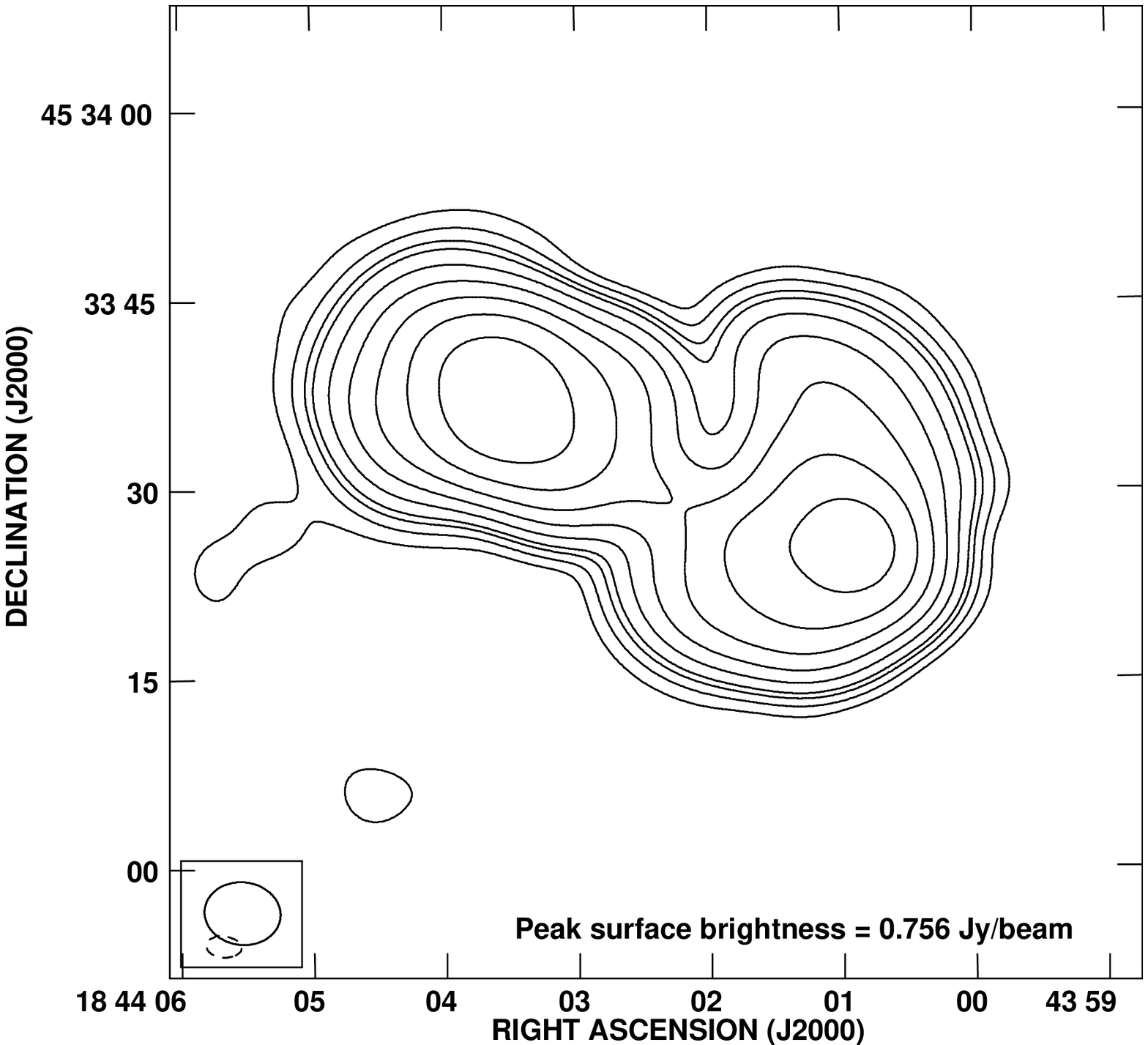} &
\hspace*{-0.2cm}\includegraphics[width=5.24cm]{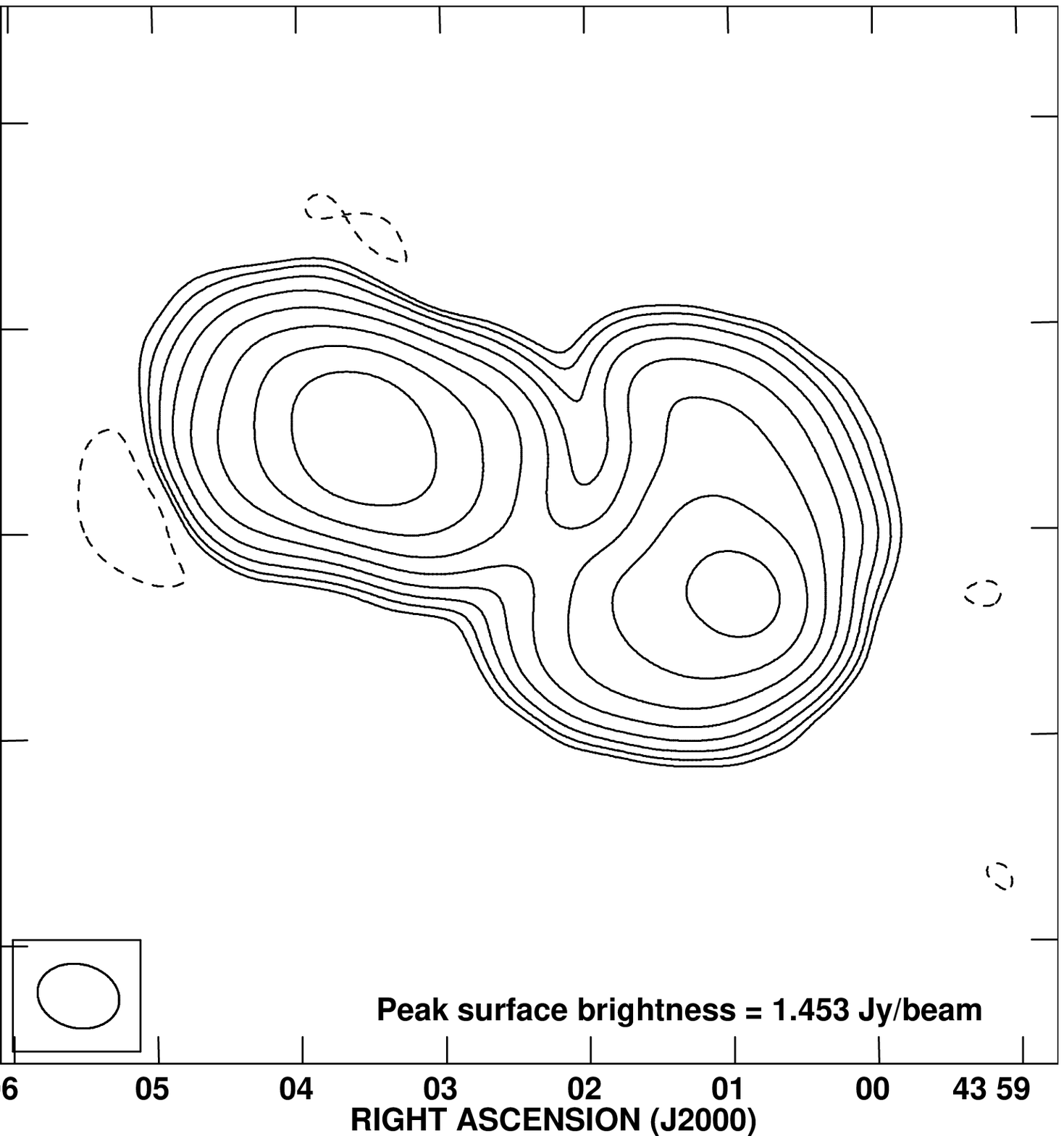} &
\hspace*{-0.2cm}\includegraphics[width=5.24cm]{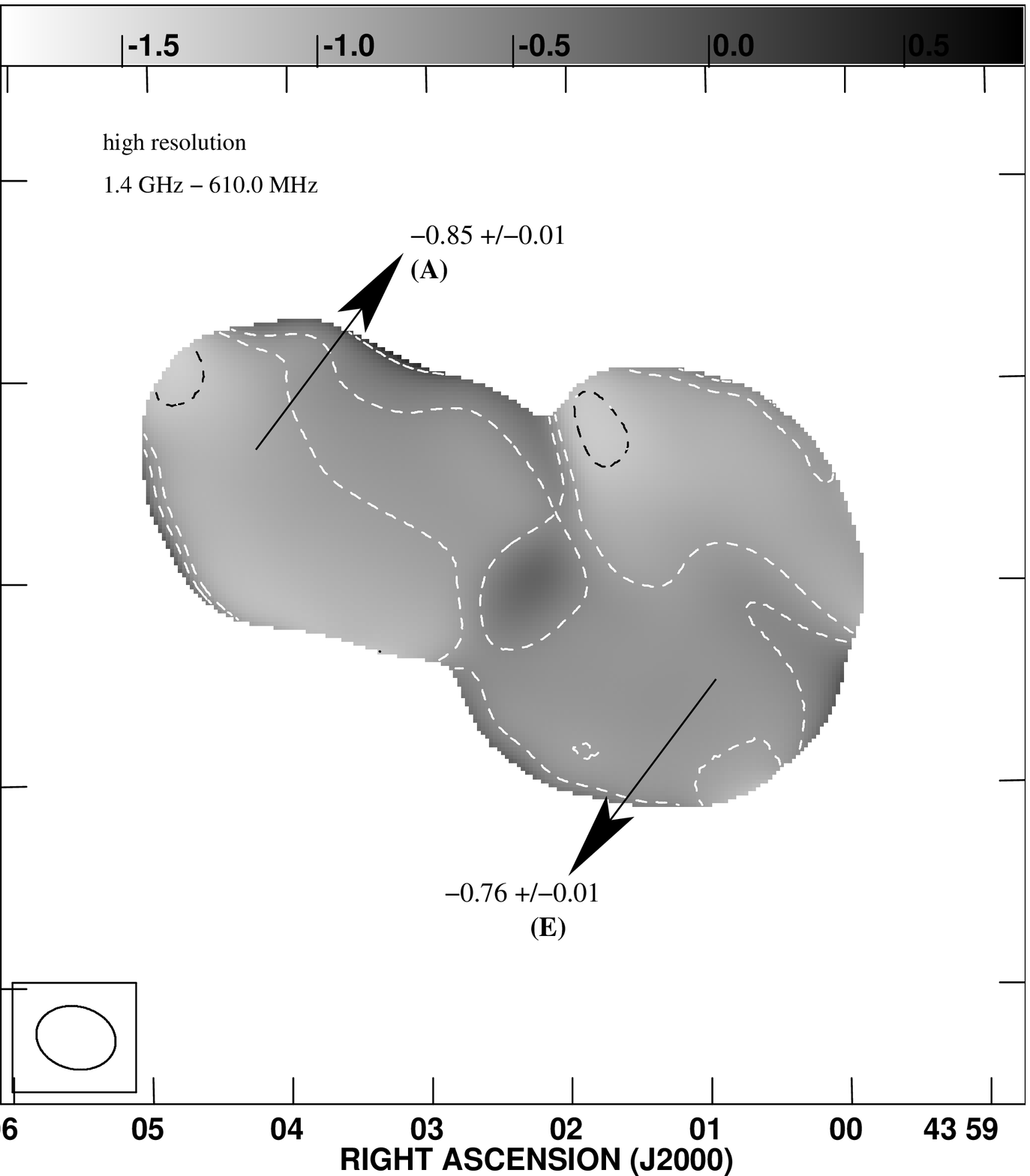} \\
\includegraphics[width=6.3cm]{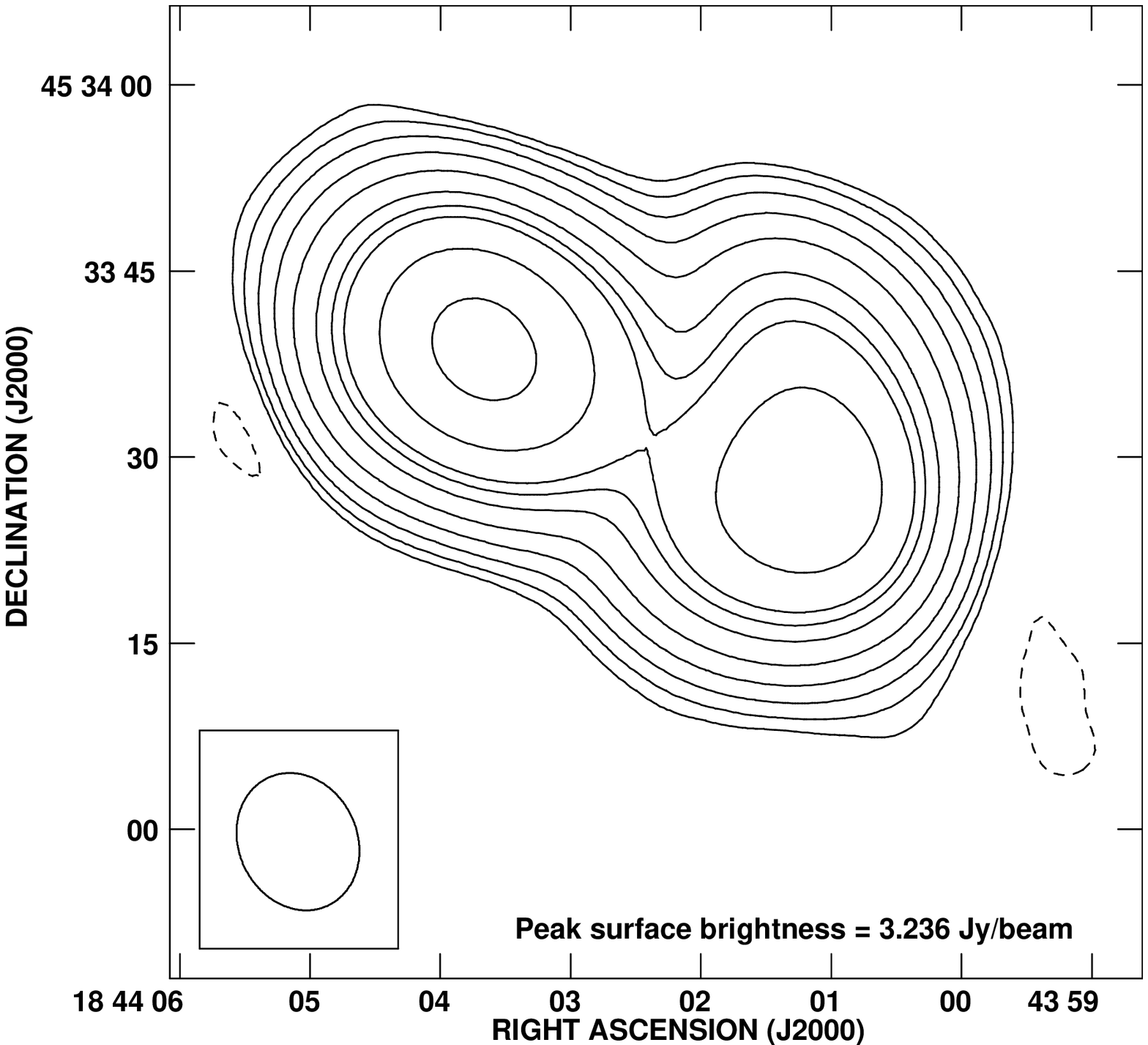} &
\hspace*{-0.2cm}\includegraphics[width=5.24cm]{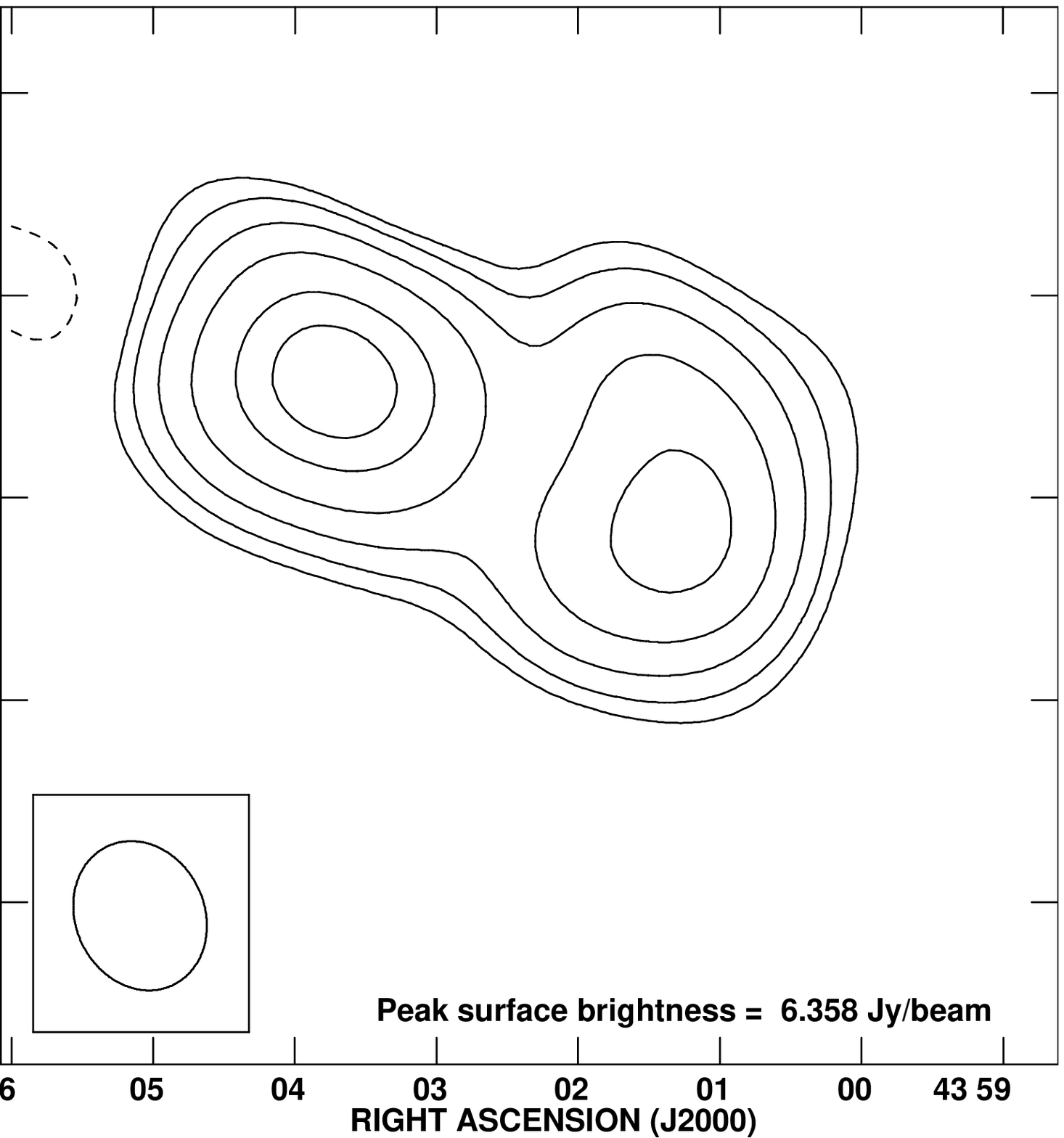} &
\hspace*{-0.2cm}\includegraphics[width=5.24cm]{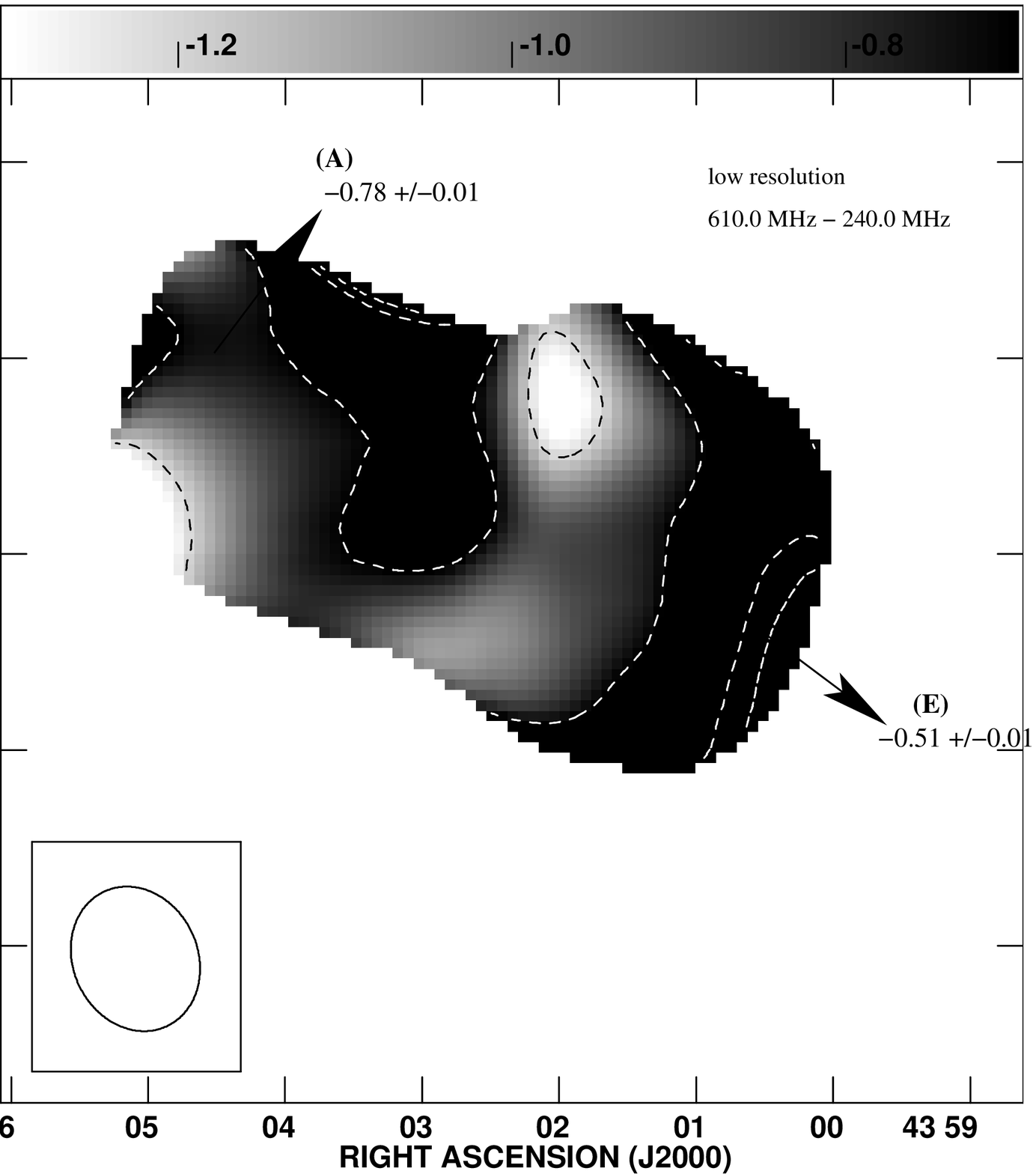} \\
\end{tabular}
\end{center}
\caption{Radio maps of 3C\,388.
Upper left: The VLA map of 3C\,388 at 1.4 GHz
matched with the resolution of 610~MHz; the contour levels in the map are
($-$1.4, 1.4, 4, 8, 12, 24, 48, 96, 192) mJy~beam$^{-1}$.
Upper middle: The GMRT map of 3C\,388 at 610 MHz; the contour levels in the map are
($-$6, 6, 12, 24, 48, 96, 192, 384, 768) mJy~beam$^{-1}$.
Lower left: The GMRT map of 3C\,388 at 610 MHz 
matched with the resolution of 240~MHz; the contour levels in the map are
($-$10, 10, 20, 40, 80, 160, 320, 480, 640) mJy~beam$^{-1}$.
Lower middle: The GMRT map of 3C\,388 at 240 MHz; the contour levels in the map are
($-$200, 200, 400, 800, 1600, 3200, 4800) mJy~beam$^{-1}$.
Upper right and Lower right panels: The distribution of the spectral index,
between 1.4~GHz and 610 MHz (upper right),
and 240~MHz and 610 MHz (lower right), for the source.
The spectral index range displayed in the two maps are
$-$1.8 and 0.8 (upper right), and $-$1.3 and $-$0.7 (lower right), respectively.
The spectral index contours are at $-$1.2, $-$0.8, $-$0.6,~0.0 and
$-$1.2, $-$0.7, $-$0.2,~0.0, respectively in the two maps.
The spectral indices listed for various regions are
tabulated in Table~\ref{fdregions}.
The r.m.s. noise values in the radio images found at a source free location
are $\sim$0.7, $\sim$1.4 and $\sim$13.9~mJy~beam$^{-1}$ at
1.4~GHz, 610~MHz and 240~MHz, respectively.
The uniformly weighted CLEAN beams for upper and lower panel maps are
6.0~arcsec $\times$~5.0 arcsec
at a P.A. of $+$83.1$^{\circ}$
and
11.4~arcsec $\times$~9.6 arcsec
at a P.A. of $+$25.,1$^{\circ}$ respectively.
}
\label{full_syn_388}
\end{figure*}

\begin{figure*}
\begin{center}
\begin{tabular}{lll}
\includegraphics[width=6.0cm]{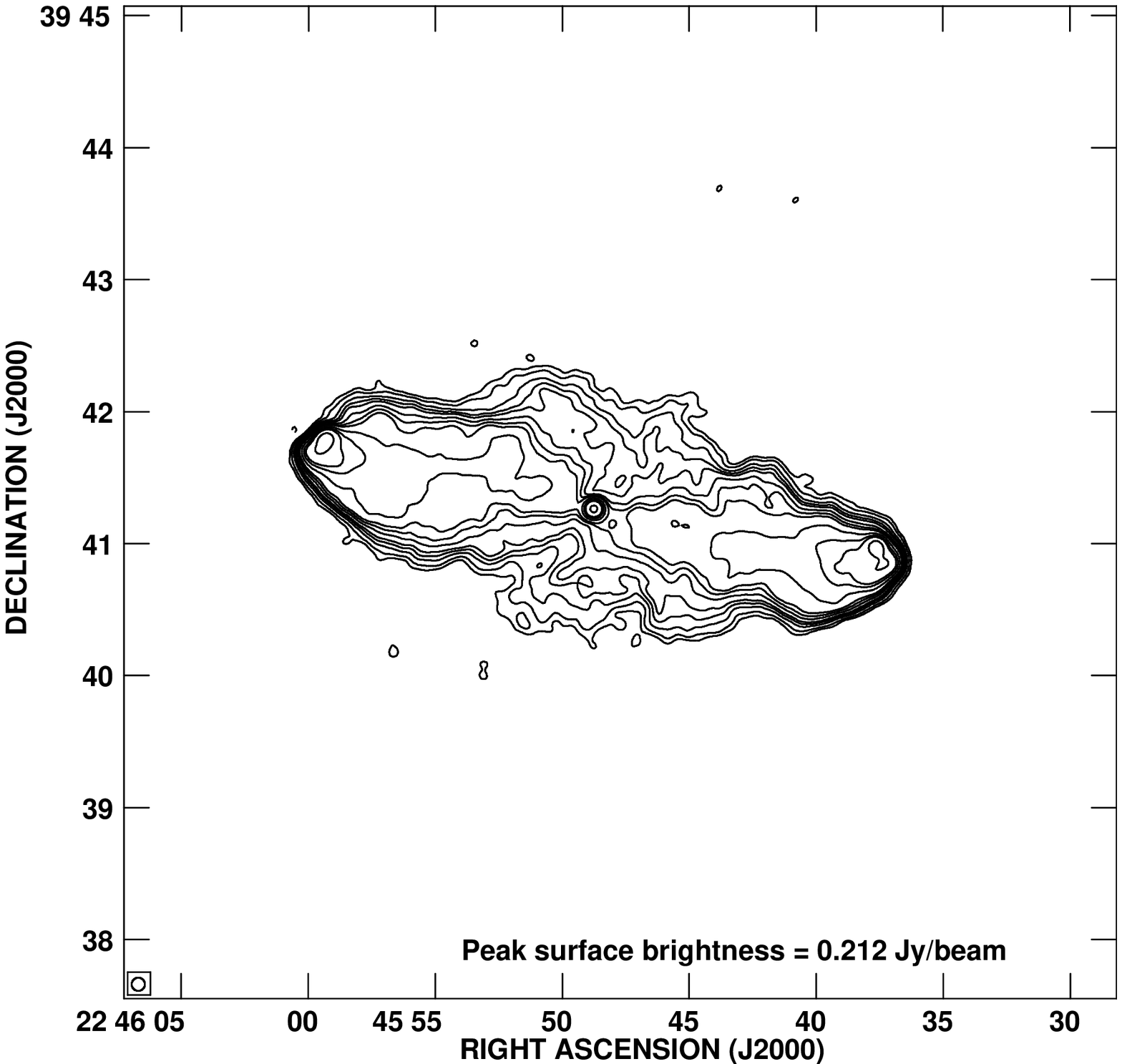} &
\hspace*{-0.2cm}\includegraphics[width=5.24cm]{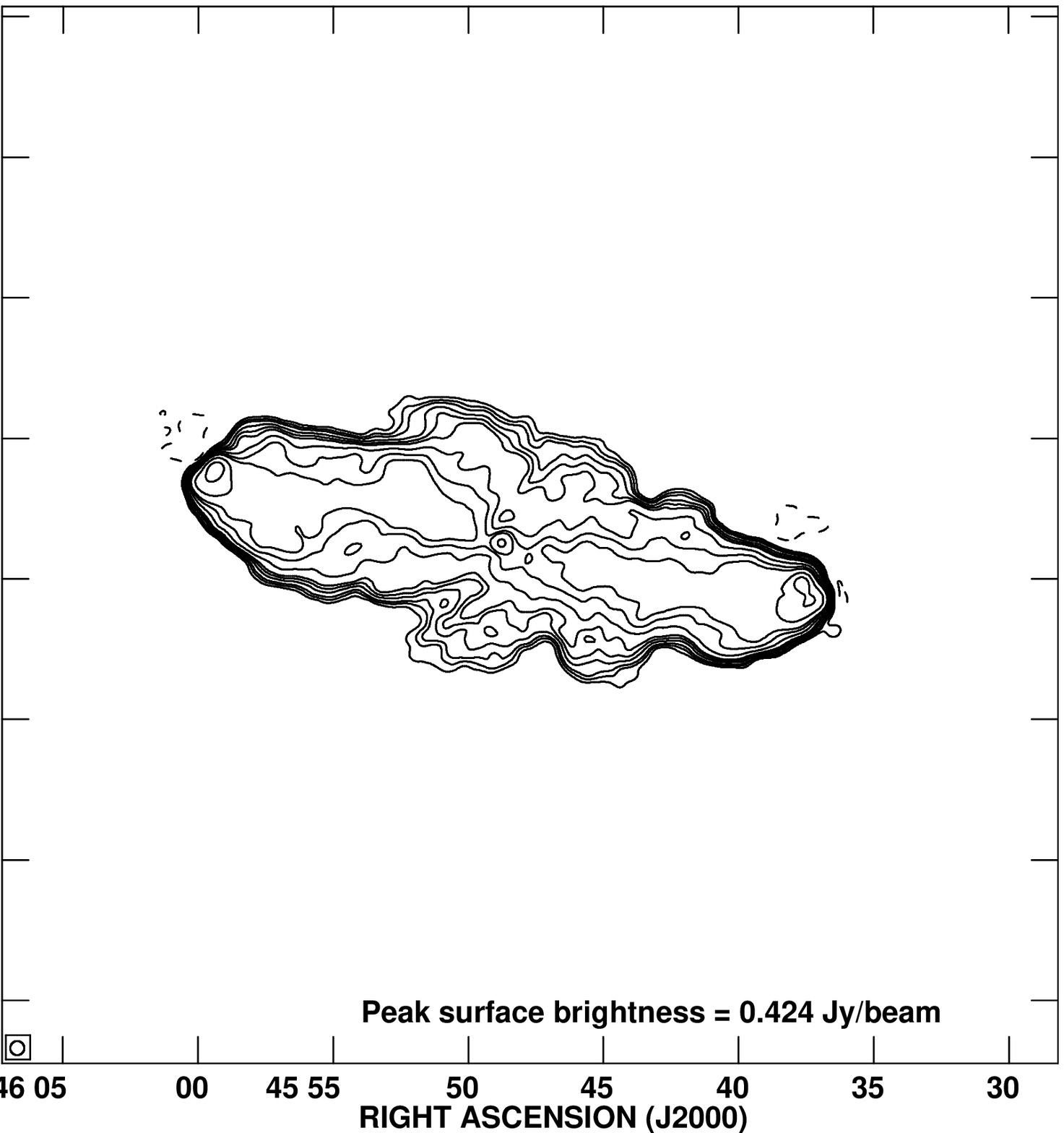} &
\hspace*{-0.2cm}\includegraphics[width=5.24cm]{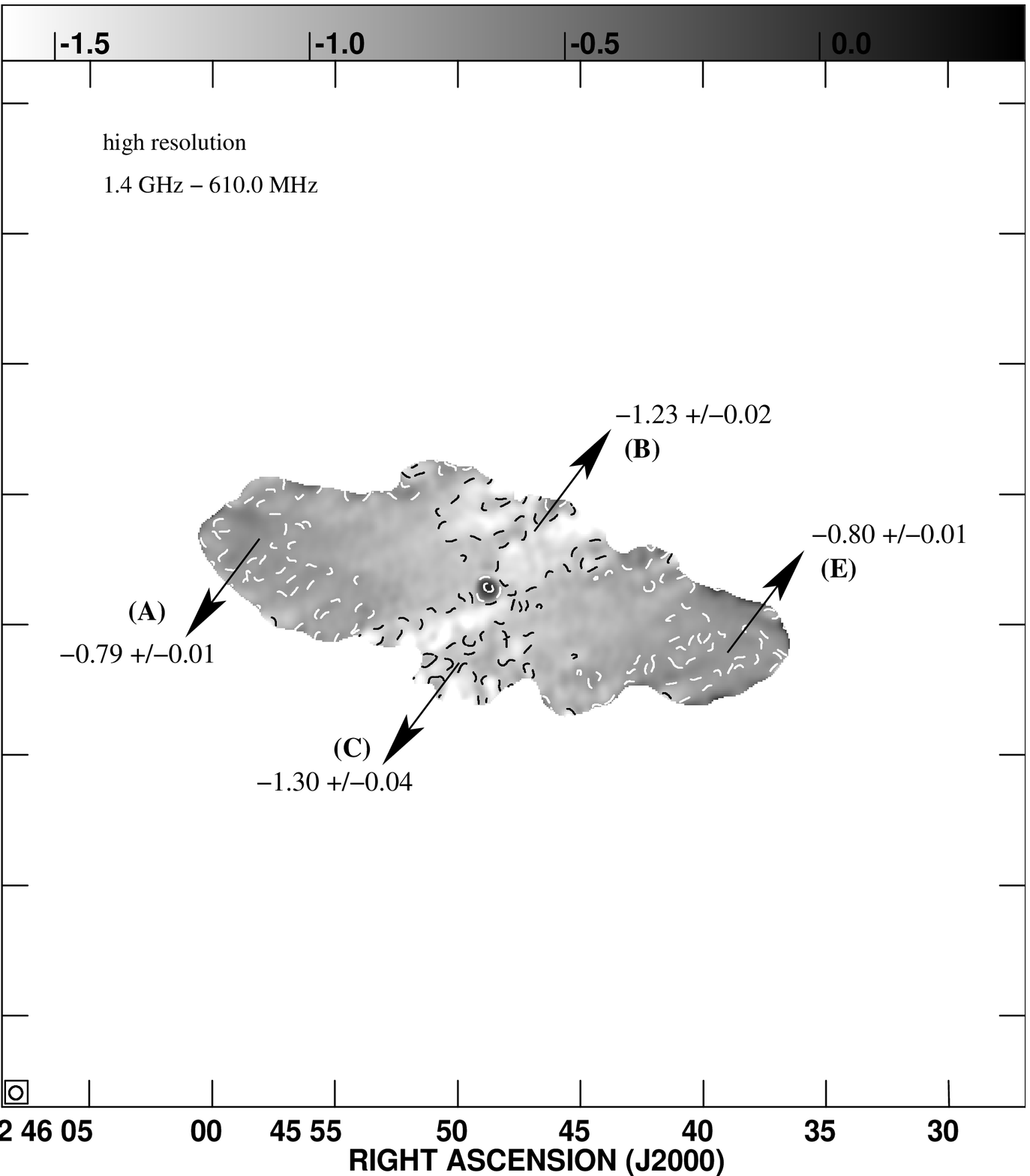} \\
\includegraphics[width=6.0cm]{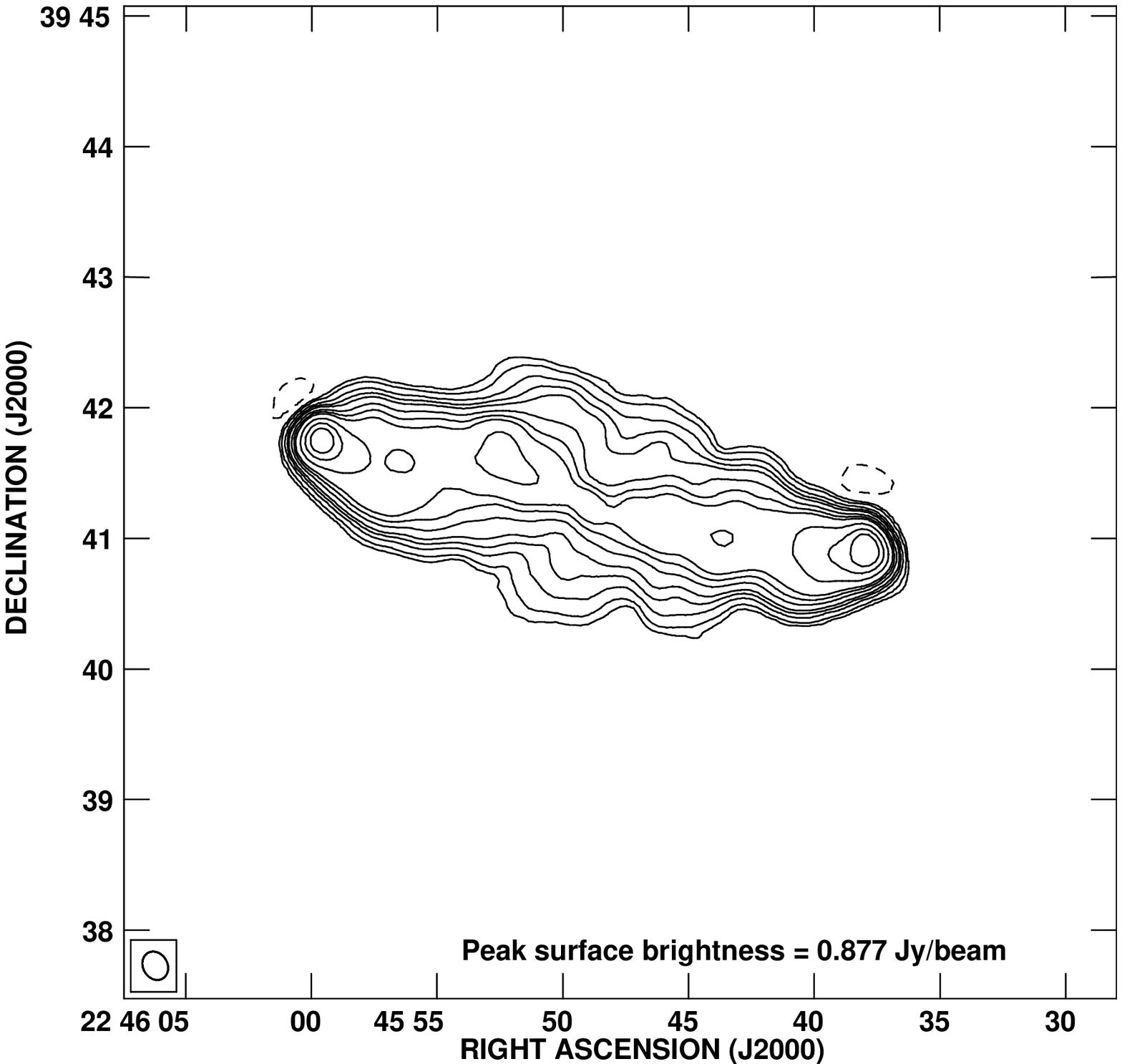} &
\hspace*{-0.2cm}\includegraphics[width=5.24cm]{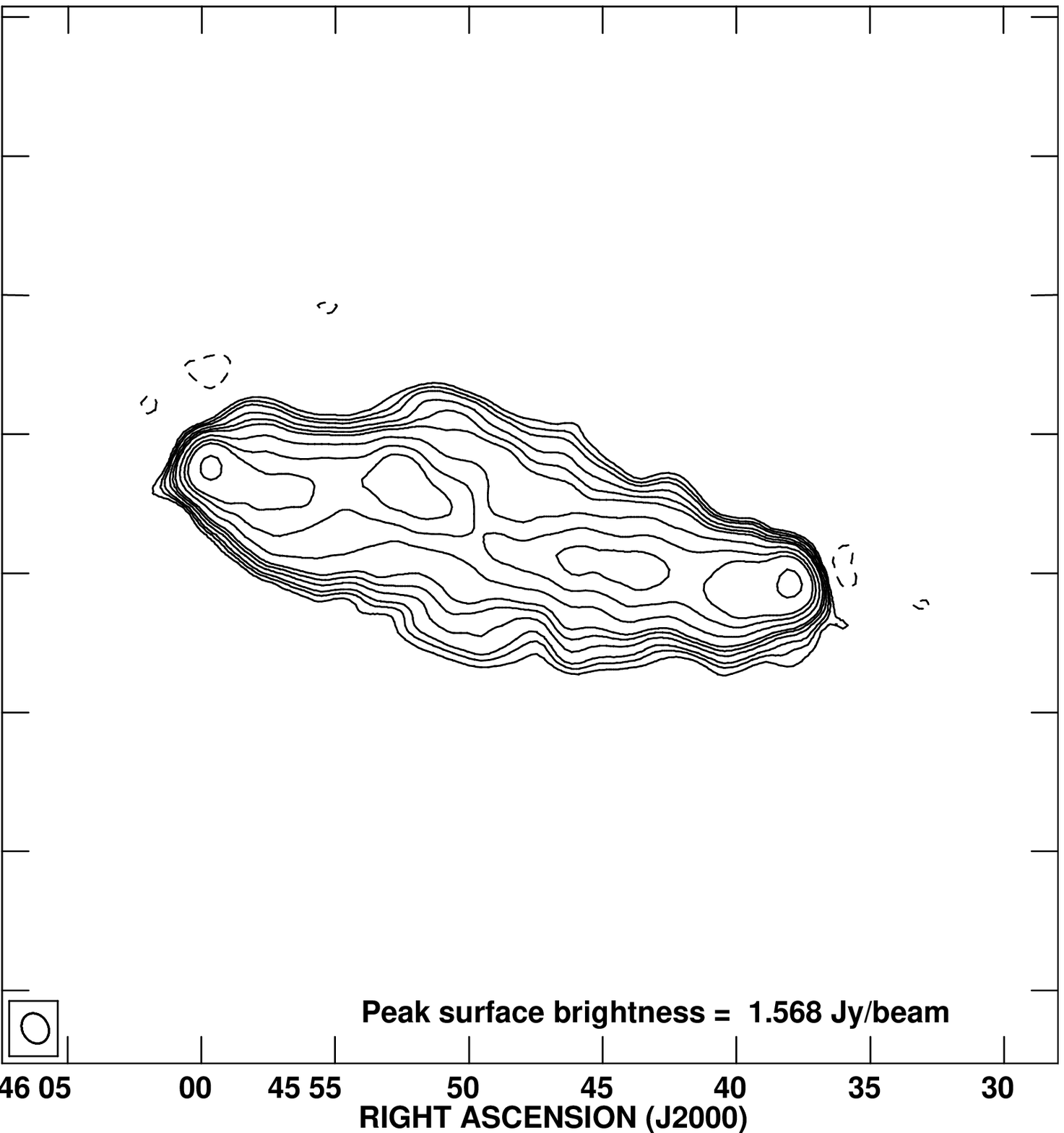} &
\hspace*{-0.2cm}\includegraphics[width=5.24cm]{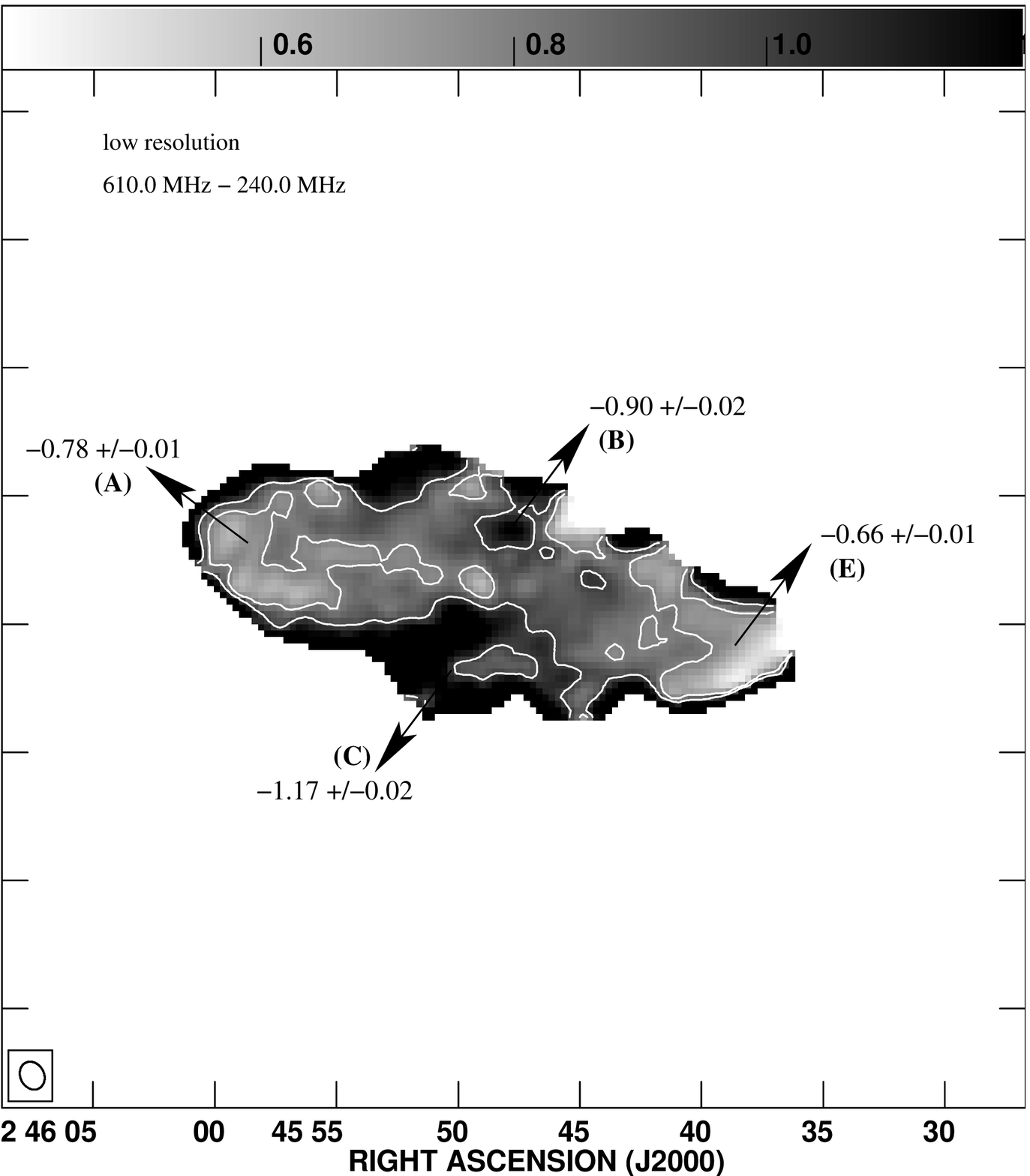} \\
\end{tabular}
\end{center}
\caption{Radio maps of 3C\,452.
Upper left: The VLA map of 3C\,452 at 1.4 GHz
matched with the resolution of 610~MHz; the contour levels in the map are
($-$1, 1, 2, 4, 6, 8, 12, 16, 20, 32, 48, 64, 128) mJy~beam$^{-1}$.
Upper middle: The GMRT map of 3C\,452 at 610 MHz; the contour levels in the map are
($-$4, 4, 6, 8, 12, 16, 20, 32, 48, 64, 128, 256) mJy~beam$^{-1}$.
Lower left: The GMRT map of 3C\,452 at 610 MHz 
matched with the resolution of 240~MHz; the contour levels in the map are
($-$10, 10, 20, 40, 60, 80, 120, 160, 200, 320, 480, 640) mJy~beam$^{-1}$.
Lower middle: The GMRT map of 3C\,452 at 240 MHz; the contour levels in the map are
($-$40, 40, 60, 80, 120, 160, 200, 320, 480, 640, 1280) mJy~beam$^{-1}$.
Upper right and Lower right panels: The distribution of the spectral index,
between 1.4~GHz and 610 MHz (upper right),
and 240~MHz and 610 MHz (lower right), for the source.
The spectral index range displayed in the two maps are
$-$1.6 and 0.4 (upper right), and 0.4 and 1.2 (lower right), respectively.
The spectral index contours are at $-$1.2, $-$0.8, ~0.0 and
0.8,~1.0, respectively in the two maps.
The spectral indices listed for various regions are
tabulated in Table~\ref{fdregions}.
The r.m.s. noise values in the radio images found at a source free location
are $\sim$0.2, $\sim$0.6 and $\sim$3.1~mJy~beam$^{-1}$ at
1.4~GHz, 610~MHz and 240~MHz, respectively.
The uniformly weighted CLEAN beams for upper and lower panel maps are
6.0~arcsec $\times$~6.0 arcsec
at a P.A. of 0.0$^{\circ}$
and
13.6~arcsec $\times$~11.2 arcsec
at a P.A. of $+$29.7$^{\circ}$, respectively.
}
\label{full_syn_452}
\end{figure*}

The flux densities at 240 MHz and 610~MHz plotted in Fig.~\ref{fluxint}
are calculated using the images shown in Figs.~\ref{full_syn_33} to
\ref{full_syn_452} (upper left and upper middle panels, and lower left
and lower middle panels), which are matched to the same resolution,
and these values are tabulated in Table~\ref{flux_density}. The flux
densities tabulated in Table~\ref{fdregions} correspond to
the regions marked on Figs.~\ref{full_syn_33} to~\ref{full_syn_452}.
Our regions are representative of the spectral indices of the radio
hotspots and lobes and the faint diffuse features, and are integrated
using the {\sc aips} task `IMEAN' over an appropriate region (a
circular region of $\sim$5 pixels radius centred at the position of
the tail of the arrows plotted on the maps). The regions used are 
at least four times the beam size  to reduce statistical errors,
and each of the regions is well within the contour level corresponding
to three times the local r.m.s. noise.
In addition, we show spectral results based on two error estimates;
(i) the estimates based on the noise at a source-free location
using a similar-sized circular region, and
(ii) conservative estimates of errors on the flux densities at each
location, which were determined from the fluctuations in the
region being averaged.
The former is generally much smaller than the latter (see the figure captions
and Table~\ref{fdregions} ). These error bars, both on spectral indices and
flux densities, do not change significantly when we increase or decrease
the size of the circles, and they also do not change significantly if
we change slightly the position of the circular region.

We have also examined the possibility that (i) the different ($u,v$)
coverages, (ii) the negative depression seen around some sources due to
undersampling or inadequate deconvolution, and (iii)
the image misalignments at two frequencies could produce some
systematic errors. Possibility (i) seems unlikely since the GMRT has good
($u,v$) coverage, and sources are only $\sim$3 arcmin to 4 arcmin across
and are much smaller than the fringe spacing of the shortest baseline lengths,
$\sim$35 arcmin ($\simeq$100 wavelengths) at 610 MHz and
$\sim$100 arcmin ($\simeq$35 wavelengths) at 240 MHz. Nevertheless, we
investigated this possibility by
Fourier transforming some of the 240~MHz CLEAN maps, sampling them with the ($u,v$)
coverage of 610~MHz and re-imaging the resulting visibility data sets. The
resultant map showed no systematic differences from the original
240~MHz map and the r.m.s. difference in the two maps was less than 4
per cent,
corresponding to an r.m.s. error in the spectral index of
$\lesssim$0.05. Furthermore, the 240~MHz map of 3C\,388
(Fig.~\ref{full_syn_388}) shows marginal evidence that this image
contains a negative depression around it. In general a negative
depression/bowl of this type is due either to inadequate ($u,v$) coverage
at short baseline lengths or to inadequate CLEANing. The former is
unlikely as explained above, while the CLEANing in our imaging is deep
precisely so as to avoid such deconvolution errors. Comparisons of the
expected integrated flux density, the total CLEANed flux density and the flux
density measured on short baselines suggest that any discrepancy is at
most 10 per cent for 3C\,388. In any case, we quantify the errors that would be
introduced due to possible negative depression for this source below.
Both these possibilities, (i) and (ii) are unlikely in case of VLA
1.4 to 1.5 GHz images; this is because of very high sensitivity images
with very high signal-to-noise ratio.
As for possibility (iii), in the low resolution GMRT images at
240 MHz and 610 MHz, we not only registered the target source,
but also registered the positions of at least five field sources
around the radio source which were seen in both maps. We estimate
that the two GMRT images are typically aligned to better than 1.6~arcsec.
In a similar manner, for the high-resolution 1.4 to 1.5~GHz VLA images and the
610 MHz GMRT images, where we could register only the target source, we
estimate that the two images are typically aligned to better than 0.6~arcsec.
Therefore, we feel confident that our spectral index maps of the
sources account well for all possible sources of systematic error.

The high-frequency radio structures of the sources in our sample are
discussed in numerous papers published in the literature; see e.g.
\citet{LeahyPerley1991,LeahyPerley95,Leahy1997,Hetal2007,Evans2008}
for a detailed account of radio morphologies. Since we do not detect
any new or remarkable features that need to be reported
(morphologically, these sources appear very similar at low and high
frequencies, as discussed by \citealt{Blundell08}),
we present
here mainly the spectral results. Analysis of the spectra, shown in
Figs.~\ref{full_syn_33} to \ref{full_syn_452} (upper right and lower
right panels), in different regions of each of these objects shows
strong variation across the sources. The lighter regions represent
relatively steep spectra as compared to the darker regions which
represent flat spectra. Although the full range of spectral index is
large, we have shown only a small range for clarity in each case.
In all our sample sources, we find that in the
hotspots/lobes there is monotonic steepening of the radio spectrum
from the lobes to the low surface brightness features, a classical
spectral signature seen in almost all normal FR\,II radio galaxies at higher
frequencies \citep{Alexander1987}. Our spectral index measurements across
several representative regions within the source are labelled in each spectral
index map from Figs.~\ref{full_syn_33} to \ref{full_syn_452} (upper
right and lower right panel maps) and the corresponding flux density
values are tabulated in Table~\ref{fdregions}. In the following
subsections we describe the spectral structure of radio sources
measured by GMRT for possible features not seen in our 240~MHz radio
maps, and comment on any unusual spectral features seen.

\paragraph*{3C\,33 (z = 0.060)}

The core has a spectral index of 0.00~$\pm$0.06 between 1.5~GHz and 610 MHz.
The 240~MHz map is dynamic range limited, so that
we detect only the north and the south hotspots at this frequency.

\paragraph*{3C\,98 (z = 0.031)}

The core has a spectral index of $-$0.69~$\pm$0.01 between 1.5 GHz and
610 MHz. The spectral index is $-$0.38~$\pm$0.03 between 610 MHz and
240 MHz at the location of the core.

\paragraph*{3C\,285 (z = 0.079)}

The core is relatively steep-spectrum as compared to the two hotspots
and has a spectral index of $-$1.02~$\pm$0.01 between 1.5~GHz and 610 MHz.
Faint features to the North-West of the East lobe and
the South-East of the West lobe in the
610 MHz maps are artefacts due to dynamic-range limitations of the data.
Though the core is not detected in the 240~MHz maps, the spectral index
is $-$0.80~$\pm$0.02 at the location of the core between 610~MHz and 240 MHz.

\paragraph*{3C\,321 (z = 0.100)}

The core has a spectral index of $-$0.99~$\pm$0.08 between 1.5 GHz and
610 MHz and $-$0.44~$\pm$0.02 between 610 MHz and 240 MHz.

This source shows unusual jet radio morphology,
which has been suggested to be due to an interaction with the companion galaxy
(Evans et~al. 2008). The low-surface-brightness feature emanating from
the core towards the north surprisingly has an unusual flat spectral
index of $-$0.30~$\pm$0.03 between 1.5~GHz and 610~MHz.
Although this faint feature is undetected in our 240~MHz map, the spectral
index between 610~MHz and 240~MHz should be flatter than $-$1.22~$\pm$0.21.

\paragraph*{3C\,382 (z = 0.058)}

The core has a spectral index of $+$0.16~$\pm$0.01 between 1.5~GHz and 610~MHz.
Although, it is not detected in the 240~MHz map,
the spectral index is $-$0.24~$\pm$0.02 at the location of core
between 610~MHz and 240~MHz.

\paragraph*{3C\,390.3 (z = 0.056)}

Most of the faint features show a steep spectral index, but
the faint feature to the north-east has a slightly flat spectral index
of $-$0.64~$\pm$0.01 between 1.5~GHz and 610~MHz.
The core has a spectral index of $-$0.79~$\pm$0.01 between 1.5~GHz and 610~MHz.
Although the core is unclear in the 240 MHz map,
the spectral index is $+$0.19~$\pm$0.02 at the location of the core
between 610~MHz and 240~MHz.

\paragraph*{3C\,388 (z = 0.092)}

This is the smallest angular sized source in our sample ($\sim$1~arcmin across)
and hence we detect only the two lobes.

\paragraph*{3C\,452 (z = 0.081)}

Although the core is unclear in all our maps, 
the spectral index is $-$0.08~$\pm$0.01, $-$0.61~$\pm$0.02
at the location of the core
between 1.4~GHz and 610~MHz, and 610~MHz and 240~MHz, respectively.

\section{Discussion}
\label{discussion}

The low frequency radio maps presented above show morphologies
that are similar to the morphologies at high frequencies
(cf. \citealt{Blundell08}).
Our spectral index mapping and measurement show that in general in
these normal FR\,II radio galaxies we do not find any evidence that
low-surface-brightness features have flatter spectral indices than the
high-surface-brightness active lobes; this is true even of the weak
`wings' seen in sources like 3C\,285, 3C\,382 and 3C\,452, and contrasts
with the population of X-shaped sources studied by \citet{LalRao2007}.

The one exception to this trend is 3C\,321, where a
low-surface-brightness feature, very similar to one `wing' of known
X-shaped sources, shows a flat spectral index between 610~MHz and
1.5~GHz as compared to the high-surface-brightness lobes. By
definition, true X-shaped sources show two symmetrical wings and the
wings are of similar angular size to the `active' lobes; marginal
differences between two axes, {\it e.g.}, in 3C\,192 and 3C\,403, may be
due to projection effects. In 3C\,321, the `wing' is somewhat smaller
than the active lobes, but again this is possibly due to projection.
The absence of a corresponding wing on the southern lobe is however harder to
account for: from the 1.4 GHz data we estimate that any non-detected
wing on this side must be at least 25 times fainter than the detected
feature.

If the `wing' of 3C\,321's northern lobe is indeed identical to the
`wings' seen in X-shaped radio sources, then its flat spectral index
suggests that it may have something in common with the class of
X-shaped sources, discussed by \citet{LalRao2007}, in which the
spectral index of the wing is flatter than that of the active lobe. In
this case the \citet{LalRao2007} model, in which the active galaxy
consists of two pairs of jets that are associated with two unresolved
AGNs, may be a possible model for 3C\,321 as well. Consistent with
this scenario, \citet{Evans2008} noted that the host galaxy of 3C\,321
and the companion galaxy are in the process of merging, and each hosts
an AGN that is luminous in the X-ray. However, the two AGNs in this
system are resolved in the radio, and only one (the one hosted by the
larger galaxy) has a radio detection, implying a current jet; in
addition, the one-sided nature of the wing in 3C\,321 is hard to
explain in this model. Other explanations are possible for the wing.
\citet{Evans2008} have shown that the inner radio morphology of the
source is atypical for an FR\,II radio source; a small-scale jet
emerges from the nucleus of its host galaxy, produces a knot of radio
emission that lies immediately to the south (in projection) of a
smaller companion galaxy, and then flares and bends into a diffuse
structure towards the north \citep{Evans2008}. Thus it is possible
that the peculiar small-scale structure is in some way related to the
large-scale wing (although Evans et al.\ argue that the small-scale
disruption of the jet is a transient phenomenon and cannot have been
going on for long enough to produce the observed wing). Although
the environment of the merging host galaxy is complex and might be
expected to produce disrupted lobe structures by purely hydrodynamical
processes, this -- as with the X-shaped
sources -- does not in itself explain the anomalously flat spectra found
in our analysis.

\section{Conclusions}
\label{conclusions}

We have presented results from a GMRT and VLA study of
a sample of eight nearby FR\,II radio galaxies that are
qualitatively matched with the sample of known X-shaped radio sources
in size, morphological properties and redshift.
The radio measurements presented here
represent most of the database that we require for rigorously testing
and understanding the formation models of X-shaped radio sources.
Our conclusions from our new observations can be summarised as follows:
\begin{enumerate}
\item[{1.}] The radio morphologies of the sample sources at low radio
frequencies are largely identical to the morphologies seen at
high frequencies using the VLA.
\item[{2.}] In almost all our sample sources we find that in the
hotspots and lobes there is monotonic steepening of the radio spectrum
from the hotspots to the low surface brightness features, a classical
spectral signature seen in almost all normal FR\,II radio galaxies.
\item[{3.}] The spectral results for FR\,II radio sources are largely
consistent with the spectral results of active lobes in X-shaped radio
sources; therefore there is something `special' about the wings of X-shaped
sources, in the sense that they do not simply behave like the
low-surface-brightness regions of more typical FR\,II sources.
\item[{4.}] For the hydrodynamic model (\citealt{LeahyWilliams};
\citealt{Worralletal}; \citealt{Capettietal}; \citealt{Kraftetal})
to remain viable in the light of this result,
the spectral flattening in the wings must be produced by some other
process, such as reacceleration at internal shocks. Detection of
shock features in the wings might support the hydrodynamic model.
The hydrodynamic model is also faced with the problem of the
rotational symmetry of X-shaped radio galaxies \citep{Rottmann}.
\item[{5.}] Since the low surface brightness feature in 3C\,321,
a classical FR\,II radio source, also
shows unusual spectral behaviour, similar to the spectral behaviour
seen in wings in some of the X-shaped sources, it seems that
3C\,321 may have some similarity to X-shaped sources.
\item[{6.}] This similarity raises the
possibility that 3C\,321 consists of two
pairs of jets, which are associated with two AGNs, a
possible formation model for known X-shaped sources
\citep{LalRao2007}, but this model has some difficulties; there is no
completely satisfactory explanation for the `wing' of 3C\,321.

\end{enumerate}

\section*{Acknowledgments}

DVL is grateful to A.P.~Rao for discussions and several useful
comments.  DVL also thanks A.L. Roy for carefully reading the manuscript
and for helpful discussions and suggestions.
MJH thanks the Royal Society for a research fellowship.
We thank the anonymous referee for helpful comments which improved the paper.
We also thank the staff of the GMRT who have made these observations
possible. GMRT is run by the National Centre for Radio Astrophysics
of the Tata Institute of Fundamental Research.
The National Radio Astronomy Observatory is a
facility of the National Science Foundation operated under cooperative
agreement by Associated Universities, Inc.
This research has made use of the NASA/IPAC Extragalactic Database,
which is operated by the Jet Propulsion Laboratory,
Caltech, under contract with the NASA, and
NASA's Astrophysics Data System.

%\bsp

\label{lastpage}

\end{document}